\documentclass[arguments, twocolumn, 10 point font, single spaced article, twocolappendix]{aastex631}

\usepackage{graphicx}	
\usepackage{amsmath}	
\usepackage{amssymb}	
\usepackage{xspace}
\usepackage[utf8]{inputenc}
\usepackage{ae,aecompl}
\usepackage{comment}
\usepackage{natbib}
\usepackage{float}
\usepackage{graphicx}
\usepackage{amssymb}
\usepackage{rotating,times,pictex,graphicx,latexsym}
\usepackage{color}
\usepackage{amsmath}
\usepackage{threeparttable}
\usepackage{lipsum}
\usepackage{booktabs}
\usepackage{amsmath}
\usepackage{bm}
\usepackage[all]{hypcap}
\usepackage[title,titletoc]{appendix}
\usepackage[]{hyperref}
\PassOptionsToPackage{pdfpagelabels=false}{hyperref}
\usepackage[T1]{fontenc}
\usepackage{changepage}
\usepackage{placeins}

\newcommand{\Ms}{\ensuremath{M_{\odot}}}

\newcommand{\beq}{\begin{equation}}
\newcommand{\eeq}{\end{equation}}

\newcommand{\hii}{\mbox{H~{\sc ii}~}}

\newcommand{\x}{\,\ensuremath{\times}\,}

\newcommand{\arcs}{\hbox{$^{\prime\prime}$}}

\newcommand{\lsun}{\mbox{\rm $L_{\odot}$}}

\newcommand{\degree}{\mbox{$^{\circ}$}}

\newcommand{\cms}{\hbox{cm$^{-2}$~}}

\newcommand{\cmq}{\hbox{cm$^{-3}$}}

\newcommand{\alf}{Alfv$\acute{\text{e}}$nic}

\newcommand{\ngc}{{\rm NGC6334I}}

\begin{document}

\title[Polarization in NGC6334I]{
\center
\Large{Magnetic Fields in Massive Star-forming Regions (MagMaR). IX. Radiative Torque Alignment and Disruption in NGC6334I}}

\author[0000-0001-6515-2863]{Vineet Rawat}
\affiliation{Korea Astronomy and Space Science Institute (KASI), 776 Daedeokdae-ro, Yuseong-gu, Daejeon 34055, Republic of Korea}

\author[0000-0003-2017-0982]{Thiem Hoang}
\affiliation{Korea Astronomy and Space Science Institute (KASI), 776 Daedeokdae-ro, Yuseong-gu, Daejeon 34055, Republic of Korea}
\affiliation{Korea University of Science and Technology, 217 Gajeong-ro, Yuseong-gu, Daejeon 34113, Republic of Korea}

\author[0000-0002-7125-7685]{Patricio Sanhueza}
\affiliation{Department of Astronomy, School of Science, The University of Tokyo, 7-3-1 Hongo, Bunkyo, Tokyo 113-0033, Japan}

\author[0000-0003-3017-4418]{Ian W. Stephens}
\affiliation{Department of Earth, Environment, and Physics, Worcester State University, Worcester, MA 01602, USA}

\author[0000-0001-5811-0454]{Manuel Fernández López}
\affiliation{Institut de Ci\`encies de l’Espai (ICE-CSIC), Campus UAB, Carrer de Can Magrans S/N, E-08193 Cerdanyola del Valles, Catalonia, Spain}

\author[0000-0002-3583-780X]{Paulo Cort\'{e}s}
\affiliation{Joint ALMA Observatory, Alonso de C\'{o}rdova 3107, Vitacura, Santiago, Chile}
\affiliation{National Radio Astronomy Observatory, 520 Edgemont Road, Charlottesville, VA 22903, USA}

\author[0000-0001-7866-2686]{Jihye Hwang}
\affiliation{Institute for Advanced Study, Kyushu University, Japan}
\affiliation{Department of Earth and Planetary Sciences, Faculty of Science, Kyushu University, Nishi-ku, Fukuoka 819-0395, Japan}

\author[0000-0002-4774-2998]{Junhao Liu}
\affiliation{School of Astronomy and Space Science, Nanjing University, 163 Xianlin
Avenue, Nanjing, Jiangsu 210023, People’s Republic of China.}
\affiliation{Key Laboratory of Modern Astronomy and Astrophysics (Nanjing
University), Ministry of Education, Nanjing, Jiangsu 210023, People’s
Republic of China.}

\author[0000-0002-6752-6061]{Kaho Morii}
\affiliation{Center for Astrophysics | Harvard \& Smithsonian, 60 Garden Street, Cambridge, MA 02138, USA}

\author[0000-0002-3829-5591]{Josep Miquel Girart}
\affiliation{Institut de Ci\'{e}ncies de l’Espai (ICE, CSIC), Can Magrans s/n, 08193, Cerdanyola del Vall\'{e}s, Catalonia, Spain}
\affiliation{
Institut d’Estudis Espacials de Catalunya (IEEC), 08860 Castelldefels, Catalonia, Spain}

\author[0000-0001-7379-6263]{Ji-hyun Kang}
\affiliation{Korea Astronomy and Space Science Institute (KASI), 776 Daedeokdae-ro, Yuseong-gu, Daejeon 34055, Republic of Korea}

\author[0000-0002-8691-4588]{Yu Cheng}
\affiliation{National Astronomical Observatory of Japan, 2-21-1 Osawa, Mitaka, Tokyo 181-8588, Japan}

\author[0000-0003-1275-5251]{Shanghuo Li}
\affiliation{Key Laboratory of Modern Astronomy and Astrophysics, Nanjing University, Ministry of Education, Nanjing 210023, People’s Republic
of China}

\author[0000-0003-2343-7937]{Luis A. Zapata}
\affiliation{Instituto de Radioastronomía y Astrofísica, Universidad Nacional Autónoma de México, 58090, Morelia, Michoacán, México}

\author[0000-0003-0014-1527]{Eun Jung Chung}
\affiliation{Korea Astronomy and Space Science Institute (KASI), 776 Daedeokdae-ro, Yuseong-gu, Daejeon 34055, Republic of Korea}

\author[0000-0002-8250-6827]{Fernando A. Olguin}
\affiliation{Center for Gravitational Physics, Yukawa Institute for Theoretical Physics, Kyoto University, Kitashirakawa Oiwakecho, Sakyo-ku,
Kyoto 606-8502, Japan}
\affiliation{National Astronomical Observatory of Japan, 2-21-1 Osawa, Mitaka, Tokyo 181-8588, Japan}

\author[0000-0002-9774-1846]{Huei-Ru Vivien Chen}
\affiliation{Institute of Astronomy and Department of Physics, National Tsing Hua University, Hsinchu 300044, Taiwan}

\author[0000-0002-0028-1354]{Piyali Saha}
\affiliation{Academia Sinica Institute of Astronomy and Astrophysics, No.1, Sec. 4., Roosevelt Road, Taipei 10617, Taiwan}

\author[0000-0003-4761-6139]{Chakali Eswaraiah}
\affiliation{Department of Physical Sciences, Indian Institute of Science Education and Research (IISER) Mohali, Knowledge City, Sector 81, SAS Nagar 140306, Punjab, India}

\author[0000-0003-2384-6589]{Qizhou Zhang}
\affiliation{Center for Astrophysics | Harvard \& Smithsonian, 60 Garden Street, Cambridge, MA 02138, USA}

\author[0009-0001-2896-1896]{O.R. Jadhav}
\affiliation{Astronomy \& Astrophysics Division, Physical Research Laboratory, Navrangpura, Ahmedabad 380009, India}
\affiliation{Indian Institute of Technology Gandhinagar Palaj, Gandhinagar 382355, India}

\begin{abstract}
Intense radiation from high-mass stars is expected to significantly affect dust grain alignment and evolution through RAdiative Torques (RATs). We investigate this effect in a massive star-forming region, NGC6334I, using 1.2~mm dust continuum polarization observations from the Atacama Large Millimeter/submillimeter Array. The polarization fraction spans from $\lesssim1\%$ to $\sim10\%$ and decreases with increasing column density, remaining below $2\%$ in dense cores despite high temperatures ($\sim100$~K), where efficient grain alignment by RATs is expected. We investigate how grain alignment, grain growth, grain disruption, B-field tangling, and local physical conditions affect the polarization properties of MM1, MM2, MM3, and their surroundings. Polarization angle dispersion shows that B-field tangling contributes to depolarization at moderate densities but cannot fully explain the lowest polarization fractions. Using RAT-based grain alignment and polarization modeling, we find that reduced alignment efficiency and high optical depth reproduce the low polarization in the densest regions. MM2 shows evidence of grain growth, with maximum grain sizes $a_{\max}\sim0.35\text{--}1.0~\mu$m, while MM1 exhibits smaller values of $\sim0.35\text{--}0.50~\mu$m. Accounting for optical depth increases the inferred grain sizes in MM1 to $\sim1.0\text{--}2.0~\mu$m. Analytical estimates of radiative torque disruption from the intense outburst suggest that micron-sized grains in high-temperature, moderate-density regions can fragment into submicron grains. Alternatively, high optical depth may also explain the low polarization in the densest regions even in the presence of micron-sized grains. Incorporating the B-field inclination effect indicates a transition from predominantly plane-of-sky fields at low densities to more line-of-sight-aligned configurations at high densities. 
      
\end{abstract}



\section{Introduction}
\label{sec:int}

Dust plays a very important role in the thermodynamics of the interstellar medium (ISM) and star formation. One of the prominent effects of dust grains is the scattering, absorption, and reemission of light from distant stars. \cite{Hall_1949} and \cite{Hilt_1949} showed that dust being non-spherical and systematically aligned in the ISM polarizes the background starlight. 
These findings provided evidence that dust polarization can be used to trace interstellar magnetic fields (hereafter B-fields) \citep{Davis_1951}. Furthermore, thermal dust emission polarization \citep{Hild_1988} opened the avenue for probing B-fields in dense and cold regions where young stars are being formed. Together with gravity and turbulence, B-fields are now established to play a crucial role in the star formation process \citep{Crutcher_2012}, spanning from the diffuse ISM to molecular clouds, filaments, and dense cores to protostellar disks. Therefore, observing B-field morphologies and measuring their strengths, along with gravity and turbulence, has now become pivotal in studying the star formation processes \citep{Pattle_2023}.

The alignment process of dust grains involves the alignment of the grain's axis of maximum moment of inertia with its angular momentum through internal relaxation (i.e., internal alignment; \citealt{Purcell.1979}), followed by the alignment of the angular momentum with a preferred direction, such as the B-field, anisotropic radiation field, or gas-flow direction (i.e., external alignment; see \citealt{And_2015, LAH.2015} for reviews). Rotating paramagnetic dust grains (i.e., grains containing unpaired electrons, e.g., silicates) become magnetized through the Barnett effect \citep{Barnett.1915}. When these magnetized grains interact with an external B-field, they undergo Larmor precession, allowing the grain's angular momentum to couple with the B-field. The paramagnetic relaxation within rotating grains dissipates the grain's rotation and gradually aligns it with the B-field \citep{DavisGreenstein.1951}. Another prominent early theory of grain alignment is the stochastic mechanical torque mechanism \citep{Gold_1952a, Gold_1952b}, which arises from random collisions between gas atoms and elongated grains moving in a supersonic gas flow, such as in the presence of outflows. 

For the ISM and star-forming regions, the most widely accepted theory for grain alignment is based on Radiative Torques \citep{Dolginov_1976, Draine_1996, LazHoang.2007}. According to the RAT-Alignment (RAT-A) theory, paramagnetic grains can efficiently align with their long axes perpendicular to B-fields after attaining suprathermal rotation due to RATs, which is referred to as the B-RAT mechanism 
\citep[see review articles by][and references therein]{And_2015, LAH.2015}. Thus, RATs play a key role in grain suprathermal rotation as well as grain alignment. Moreover, superparamagnetic grains with embedded iron inclusions could get perfectly aligned via the Magnetically Enhanced RAT (MRAT) mechanism due to enhanced magnetic relaxation and collisional excitations \citep{Hoang_Laz_2016}, except in the case of intense radiation, where grain trapping by RATs is efficient \citep{Hoang_2025}. Regular mechanical torques arising from gas flow-irregular grain interaction \citep{LazHoang.2007b, Hoangetal.2018MET, Reissl.2023} can also be important for grain alignment in some limited astrophysical environments \citep{Hoangetal.2022}. 


The plane-of-the-sky (POS) component of B-fields in star-forming regions is commonly inferred from thermal dust polarization produced by asymmetric grains aligned with the B-field. While the polarization angle reveals the B-field morphology, the polarization degree provides insights into grain-alignment mechanisms and dust grain properties. As per the RAT-A theory, the dust grain alignment efficiency mainly depends on local gas density, radiation, and grain properties (size, shape, and composition) \citep{Hoang_2008, Hoang_2021a}. In particular, \cite{Hoang_2019a} showed that dust grains subject to an intense radiation field can be spun up to extremely high rotational speeds and undergo centrifugal breakup, a process known as Radiative Torque Disruption (RAT-D). Basically, when the centrifugal stress exceeds the tensile strength of dust grains, they can fragment into smaller grains. A decrease in the population of large, efficiently aligned grains results in a lower polarization degree in regions exposed to stronger radiation (i.e., higher dust temperatures) \citep{Hoang_2019a, Lee_2020}. The combination of RAT alignment and disruption establishes the RAT paradigm \citep{Tram_2022}. To achieve a comprehensive testing of the RAT paradigm, it is necessary to observationally study a variety of star-forming environments across all spatial scales, from diffuse clouds to dense molecular clouds, cores, and protostellar environments. 

The RAT paradigm has been tested in the molecular clouds Musca and OMC-1 \citep{Ngoc_2024}, the interstellar filament G34.43+0.24 \citep{Pravash_2025}, and the photodissociation regions $\rho$ Ophiuchi-A, 30 Doradus, and M17 \citep{Tram_2021, Tram_2022} using single-dish dust polarization observations from the Stratospheric Observatory for Infrared Astronomy (SOFIA) and the James Clerk Maxwell Telescope (JCMT). The results of synthetic polarization modeling of ALMA observations for low- and intermediate-mass protostellar environments are consistent with grains aligned with MRAT \citep{Gouellec_2020, Giang.2024, Giang.2025}.


In star-forming regions, observations show that the polarization fraction tends to decrease with increasing density (or visual extinction), a trend commonly referred to as depolarization \citep[e.g.,][]{Tang_2013, Liu_2019, Liu_2020, Rawat_2024c}. This observed depolarization can be attributed to the loss of grain alignment in dense regions \citep{Hoang_2021a} or to tangled B-fields and the inclination of B-fields with respect to the line-of-sight (LOS) \citep{Planck_2015a, Hull_2017, Hoang_2024}. Along with that, multiple layers with different field morphologies along the LOS, optical depth effects, and flux loss in the sub-millimeter/far-infrared interferometric observations can also contribute to depolarization. Under the RAT paradigm, the polarization fraction is expected to decrease with density due to gas randomization, but to increase with dust temperature as a result of an enhanced radiation field. \cite{Hoang_2021a} predicted that, after the initial decrease in polarization fraction at high density, it can subsequently increase or become nearly flat if an enhanced radiation source is present in the dense region, such as a protostar. We refer to such a rise in polarization degree at high column density and elevated temperature (e.g., near an embedded protostar), following an initial depolarization, as “Repolarization.” Observational evidence for the repolarization trend has been found in some studies \citep[][Ngan et al., in preparation]{Fernandez_2021, S_Kumar_2026}. 

The hot cores of massive star-forming regions, with their high densities and temperatures driven by intense protostellar radiation, provide ideal environments where the aforementioned factors can influence dust properties, including grain alignment, size distribution, and shape. In dense star-forming regions, such as dense clumps and cores, dust grain properties, such as grain size, shape, composition, and porosity, are expected to evolve from those in the diffused ISM due to grain coagulation and gas accretion on grains \citep[e.g., see][]{Chokshi_1993, Ossenkopf_1994}. These changes affect both the dust opacity \citep{Ossenkopf_1994} and grain alignment efficiency \citep{Hoang_2021a}, which leads to an impact on the polarized dust emission \citep{Hild_1988}.  \citet{Hoang.2021} theoretically investigated the impact of massive protostellar radiation on grain alignment and disruption in hot cores, finding that larger grains ($>0.5~\mu$m) can be disrupted by RAT-D near massive protostars. 
Therefore, understanding how dust grains align and evolve in regions under massive stellar feedback is key to interpreting polarization signals, probing B-fields, and surface astrochemistry. The main goal of this paper is to study grain alignment and disruption by RATs in a massive star-forming region, using ALMA dust emission polarization data combined with detailed physical modeling based on the RAT paradigm.

The MagMaR papers so far have primarily focused on studying the B-field morphology and its relative importance compared to gravity and turbulence \citep[e.g.,][]{Sanhueza_2021, Fernandez_2021, Cortes_2021, Cortes_2024, Saha_2024, Zapata_2024, Sanhueza.2025, Hwang_2026, Xu_2026, Fernandez_2026, Liu_2026}, while dust grain properties and alignment physics remain largely unexplored. Here, we extend the scientific scope of MagMaR by presenting the first systematic study of dust properties and dust alignment physics across the sample.

\subsection{Target: \ngc}
We focus on \ngc, an active and nearby massive star-forming region in the Milky Way, which is part of the giant molecular cloud NGC 6334 and is located at a distance of 1.3 kpc \citep{Chibueze_2014}. \ngc~hosts multiple substructures, including MM1—a protocluster region containing hot molecular cores, an outburst source, and prominent outflows; MM2—a hot molecular core and outflows; MM3—a cometary ultracompact \hii region; and MM4—a protostellar core with bipolar outflows, and other substructures (MM5, MM6, MM7, MM8, and MM9) \citep[see][and references therein]{Brogan_2016, Hunter_2021}. These sources are marked in Figure \ref{P_Intensity}. In fact, multiple outflows and maser emissions (OH, CH$_3$OH, and H$_2$O) have been identified throughout the \ngc~region, indicating active star formation and the presence of young stellar objects (YSOs), as well as possible accretion events \citep{Bachiller_1990, McCutcheon_2000, MacLeod_2018, Jayender_2026}. This diversity makes NGC6334I an excellent laboratory for studying the effects of stellar feedback on dust grain alignment. 

The B-field morphology in \ngc~has been studied multiple times at different spatial resolutions, e.g., through SMA \citep[][]{Zhang_2014, Hua-Bai_2015}, JCMT \cite[$\sim14$\arcsec;][]{Arzou_2021}, ALMA 1.4 mm \citep[$\sim0.6$\arcsec;][]{Liu_2023b}, and ALMA 1.2 mm \citep[][]{Cortes_2024}. The B-field properties of some specific regions of \ngc~based on Zeeman observations have also been studied \citep{Hunter_2018, Chanapote_2019, MacLeod_2023}. 
The total mass and luminosity of \ngc~is found to be around 700~\Ms~and 1.5 $\times$ 10$^5$ \lsun, respectively \citep{Brogan_2016}. \cite{Cortes_2024} reported an average B-field strength of 1.9 mG and a total outflow energy (traced by CS ($J=5-4$)) of $\sim3.5 \times 10^{45}$ erg, and found that most of the region is in a trans-\alf/super-\alf\ regime, with magnetic energy lower than the gravitational, thermal, turbulent, and outflow energies. Thus, protostellar outflows can significantly impact the B-field morphology and potentially the dust-grain alignment efficiency in \ngc. 

The paper is organized as follows. In Section \ref{sec:obs}, we present the observational data used in this work. In Section \ref{sec:analysis}, we analyze the dust emission polarization and its variation with polarization angular dispersion, column density, and temperature. In Section \ref{sec:model_theory}, we present the modeling of thermal dust polarization. In Section \ref{sec:discuss}, we discuss the factors responsible for the variation in polarization fraction and the results of the polarization modeling, along with the relevant dust grain alignment physics. We also discuss the limitations of this study and summarize our findings with conclusions in Section \ref{sec:conc}.

\section{Observations}
\label{sec:obs}
We use the ALMA Band 6 (1.2 mm) full Stokes polarization observations of \ngc, which were observed as part of the Magnetic Fields in Massive star-forming Regions (MagMaR) project (code 2017.1.00101.S and 2018.1.00105.S; PI: P. Sanhueza). The MagMaR survey has mapped B-field morphologies in 30 high-mass star-forming regions at an angular resolution of $\sim 0\farcs3$ \citep[for details, see][]{Sanhueza_2021}. The target was observed on 2018 December 13 under conﬁguration C43-4 (12m array), providing baseline lengths from 15 to 783
m. The observational and data reduction details can be found in \cite{Sanhueza_2021}. The final Stokes $I$ image has an overall rms noise level of 1.54 mJy beam$^{-1}$, with a synthesized beam of $\sim0\farcs6$ $\times$ $0\farcs5$, a position angle of $\sim-86\degree$, and a pixel size of $\sim 0\farcs05$. The data were primary beam corrected and debiased on a pixel-by-pixel basis following \citet{Wardle_1974, Hull_2015}. We also use the $\rm{^{13}CH_3OH}$ multiple transition lines with upper-state energies of $\sim 114.94$, 122.33, 287.84, and 321.79 K to derive the gas temperature. 

The debiased polarization intensity ($PI$) and polarization fraction ($P$) are calculated following \cite{Wardle_1974}, and the polarization position angle ($\theta$) and the corresponding uncertainties are calculated as follows
\begin{eqnarray}
    PI &&= \sqrt{Q^{2}+U^{2} - 0.5(\delta Q^2 + \delta U^2)},\\
    \delta PI &&= \sqrt{\frac{Q^2 \delta Q^2 + U^2 \delta U^2}{Q^2 + U^2}},\\
    P &&=  \frac{PI}{I}, \\
    \delta P &&= \sqrt{\frac{\delta P I^2}{I^2} + \frac{\delta I^2 (Q^2 + U^2)}{I^4}},\\
    \theta&& =\frac{1}{2}\mathrm{atan2}\left(\frac{U}{Q}\right),\\
    \delta \theta &&= \frac{1}{2} \times \frac{\sqrt{U^2 \delta Q^2 + Q^2 \delta U^2}}{(Q^2 + U^2)}.
\end{eqnarray}
For further analysis in this paper, we selected the data using the following criteria: $I/\delta I > 5$ and $P/\delta P > 3$, where $\delta I$ = 1.54 mJy beam$^{-1}$.

\section{Analysis and Results}
\label{sec:analysis}
\subsection{Dust Emission Polarization and Polarization Angle Dispersion}
\label{sec:pol}

The polarized intensity map of the ALMA 1.2 mm dust emission is shown in Figure \ref{P_Intensity}, and the corresponding dust continuum emission is shown in Figure \ref{intensity} in Appendix \ref{Total_intens}. From Figure \ref{P_Intensity}, it can be seen that there are multiple resolved compact structures within MM1, along with MM4, that have high polarization intensities (3$-$5 mJy beam$^{-1}$). The sources identified in \cite{Brogan_2016} are marked in the figures. 

Figure \ref{pol_map}(a) shows the inferred B-field structure of the region obtained by rotating the Polarization Angles (PAs) by 90\degree~and sampling the pixels within half a beam (beam size $\sim0.5$\arcsec). From the figure, a pinched or spiral-like B-field morphology is apparent in the MM1 protocluster region, together with an almost 90-degree change in the B-field orientation near the western edge of MM1, between MM1 and MM3, and in the MM2 and MM4 regions, as initially reported by \cite{Cortes_2024}. This change in the orientation of the POS B-fields can be attributed to B-field fluctuations driven by gravity or turbulence induced by outflows, as strong outflows are present in \ngc. 

Figure \ref{pol_map}(b) shows the polarization fraction map, which reveals that although the polarized intensities are high in dense regions (see Figure \ref{P_Intensity}), the polarization fractions are still much lower at these locations. The polarization degree, $P$ (\%), ranges from $\sim0.04$\% to $\sim10.5$\% in \ngc, with values relatively very low in MM1 ($\sim0.04$\%--1.6\%), MM2 ($\sim0.06\%-1.2$\%), MM3 ($\sim0.17\%-8.0$\%), and MM4 ($\sim0.16\%-4.5$\%). Thus, a depolarization trend is evident in \ngc, as shown in Figure \ref{pol_map}(b). However, since the ALMA accuracy is $\sim 0.1\%$, the exact $P$ values below 0.1\% are less reliable.  

\begin{figure*}
    \centering
    \includegraphics[width=12cm]{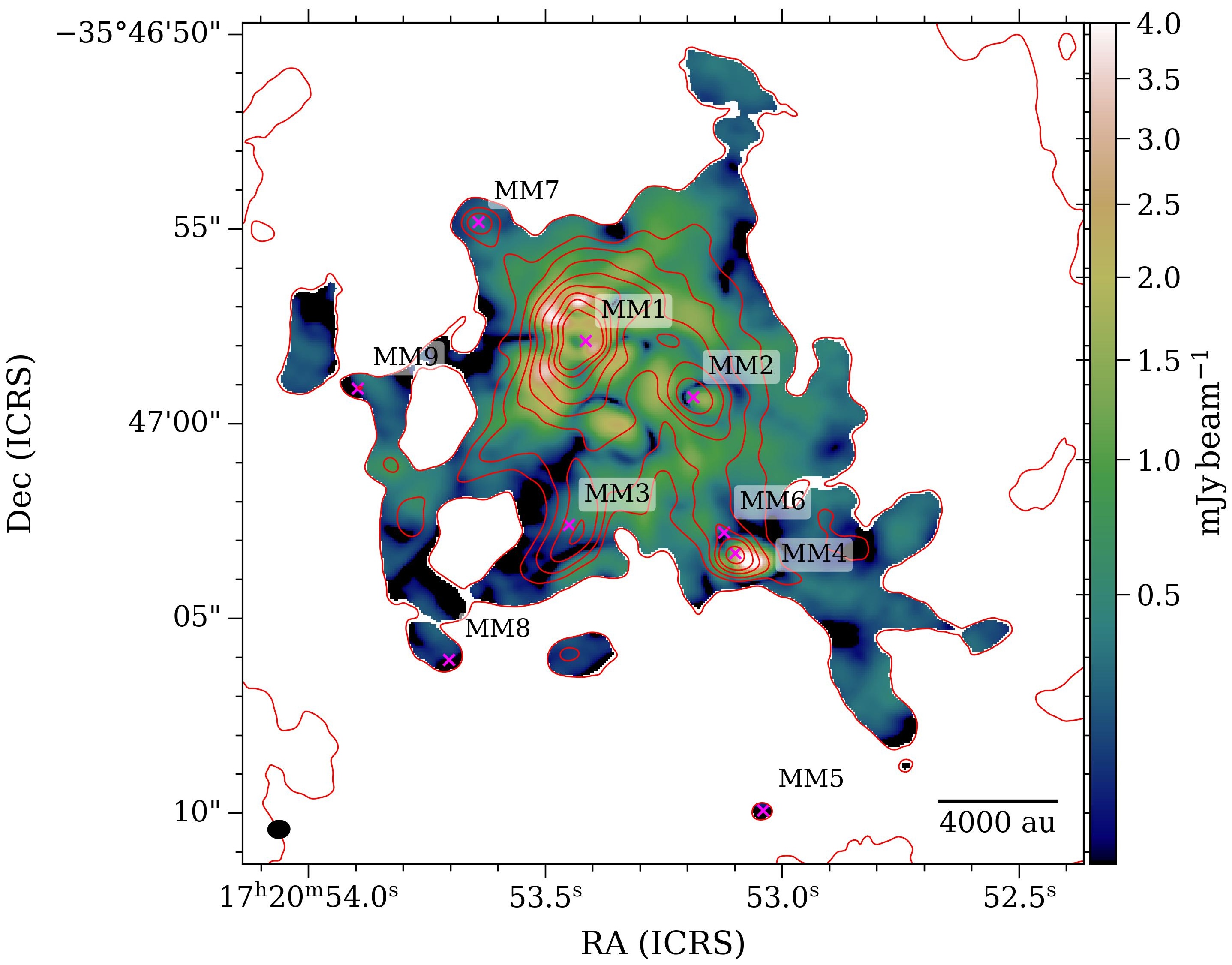}    
 \caption{1.2 mm polarized intensity map, over which the millimeter sources detected by \cite{Brogan_2016} are shown by magenta crosses. The contour levels of Stokes $I$ total intensity are shown at 0.0043, 0.018, 0.036, 0.072, 0.120, 0.240, 0.480, 0.720, and 0.960 Jy $\mathrm{beam}^{-1}$. The black ellipse at the bottom left shows the synthesized beam of the ALMA observations, with a size of $\sim 0\farcs6 \times 0\farcs5$ and a position angle of $\sim-86\degree$, and the black line at the bottom right shows the linear scale size of $\sim4000$ au.}
    \label{P_Intensity}
\end{figure*}

\begin{figure*}
    \centering
    \includegraphics[width=18cm]{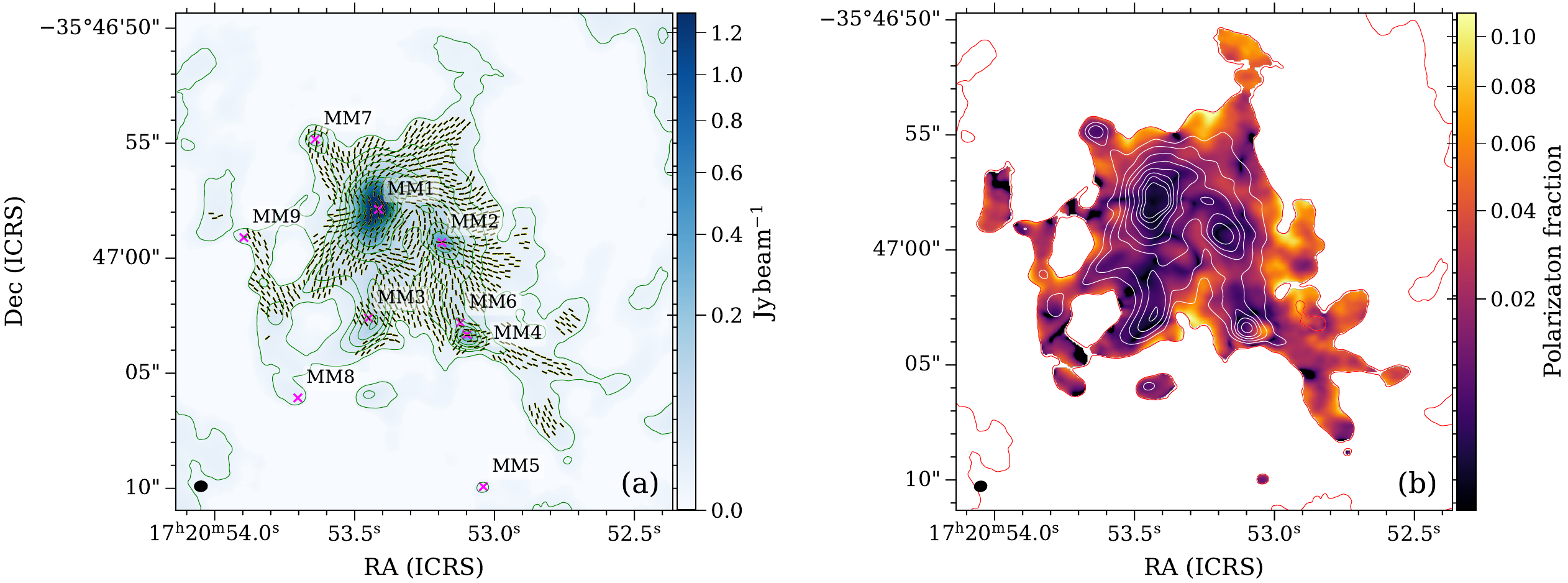}    
 \caption{(a) B-field morphology in NGC6334I inferred from dust polarization angles rotated by 90° ($P/\delta P > 3$) and sampled at every 5 pixels. The background map shows the Stokes $I$ total intensity.
(b) Polarization fraction map. The contours in both panels are the same as in Figure \ref{P_Intensity}.}
    \label{pol_map}
\end{figure*}

To quantify fluctuations in the B-fields, we compute the polarization angle dispersion function as the root-mean-square of the polarization angle differences between a reference pixel, $\bm{r}$, and pixel $i$, evaluated over a spatial scale, lag ($\delta$) \citep[see][]{Planck_2020}:

\begin{equation}
\label{eqn:S1}
    S(\bm{r},\delta) = \sqrt{\frac{1}{N} \sum_{i=1}^{N} \left[\theta(\bm{r}+\delta_i) - \theta(\bm{r})\right]^2},
\end{equation}
where N is the number of pixels, indexed by $i$, lying within an annulus of lag $\delta$/2 and 3$\delta$/2 from the reference pixel. 
In this work, we have taken the value of $\delta$ to be around half the beam size (FWHM) of the observational data. Due to the similar fact that $P$ is defined to be positive, the noise in Stokes $Q$ and $U$ biases $S$ towards larger values; a debiased $S$ is calculated by subtracting the variance of the angle dispersion function ($\sigma_S$) as

\begin{equation}
\label{eqn:S2}
    S_{db}^2 = S^2 - \sigma_S^2, 
\end{equation}
\begin{equation}
\begin{aligned}
\sigma_S^2(\bm{r},\delta) &=
\frac{\sigma_\theta^2(\bm{r})}{N^2 S^2}
\left(\sum_{i=1}^{N} \theta(\bm{r}+\delta_i) - \theta(\bm{r})\right)^2 \\
&\quad + \frac{1}{N^2 S^2} \sum_{i=1}^{N}
\sigma_\theta^2(\bm{r}+\delta_i)\,(\theta(\bm{r}+\delta_i)-\theta(\bm{r}))^2.
\end{aligned}
\end{equation}

\begin{figure}
    \centering
    \includegraphics[width=8.53cm]{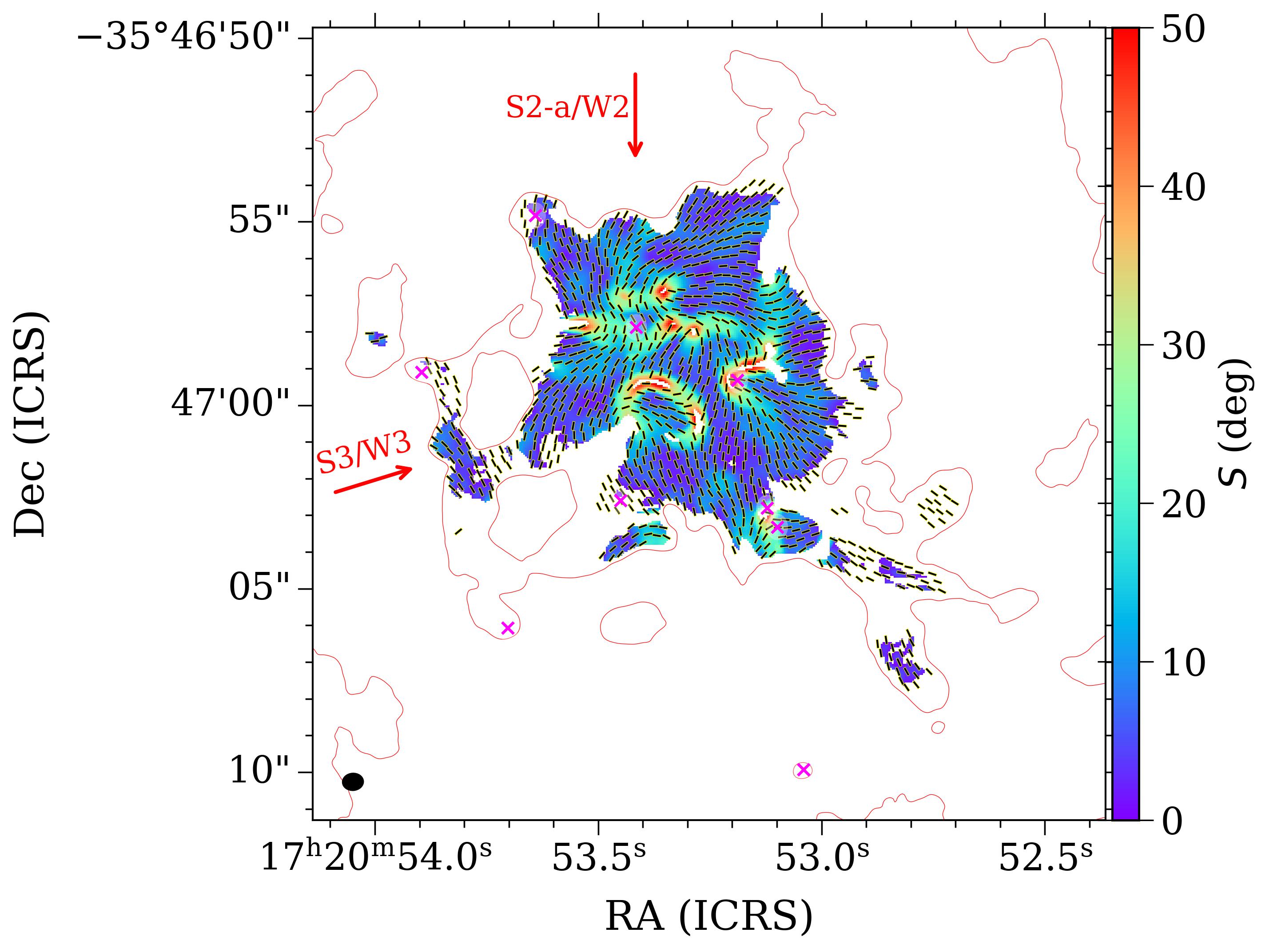}
     \caption{
     Angular dispersion of polarization position angles ($S$), over which the B-field segments are shown. It shows that the dispersion is higher where the B-field lines are flipped. The red-colored contour shows the total intensity at the level of 0.0043 Jy beam$^{-1}$. The magenta crosses are the same as those shown in Figure \ref{P_Intensity}. The red arrows indicate the approximate expansion directions of the \hii regions S2-a/W2 and S3/W3 \citep{Tahani_2023}.}
    \label{S_map}
\end{figure}

For convenience, we denote $S_{db}$ as $S$ hereafter. We note that pixels with $S \leq \sigma_S$ are excluded when computing the map of $S$. The value of $S$ is smaller for a uniform B-field and increases for a perturbed B-field, reaching a maximum of $\approx 52^\circ$ for a completely random field. The $S$ map is shown in Figure \ref{S_map}, where higher values of $S$ (i.e., 20\degree$-$50\degree) are observed in regions where the B-field orientation changes by $90^\circ$, as seen in Figure \ref{pol_map}(a). The median and mean $S$ values are $\sim7.4 \degree$ and $\sim 10.1 \degree$, respectively. We acknowledge that the value of $S$ is affected by the lag size, as the coherence gradually decreases with a larger lag, i.e., by going further away from the reference point. While at small lags, the function may become steeper owing to turbulent fluctuations or angular resolution effects in the dataset \citep{Planck_2015a}. To investigate this, we examine the effect of increasing the lag to one and two times the beam size, and the comparison is shown in Figures \ref{S_different}(a), (b), and \ref{S_distribution} in Appendix \ref{S_diff}. 
We find that the value of $S$ increases with the lag, consistent with earlier Planck studies at different data resolutions \citep{Planck_2015a, Planck_2015b} and with results from magnetohydrodynamic simulations by \cite{King_2018}. The median of $S$ map with $\delta=1$ and $\delta = 2$ beam sizes are $\sim16\degree$ and $\sim32\degree$, respectively. So, a lag of 2 beam sizes is smearing out the dispersion in the \ngc~region due to a larger annulus. From Figure \ref{S_distribution}(a) in Appendix \ref{S_diff}, it is evident that $S$ values larger than $\approx 52^\circ$, although very few in number, are present in the tails. Furthermore, the number of points approaching this limit increases with lag size. The values exceeding the saturation limit of $S$ could be due to low-polarization and high-noise regions, or to locations where the B-field geometry changes sharply.  

Alternatively, we also compute the polarization angular dispersion using regridded maps of the $I$, $Q$, and $U$ Stokes parameters to ensure the statistical independence of neighboring pixels in the high-resolution ALMA data, following the method described in \cite{Gouellec_2020}. We regrid the maps considering the Nyquist sampling, with around 4 pixels per beam area. Using the regridded maps (pixel size $\sim 0.25\arcs$), we calculate $S$ at each pixel $i$ by taking the $n = 8$ nearest neighboring pixels $j$, as follows:

{\scriptsize
\begin{equation}
\label{eqn:S3}
\begin{aligned}
S(\delta)_i = \left[
\frac{1}{n} \sum_{j=1}^{n}
\left(
\frac{1}{2}
\arctan\left(
\frac{Q(j)U(i) - U(j)Q(i)}
{Q(i)Q(j) + U(i)U(j)}
\right)
\right)^2
\right]^{1/2}.
\end{aligned}
\end{equation}}

The resulting $S$ map, denoted as $S$ (regridded), and its corresponding distribution are shown in Figures \ref{S_different}(c) and \ref{S_distribution}(b), respectively, in Appendix \ref{S_diff}. 
From the figure shown in Appendix \ref{S_diff}, it can be seen that the $S$ (regridded) map and $S$ ($\delta=0.5$ beam) look similar in terms of distribution and median values ($9.2\degree$ and $7.4\degree$, respectively). We further examine the correlation between $S$ and intensity for $S$ ($\delta = 0.5$ beam), $S$ ($\delta = 1$ beam), $S$ ($\delta = 2$ beam), and $S$ (regridded), which is shown in Figure \ref{S_distribution}(c) in Appendix \ref{S_diff}. We find that, in our particular case, $S$ (regridded) is similar to the $S$ map computed using the Planck method with a lag of half the beam size. Thus, we opted to consider the $S$ ($\delta = 0.5$ beam) for further analysis in this paper.   

Figure \ref{S_map} also shows the approximate directions of two nearby expanding \hii regions, the S3/W3 and S2-a/W2 bubbles \citep[][]{Simpson_2012, Anderson_2014}, among the multiple \hii regions associated with the NGC~6334 complex \cite[see][]{Tahani_2023}. The expansion of \hii regions can trigger star formation in nearby clouds \citep[e.g.][]{Elmegreen_1998, Zavagno_2010, Samal_2014} and significantly alter the B-field geometry, which may in turn resist their expansion. In the NGC 6334 complex, \citet{Tahani_2023} found ordered tangential B-field lines around the boundaries of several bubble shells (e.g., W5), resulting in higher polarization fractions ($\sim$7\%). However, toward W3/S3 and W2/S2-a, which encompass the \ngc~region, they found no such tangential B-field morphology and nearly constant polarization fractions, suggesting that the B-field geometry is instead dominated by the dense cores within \ngc. However, we cannot rule out the influence of other \hii regions on \ngc, either through photoionizing radiation or their potential impact on the B-field geometry. Quantifying these effects would require further analysis in future work.


\subsection{Gas Density and Dust Temperature}
\label{sec:Temp}
To obtain the gas density and dust temperature of the \ngc~region, we use $^{13}$CH$_3$OH emission as a tracer of the gas temperature. We assume that the $^{13}$CH$_3$OH line emission is optically thin and that the dust and gas are thermally coupled in the dense regions (density $\gtrsim10^{4.5}$ \cmq) \citep{Goldsmith_2001}. Under these assumptions, the gas temperature derived from $^{13}$CH$_3$OH rotational transitions can be taken as a representative of the dust temperature in the same region. The details about the derivation of rotational temperature are given in Appendix \ref{Trot}. For pixels that do not have a detection in $^{13}$CH$_3$OH and are lying within the spatial coverage of the 1.2 mm dust continuum emission (up to 3$\sigma$ level, see contours in Figure \ref{P_Intensity}), we have assumed that dust temperature would be around 30 K. This assumption is based on the results from \cite{McCutcheon_2000}, where a temperature of 30 K is reported for \ngc~at a density of $\sim10^6 - 10^7$ \cmq. However, for a smooth convergence to 30 K, missing pixels were filled using a distance-weighted nearest-neighbor interpolation with an exponential taper to minimize discontinuities at mask boundaries. 

\begin{figure*}
    \centering
    \includegraphics[width=17cm]{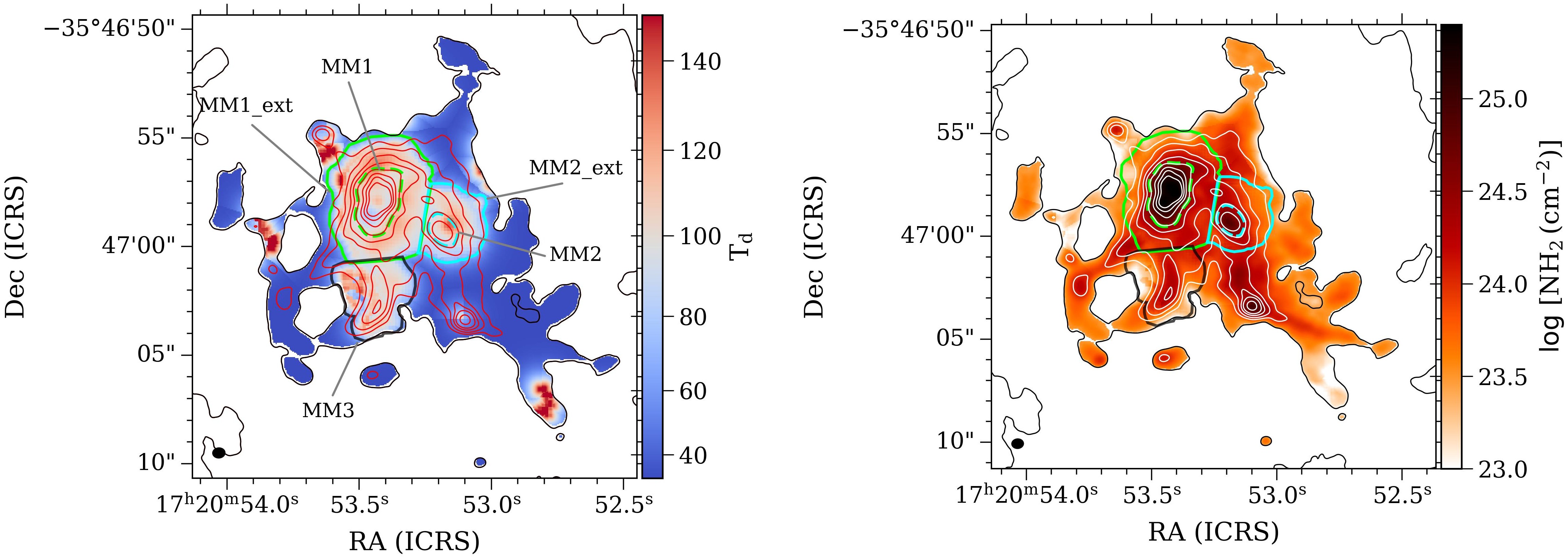}    
 \caption{(a) Dust temperature ($T_{\rm d}$) map based on rotational temperature from $^{13}\rm{CH_3OH}$. The regions studied in detail are shown as follows: MM1 and MM2 are outlined by dashed green and cyan polygons, respectively, while MM1\_ext, MM2\_ext, and MM3 are outlined by solid green, cyan, and black polygons, respectively. (b) Molecular hydrogen column density ($N_{\mathrm{H_2}}$) map. The contours in both panels are the same as in Figure \ref{P_Intensity}. }
    \label{Cdens_map}
\end{figure*}

Figure \ref{Cdens_map}(a) shows the dust temperature map of \ngc~based on $^{13}$CH$_3$OH rotational temperatures, where the temperature ranges from $\sim30.0$ K to $\sim180$ K. To examine the properties of the millimeter sources discussed in Section \ref{sec:int}, we defined their boundaries based on their closed contours of total-intensity Stokes $I$ emission, which are marked by dashed (MM1 and MM2) and solid (MM3) lines in the figure. The temperatures in these substructures range from around 80 K to 131 K, 88 K to 138 K, and 54.5 K to 128 K, respectively. Note that the peak temperature in MM2 is slightly offset from the total-intensity peak, most likely due to the presence of the disk candidate NGC6334I ALMA e4, resolved in \citet{Olguin_2026}. 
The temperature values obtained here are approximately a factor of two lower than those estimated in \cite{Cortes_2024}, as they used a relatively optically thick CH$_3$OH emission line, the limitations of which are also discussed in their paper. 
In the figure, the extended regions MM1\_ext and MM2\_ext are also marked by solid polygons of green and cyan colors, respectively, that, along with MM3 (black polygon), fill up the full coverage detected in $^{13}$CH$_3$OH (i.e., detected pixels without extrapolation to 30 K).
The remaining sources marked in Figure \ref{P_Intensity} are either too small (i.e., lacking sufficient pixels) or not fully resolved. Therefore, we restrict the subsequent analysis to the MM1, MM2, and MM3 regions unless otherwise specified. In Figure \ref{Cdens_map}(a), some high-temperature pixels appear in low-intensity regions. The rotational temperatures at these locations were derived using only three of the four available $^{13}$CH$_3$OH transitions, as the highest-excitation transition is not detected in the diffuse regions (see Appendix \ref{Trot} for details). Consequently, the corresponding fits have relatively large $\chi^2$ values, suggesting that the temperature estimates at these pixels are subject to larger uncertainty.   

The column density is derived from the dust continuum emission, assuming it to be optically thin, and the dust temperature map, following the formulation given in \cite{Hildebrand_1983} and the modified relation in \cite{Kauffmann_2008}:

{\scriptsize
\begin{equation}
\begin{aligned}
N(\mathrm{H}_2) = 2.02 \times 10^{20}\,\mathrm{cm^{-2}}
\left[
\exp\!\left(1.439
\left(\frac{\lambda}{\mathrm{mm}}\right)^{-1}
\left(\frac{T_{\rm d}}{10\,\mathrm{K}}\right)^{-1}
\right) - 1
\right]\\
\left(\frac{\kappa_\nu}{0.01\,\mathrm{cm^2\,g^{-1}}}\right)^{-1}
\left(\frac{S_\nu}{\mathrm{mJy\,beam^{-1}}}\right)
\left(\frac{\Theta_{\mathrm{HPBW}}}{10\,\mathrm{arcsec}}\right)^{-2}
\left(\frac{\lambda}{\mathrm{mm}}\right)^3, 
\end{aligned}
\end{equation}}
where $\lambda$ is the wavelength (1.2 mm), $T_{\rm d}$ is the dust temperature in K, $S_\nu$ is the dust continuum flux density in mJy~beam$^{-1}$, $\kappa_\nu$ = 0.1 $\times$ ($\nu$/1000 GHz)$^\beta$ is the dust opacity (per unit mass of gas $+$ dust) in \cms~g$^{-1}$ \citep{Hildebrand_1983}, $\nu$ is the frequency in GHz, $\beta$ is the dust opacity index, and $\Theta_{\mathrm{HPBW}}$ is the beam size ($\sim0.^{\prime\prime}5$). Here, we have taken $\beta$ = 1.6  \citep{Draine_2006} to obtain $\kappa_\nu$ = 0.0109 \cms~g$^{-1}$ for 1.2 mm. 

Figure \ref{Cdens_map}(b) shows the gas column density ($N_{\mathrm{H_2}}$) map, which clearly highlights the millimeter sources. The $N_{\mathrm{H_2}}$ value in \ngc~ranges from $\sim4.5 \times 10^{22}$~\cms to 2.8 $\times$ 10$^{25}$~\cms, with a mean around 1.3 $\times$ 10$^{24}$~\cms. Calculating the number density ($n_{\rm{H_2}}$) is complicated because it requires assuming a depth or thickness for the target, which is difficult to constrain from its projection on the POS. Additionally, \ngc\ consists of intricate structures and, therefore, is neither spherical nor cylindrical in shape. 

Assuming the same depth based on the cloud/filament width in the POS is not robust, especially if the cloud is highly fragmented and contains multiple substructures (e.g., clumps and cores). The dense cores are likely to exhibit a power-law density profile in comparison to extended filamentary features or the low-density surrounding regions. In this work, we use a 3D density reconstruction tool, \emph{Volume Density Mapper}\footnote{\href{https://github.com/gxli/volume-density-mapper} {https://github.com/gxli/volume-density-mapper}}, which decomposes the input column density map into multiple components based on their scale sizes \citep{Guang_2025, Zhao_2026}. It uses the method of constrained diffusion to decompose the 2D density distribution (e.g., column density or surface density) into different structures using multiple scales, known as the multi-scale decomposition reconstruction (MDR) method \citep[][]{Zhao_2026}. These scales correspond to the characteristic width ($l_c$) of the structures, which are then used to construct the mean density and the 3D density distribution. More details on the method are given in Appendix \ref{VDM}. Figure \ref{width} shows the characteristic width and mean density maps of \ngc. The width in the region varies from $\sim0.002$ pc to 0.041 pc. The maximum width obtained here is similar to the width of the whole \ngc~($\sim0.03$ pc) obtained in \cite{Cortes_2024}. The derived $n_{\mathrm{H_2}}$ value ranges from  $\sim3 \times 10^{5}$ to $1.5 \times 10^9$ \cmq. 

However, this tool also estimates the width of the structures based on their projection onto the sky, but provides a variable width rather than a constant one across the whole region. Moreover, the decomposition cannot recover sub-beam physical information, so characteristic widths comparable to or smaller than the beam should be regarded as unresolved. Although beam-limited data contain little power on scales below the beam, making the contribution of these small-scale channels to the intensity-weighted $l_c$ negligible, the measured $l_c$ at unresolved pixels is effectively limited by the beam size. Consequently, if the true structure is smaller than the beam, the measured $l_c$ overestimates its size, and the derived volume density should be regarded as a lower limit. Only $\sim$1.6\% of valid pixels have $l_c < 0.003$ pc (approximately the beam size), and these are confined to the outer parts of the cloud.

\begin{figure*}
    \centering
    \includegraphics[width=17cm]{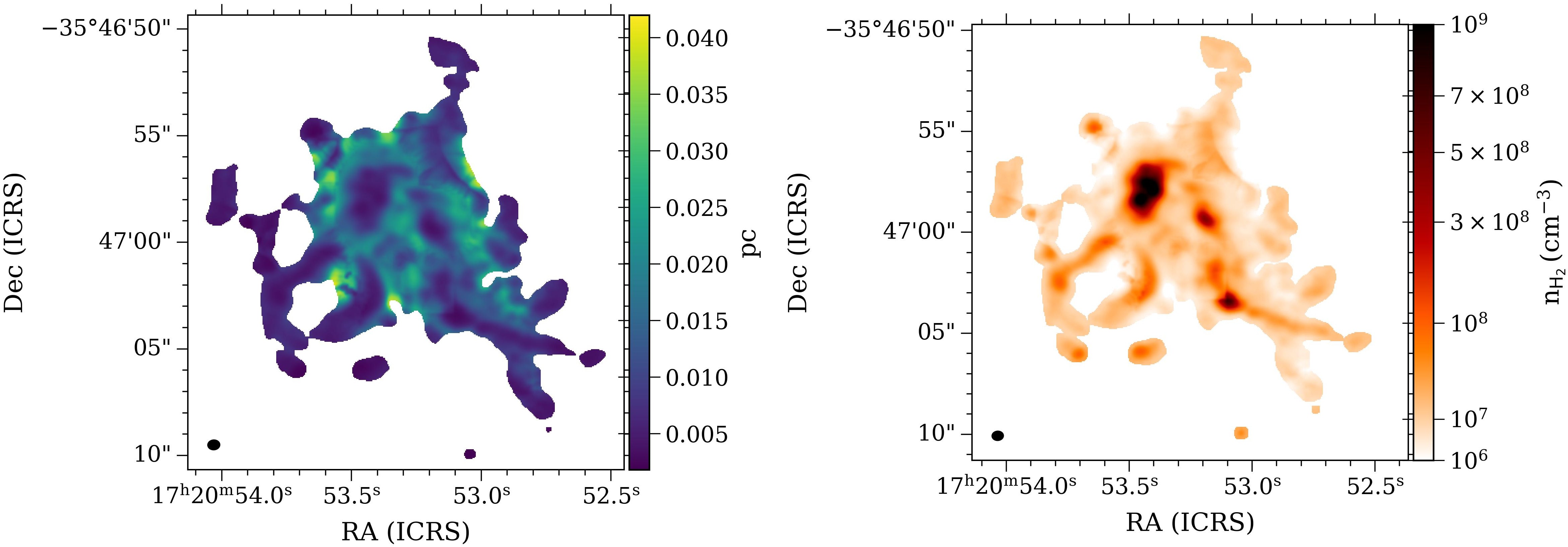}    
 \caption{(a) Characteristics width ($W$) map. (b) Molecular hydrogen mean density ($n_{\mathrm{H_2}}$) map.}
    \label{width}
\end{figure*}

\subsection{Variation of Polarization Fraction with Intensity, Dust Temperature, and Column Density}
The polarization fraction versus intensity (i.e., $P$ vs. $I$) is commonly used to analyze the extent of dust grain alignment. This relation is generally expected to follow a power-law form, $P \propto I^{-\alpha}$, where $\alpha = 0$ indicates a uniform alignment throughout the region and $\alpha = 1.0$ indicates an alignment limited in the outer regions only \citep{Whittet_2008}. But observationally, recent studies discuss that this behavior cannot be described by a single power law \citep[e.g.,][]{akshaya_2023, Pravash_2025}, for the reasons outlined in the introduction: enhanced alignment from increased RATs, RAT-D, gas randomization, and optical depth. Therefore, in addition to intensity, it is important to examine the variation of $P$ with dust temperature and gas density. 
We acknowledge that all the correlations discussed below are based on the running median of the trends; however, in cases where the data points are much scattered, further analysis is required to understand the cumulative effects of temperature, density, dust grain properties, and B-field tangling.

\begin{figure*}
    \centering
    \includegraphics[width=16cm]{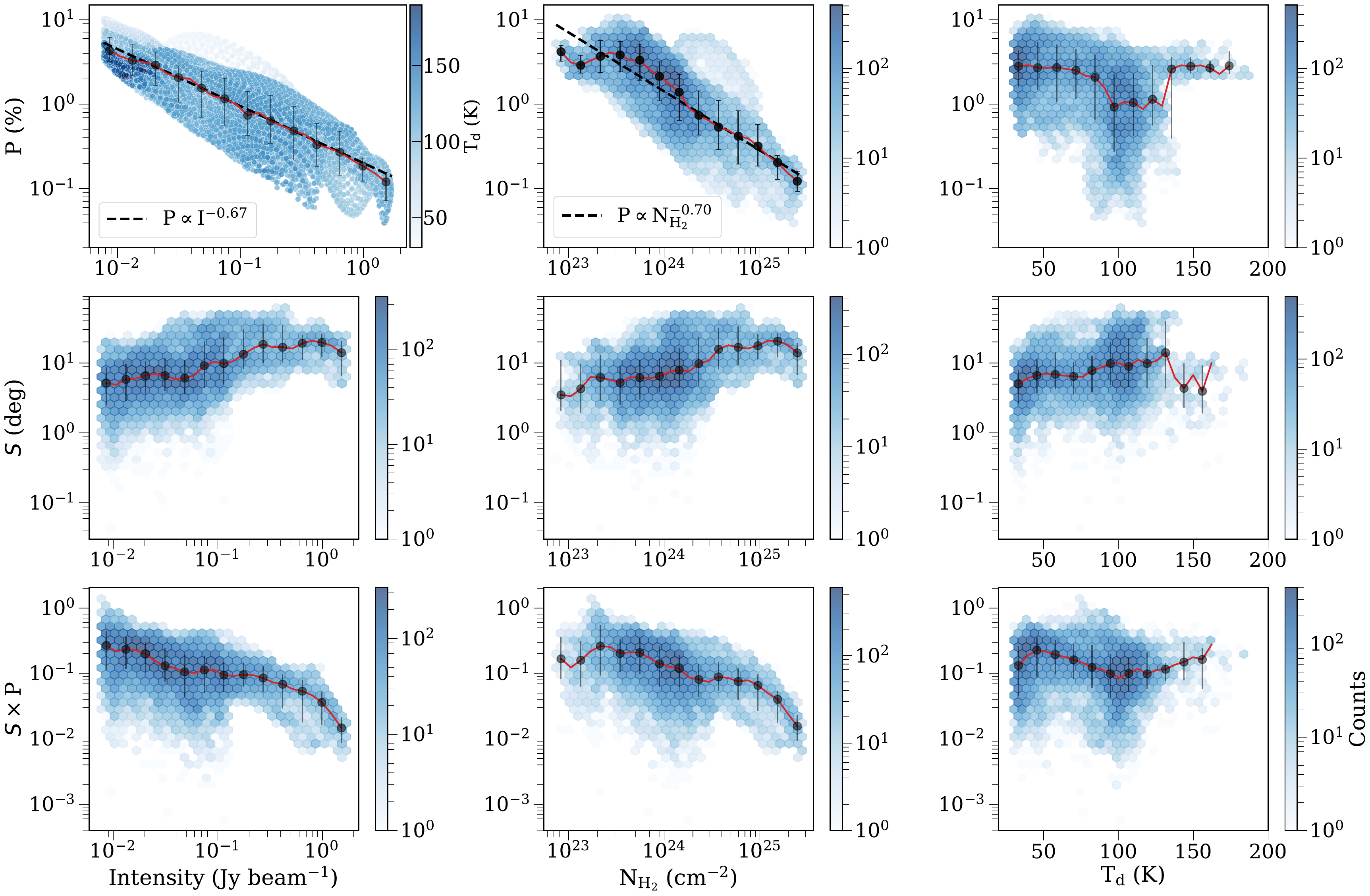}    
 \caption{Variation of (\emph{top}): polarization degree ($P$), (\emph{middle}): polarization angle dispersion ($S$), and (\emph{bottom}): alignment efficiency ($S \times P$) as functions of total emission intensity ($I$), column density ($N_{\mathrm{H_2}}$), and dust temperature ($T_{\rm d}$) in \ngc. Only data points satisfying $I/\delta I > 5$ and $P/\delta P > 3$ are shown, where $\delta I$ = 1.54 mJy $\rm{beam}^{-1}$ is the rms noise level in the total intensity. In the top-left panel ($P$ vs. $I$), individual data points are colour-coded by $T_{\rm d}$, and the dashed black line shows a power-law fit with index = $-$0.67 $\pm$ 0.01. Please note that, except for the top-left panel, the data points in all other panels are shown using hexbinning with a grid size of 28, with the number of data points in each hexbin indicated by the color bar. In the top-middle panel ($P$ vs. $N_{\rm{H_2}}$), the dashed line shows a power-law fit with index = $-$0.70 $\pm$ 0.01. The red curve and black dots with error bars show the running median and 16th–84th percentile range computed using 25 equally spaced bins: logarithmic bins for $I$ and $N_{\mathrm{H_2}}$ (bin widths of 0.094 and 0.103 dex, respectively) and linear bins for $T_{\rm d}$ (bin width of $\sim$6 K). The black dots are shown for every other bin for clarity, and each bin contains at least 20 data points.}
    \label{corr_full}
\end{figure*}

\begin{figure*}
    \centering
    \includegraphics[width=17cm]{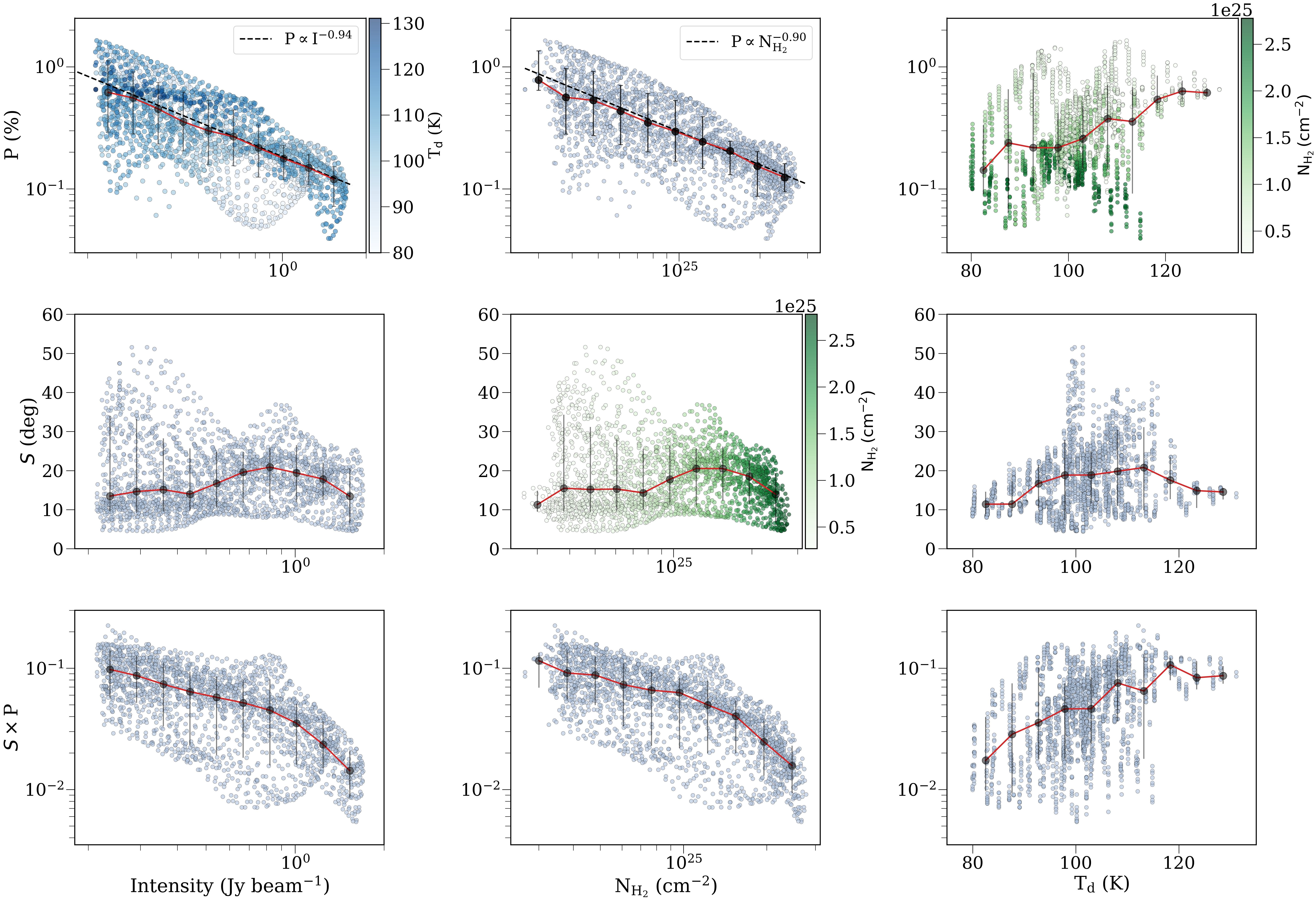}    
 \caption{Same as Figure \ref{corr_full}, but for the MM1 region. The running median and 16th–84th percentile ranges are computed using 10 equally spaced bins: logarithmic bins for $I$ and $N_{\mathrm{H_2}}$ (bin widths of 0.090 and 0.102 dex, respectively) and linear bins for $T_{\rm d}$ (bin width of $\sim$5 K), with a minimum of 10 data points per bin.}
    \label{corr_MM1}
\end{figure*}

\begin{figure*}
    \centering
    \includegraphics[width=17cm]{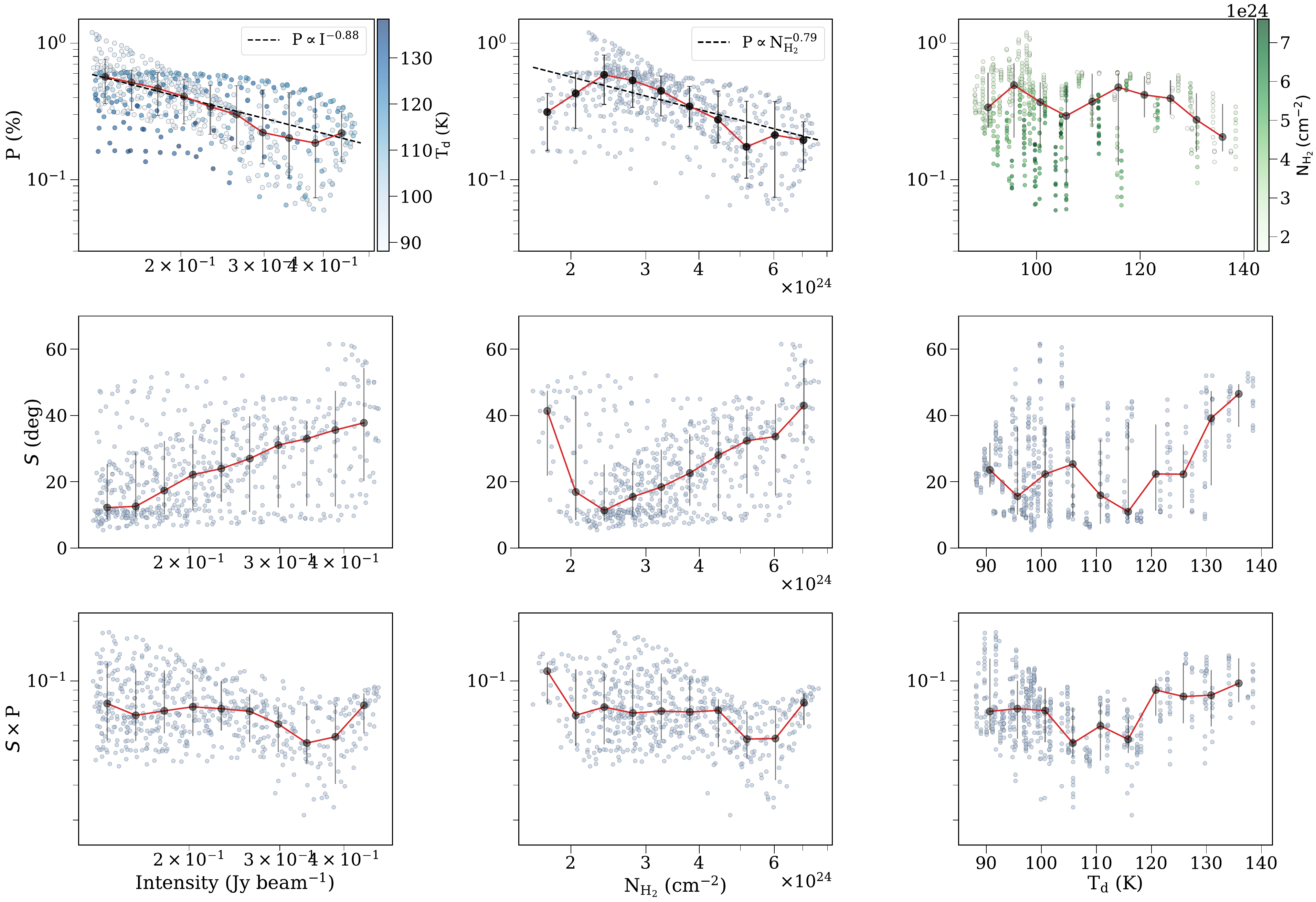}    
 \caption{Same as Figure \ref{corr_full}, but for the MM2 region. The running median and 16th–84th percentile ranges are computed using 10 equally spaced bins: logarithmic bins for $I$ and $N_{\mathrm{H_2}}$ (bin widths of 0.056 and 0.068 dex, respectively) and linear bins for $T_{\rm d}$ (bin width of $\sim$5 K), with a minimum of 10 data points per bin.}
    \label{corr_MM2}
\end{figure*}

\begin{figure*}
    \centering
    \includegraphics[width=17cm]{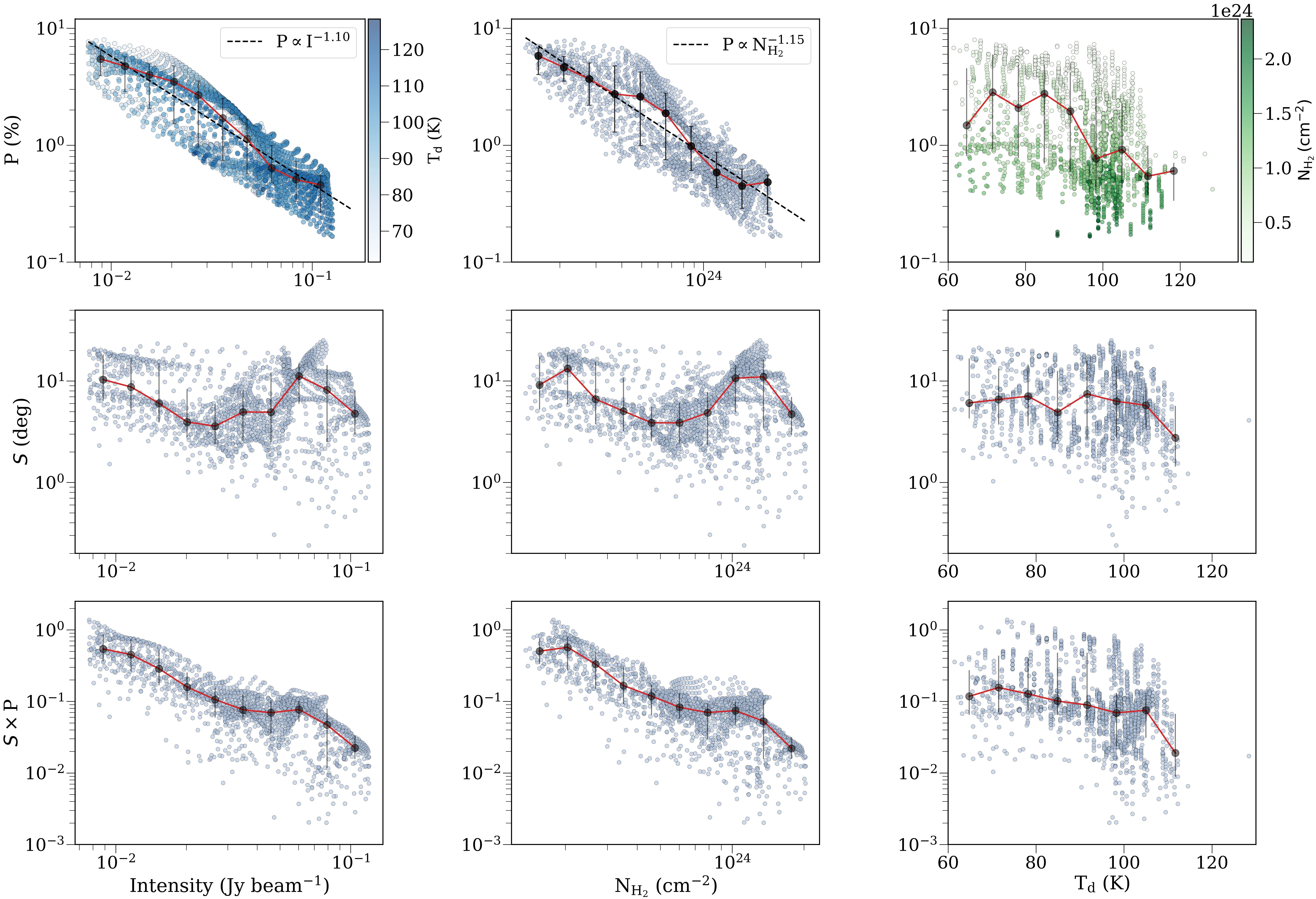}    
 \caption{Same as Figure \ref{corr_full}, but for the MM3 region. The running median and 16th–84th percentile ranges are computed using 10 equally spaced bins: logarithmic bins for $I$ and $N_{\mathrm{H_2}}$ (bin widths of $\sim0.122$ and $\sim0.128$ dex, respectively) and linear bins for $T_{\rm d}$ (bin width of $\sim$6.7--8.6 K), with a minimum of 10 data points per bin.}
    \label{corr_MM3}
\end{figure*}

Figure \ref{corr_full} shows the variation of $P$ with intensity, $T_{\rm d}$, and $N_{\mathrm{H_2}}$ for the whole \ngc~region. 
The depolarization trend is clearly visible, and although a single power-law fit over $P-I$ and $P-N_{\mathrm{H_2}}$ plots, obtained via weighted least-squares fitting, yields $\alpha \approx 0.67$ and $\approx$ 0.70, respectively, the dense regions might have a steeper slope. This can easily be verified from the individual plots of MM1, MM2, and MM3, shown in Figures \ref{corr_MM1}, \ref{corr_MM2}, and \ref{corr_MM3}, respectively, where power-laws with higher $\alpha$ values close to 1 are obtained. Since the observed data points are scattered and non-uniformly distributed, extreme and noisy $P$ values can significantly bias the running mean. Therefore, to robustly
trace the underlying trend and obtain a smoother variation of $P$, we adopt the running median, which is less sensitive to outliers and skewed distributions. Additionally, we show the 16th to 84th percentile range in the plots to illustrate the intrinsic scatter. For the running median, the bins are defined uniformly in logarithmic space for intensity and column density, and uniformly in linear space for dust temperature. However, the power-law fits shown in the top-left and middle panels of the figures \ref{corr_full}, \ref{corr_MM1}, \ref{corr_MM2}, and \ref{corr_MM3} are performed using all data points. 

The $P-T_{\rm d}$ plot for the full region shows that $P$ remains nearly flat till $\sim65$ K. From there, $P$ decreases with $T_{\rm d}$ until $\sim100$ K, after which it increases beyond $T_{\rm d}\sim130$ K and subsequently remains nearly flat. 
In the case of the MM1 region, an increase in $P$ with temperature is evident above $\sim80$ K, which could be the effect of enhanced RAT-A, as this region contains a cluster of YSOs. For MM2, the trend remains nearly flat. In MM3, the behavior is more complex, and it is difficult to see any trend due to the large scatter in the data. However, $P$ decreases with increasing column density in MM3 (see the top-middle panel of Figure \ref{corr_MM3} and the color bar in the top-right panel).

To investigate the effect of B-field fluctuations on polarization fraction, we calculated the $S$ parameter whose variation with intensity, $T_{\rm d}$, and $N_{\mathrm{H_2}}$ is shown in the middle panels of Figures \ref{corr_full}, \ref{corr_MM1}, \ref{corr_MM2}, and \ref{corr_MM3} for the whole \ngc, MM1, MM2, and MM3 regions, respectively. Another quantity, $S \times P$, can serve as an indicator of the net alignment efficiency inferred from observations \citep{Planck_2020}, enabling an assessment of whether the decrease in $P$ arises from polarization angle dispersion (i.e., $S$) or changes in dust properties. Therefore, an increase in $S \times P$ would suggest more efficient grain alignment. However, the reliability of this quantity as a tracer of alignment efficiency also depends on the inclination angle of the B-field with respect to LOS \citep{Hoang_2024}.
 
The variation of $S \times P$ is shown in the bottom panels of Figures \ref{corr_full}, \ref{corr_MM1}, \ref{corr_MM2}, and \ref{corr_MM3} for the aforementioned regions. It is noted that an increase in $S$ is expected to correspond to a decrease in $P$, and vice versa, such that $S \times P$ remains nearly flat and mitigates the effect of dispersion. Therefore, any deviation from this behavior reflects the influence of dust properties. In Figure \ref{corr_full}, the $S$–$I$ and $(S \times P)$–$I$ plots show that $S$ gradually increases while $P$ decreases with $I$, but $S \times P$ is not flat and exhibits a shallower slope. Beyond $\sim0.3$~Jy beam$^{-1}$, $S$ becomes relatively flat, whereas $P$ continues to decrease sharply, leading to a decline of $S \times P$ with intensity. Hence, both B-field tangling and dust properties may contribute to the depolarization. 
With increasing $T_{\rm d}$, $S$ remains fairly flat up to $\sim75$ K, and $P$ is also nearly flat. Then, $S$ gradually increases up to $\sim130$ K and $P$ correspondingly decreases to $\sim1$\%. Beyond this point, $S$ slightly decreases, while $P$ increases to $\sim3.0$\%. Thus, over most of the temperature range, $P$ seems to be governed by dispersion. However, the continued decrease in $P$ even when $S$ becomes relatively flat at the high-density end ($S$ vs. $N_{\rm{H_2}}$) may also result from reduced alignment efficiency due to gas randomization or optical depth (see section \ref{Grain_growth} for more discussion). 
  
For MM1, $S$ gradually increases with intensity up to $\sim0.8$ Jy beam$^{-1}$ and subsequently decreases, but its overall variation remains relatively shallow. In contrast, $P$ exhibits a pronounced decline with increasing intensity, resulting in a monotonic decrease of $S \times P$ over the full intensity range. From the $S$–$T_{\rm d}$ and $(S \times P)$–$T_{\rm d}$ plots, it can also be seen that $S$ gradually increases with temperature up to $\sim113$ K. Consequently, $P$ would be expected to decrease, but here, $P$ instead increases. As a result, $S \times P$ increases almost throughout the temperature range. This trend indicates an increase in dust grain alignment with temperature (see discussion in section \ref{diss_1}). The low polarization at relatively low temperatures and high densities can be because of effective gas randomization, as seen from the colorbar of the $P-T_{\rm d}$ plot and $P-N_{\mathrm{H_2}}$ plot, or due to optical depth. 
This suggests that although B-field tangling is present, it is not sufficient to fully account for the depolarization in MM1. Dust grain properties (e.g., shape and size; see Section \ref{ideal_model}), gas randomization, and optical depth are also likely contributing here.

For MM2, $S$ increases while $P$ decreases with intensity, keeping $S \times P$ nearly flat. Although the number of data points in the $S$–$T_{\rm d}$ plots is limited to draw a firm conclusion, a broad variation is observed: as $S$ increases, $P$ decreases with temperature. A similar correlation is also observed with $N_{\mathrm{H_2}}$. Since $S \times P$ remains nearly flat with increasing intensity, $T_{\rm d}$, and $N_{\mathrm{H_2}}$, the observed decrease in polarization fraction in MM2 is not dominated by significant changes in dust properties. Instead, it could be due to variations in $S$, or equivalently, B-field tangling or inclination effects. For MM3, from Figure \ref{corr_MM3}, it can be seen that $S \times P$ decreases with intensity and $N_{\mathrm{H_2}}$, while the running median of $S$ fluctuates in the range $\sim4\degree-13\degree$ with $I$ and $N_{\mathrm{H_2}}$. It suggests that the effects of dust grain alignment loss are more dominant here. 

To better distinguish the aforementioned cumulative effects, in the next sections, we analytically calculate the dust grain alignment and disruption sizes and model dust grain polarization based on the observed dust temperatures and gas densities in \ngc.

\section{Modeling Grain Alignment and Dust Polarization using the RAT Paradigm}
\label{sec:model_theory}
Our data analysis in the previous section reveals general trends in the correlation between the polarization degree and the observed physical properties of the local environment, such as the gas column density, dust temperature, and polarization angle dispersion. To understand the physical origins of such correlations, here we perform a detailed modeling of grain alignment and dust polarization using the RAT paradigm, including RAT-A and RAT-D effects.
\subsection{Grain Alignment and Disruption}
\label{sec:align_disr}
According to the RAT-A theory, the polarization fraction depends on the distribution of aligned dust grains. The dust grains can only efficiently align if they rotate superthermally, i.e., their angular velocity due to radiative torques ($\omega_{\mathrm{RAT}}$) exceeds their thermal angular velocity ($\omega_{\rm T}$). Therefore, the minimum size of aligned grains can be determined using the following equation based on a threshold criterion of $\omega_{\mathrm{RAT}}\, (a_{\mathrm{align}})\, = 3\, \omega_T$ \citep{Hoang_2008}: 

\begin{equation}
\label{eqn:align}
\begin{aligned}
a_{\rm align} &\simeq0.055\, \hat{\rho}^{-1/7}\left(\frac{\gamma_{-1} U}{n_3 T_1} \right)^{-2/7}\\ &\times \left(\frac{\bar{\lambda}}{1.2\, \mu{\rm m}}
\right)^{4/7}(1 + F_{\rm IR})^{2/7}\, \mu{\rm m},\\
F_{\rm IR} &\approx \left(\frac{0.038}{a_{-5}}\right) \left(\frac{U^{2/3}}{n_3 T_1^{1/2}}\right),\\
U &= {(T_{\rm d}/16.4{\rm K})}^6 \times a_{-5}^{6/15},
\end{aligned}
\end{equation}
where $F_{\mathrm{IR}}$ is the ratio of the IR re-emission to collisional damping times. The grains are spun up by RATs but damped down by gas collisions and re-emission, such that the effective rotation, $\omega_{\mathrm{RAT}}$, depends on these damping timescales. Here, $\hat{\rho}$ = $\rho/$(3 g \cmq) is the normalized dust mass density, $n_3$ = $n_{\rm H}/ (10^3$~\cmq) is the number density of hydrogen atoms ($n_{\rm H}$ $\approx 2\times n_{\mathrm{H_2}}$), $T_1$ = $T_{\rm {gas}}/(10$ K) is the gas temperature, $\gamma$ and $\bar{\lambda}$ are the anisotropy degree and the mean wavelength of the radiation field, respectively, and $a_{-5}$ = $a/(10^{-5}$~cm) is the dust grain size. Taking $\rho$ = 3 g \cmq~\citep{Draine_1984}, $\gamma$ = 0.3, as there are multiple radiation sources within \ngc, $T_{\rm {gas}}$ $\approx$ $T_{\rm d}$, $\bar{\lambda} = 1.0\, \mu$m, $a$ = 10$^{-5}$ cm, and using temperature and number density maps, we calculated $a_{\rm align}$ at every pixel. 

\begin{figure*}
    \centering
    \includegraphics[width=17cm]{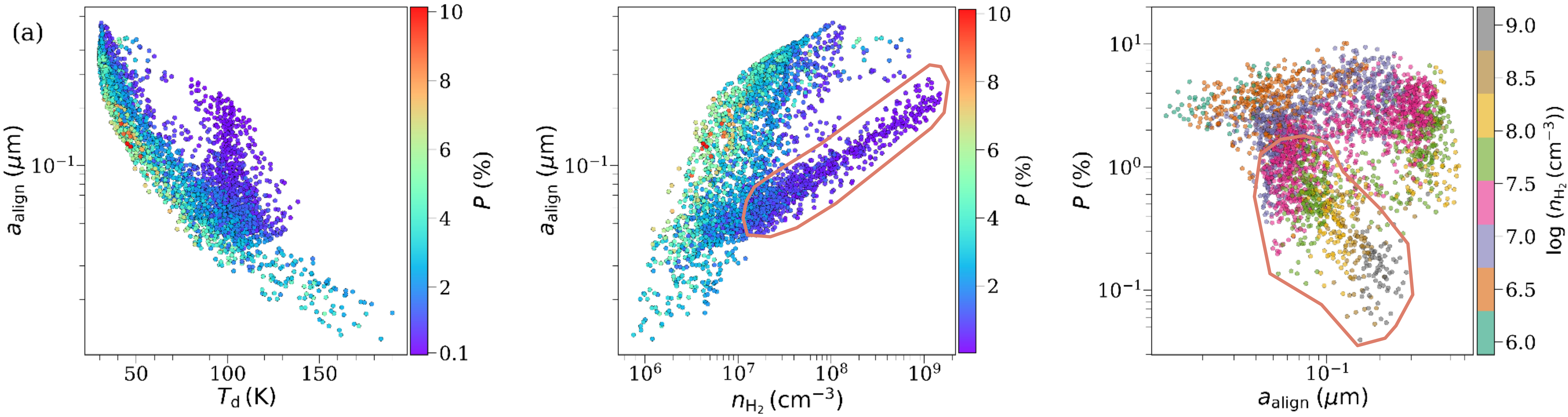}
    \includegraphics[width=17cm]{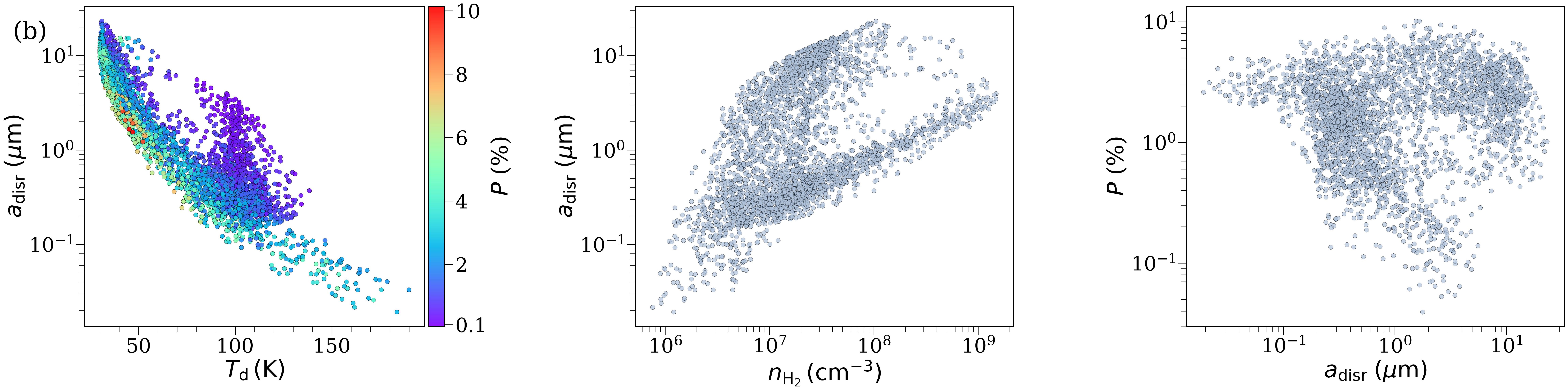}
 \caption{Minimum size of (a) aligned grains, $a_{\mathrm{align}}$ and (b) disrupted grains, $a_{\mathrm{disr}}$, shown as a function of $T_{\rm d}$ (\emph{left-panels}) and $n_{\mathrm{H_2}}$ (\emph{middle-panels}), for \ngc. The \emph{right-panels} show the variation of P with  $a_{\mathrm{align}}$ (\emph{top}) and $a_{\mathrm{disr}}$ (\emph{bottom}). The values are calculated using Equations \ref{eqn:align} and \ref{eq:adisr}, respectively. For better visualization, the data points are shown at every third pixel. The low-polarization ($\lesssim1\%$) branch is marked with an orange polygon.}
    \label{a_align_dis}
\end{figure*}

Figure \ref{a_align_dis}(a) shows the variation of $a_{\rm align}$ with $T_{\rm {d}}$, $n_{\rm {H_2}}$, and $P$. The left panel of Figure \ref{a_align_dis}(a) shows that $a_{\rm align}$ decreases with increasing temperature, indicating that smaller grains, even down to sizes of $\sim0.015\, \mu$m, can be aligned at higher temperatures of around 150 K. As a result, the distribution of aligned grains becomes broader for a fixed maximum grain size ($a_{\rm{max}}$), and thus, $P$ increases (see the corresponding color bar and right panel in Figure \ref{a_align_dis}(a)). While with density, the $a_{\rm align}$ increases up to $\sim0.5\, \mu$m at densities of $\gtrsim5 \times 10^7$ \cmq~due to gas randomization, i.e., disalignment due to collisions (see middle panel of \ref{a_align_dis}a), observed in \ngc~also.  

While higher temperatures (i.e., stronger radiation fields) favor the alignment of smaller grains, an overly intense radiation field can lead to rotational disruption, thereby determining the maximum grain size that can be aligned. So, when the centrifugal stress of a rotating grain becomes greater than its tensile strength, $S_{\rm{max}}$, the grain disrupts into fragments. Based upon this criterion, the size of disrupted grains can be estimated using the following equation \citep{Hoang_2021a}:

\begin{equation}    
\begin{aligned}
a_{\rm disr} &\simeq 1.7 \left(\frac{\gamma_{-1} U}{n_3 T_1^{1/2}}\right)^{-1/2}
\left(\frac{\bar{\lambda}}{1.2\,\mu{\rm m}}\right)\hat{\rho}^{-1/4}\\
&\times S_{{\rm max},7}^{1/4}\, (1 + F_{\rm IR})^{1/2}\,\mu{\rm m},
\end{aligned}
\label{eq:adisr}
\end{equation}
where $S_{{\rm max},7}$ = $S_{{\rm max}} / (10^7$ erg \cmq). Here, we assumed $S_{\rm{max}}$ = 10$^5$ erg \cmq~for porous composite grains \citep{Hoang_2019b} that are expected in dense cores due to grain growth. Other parameters are taken as the same as in the calculation of $a_{\rm align}$. The resulting plots of $a_{\rm{disr}}$ with observed parameters ($T_{\rm d}$, $n_{\rm{H_2}}$, and $P$) are shown in Figure \ref{a_align_dis}(b). Similar to $a_{\rm{align}}$, $a_{\rm{disr}}$ also decreases with increasing temperature. However, in this case, it implies that at higher temperatures, even smaller grains (such as size $\lesssim 0.5 \mu$m) can be rotationally disrupted into fragments, thereby imposing an upper limit on the maximum grain size that can remain aligned. Thus, for a fixed $a_{\rm{align}}$, an increase in temperature decreases $a_{\rm{max}} \rightarrow a_{\rm{disr}}$, and consequently $P$ decreases. However, here, conversely, $P$ is observed to be higher because $a_{\rm align}$ is not fixed in reality, as discussed above. The $a_{\rm{disr}}$ increases with the density, meaning that high-density regions tend to suppress the disruption of dust grains. Therefore, the combined effect of local temperature and density determines the distribution of aligned grains and, hence, the resulting polarization degree. The color maps of $a_{\rm{align}}$ and $a_{\rm{disr}}$ are shown in Figure \ref{align_disr_maps} in Appendix \ref{grain_size_maps}. 

Figure \ref{a_align_dis} also shows two distinct branches in the distributions of the minimum aligned and disrupted grain sizes. One branch shows relatively higher polarization ($P\sim1\%-10\%$), with most values $\gtrsim$ 2\% and only decreasing below 2\% at higher densities. While the other branch shows very low polarization, i.e., $P \lesssim 1\%$. This second branch (marked in Figure \ref{a_align_dis}(a) by an orange polygon) mostly shows the regions with high density ($n_{\rm H}\sim10^8 - 10^9$~\cmq), MM1 and MM2, and highlights the effect of gas randomization at high density, even at high temperatures (i.e., 100 K). Whereas the first branch includes mostly the low-density outer regions in \ngc, and thus shows a higher polarization degree. However, in the low-polarization branch, there are some points that have $a_{\rm{align}}$ $\sim0.06~\mu$m but still $P \lesssim 1\%$ at $n_{\rm H}\sim10^7 - 10^{7.5}$~\cmq. From Figure \ref{align_w_S} given in Appendix \ref{S_diff}, which shows the $P$ vs. $a_{\mathrm{align}}$ relation with points having $S > 25^\circ$ highlighted, it is evident that these points are associated with higher $S$ values. This suggests that the low polarization observed even at high temperatures ($\sim100$~K) may be caused by collisional randomization of dust grains at high densities, as well as B-field tangling.

\subsection{Modeling Thermal Dust Polarization and Comparison to Observations}
\label{sec:mod}
\subsubsection{Model Setup and Assumptions}
We next model the polarization fraction of thermal dust emission induced by aligned grains and compare it with the observational data. According to the RAT paradigm, the polarization degree depends on the alignment size, $a_{\rm align}$; the size-dependent alignment degree, $f_{\rm align}(a)$, which is determined by $a_{\rm align}$ and the maximum degree of grain alignment ($f_{\rm max}$); and the angle between the B-field and the LOS (see \citealt{Hoang_2024} for details).

The value of $f_{\rm max}$ depends on the local gas density and radiation, grain size, and magnetic susceptibility \citep{Hoang_Laz_2016, Hoang_2025}. \cite{Hoang_Laz_2016} suggested that grains with iron inclusions or clusters (i.e., superparamagnetic grains) can achieve nearly perfect alignment due to magnetically enhanced RAT. The iron inclusions significantly increase the magnetic properties of grains \citep{Hoang_2025}. Based on dust evolution models, it has been suggested that a large fraction of iron in the ISM likely resides as inclusions within silicate grains \citep{Kohler_2014, Zhukovska_2018}. The studies of the ``search for interstellar iron'' show that the formation of pure Fe grains has a very low probability, even in Fe-dominant environments such as type Ia Supernovae \citep[e.g.,][]{Kimura_2017, Moutard_2026}. Moreover, \cite{Hensley_2023} shows that the thermal dust polarization observed through Planck can be reproduced by assuming a perfect alignment of dust grains. Therefore, here we considered $f_{\rm{max}}$ = 1 for composite grains consisting of iron inclusions in this collisional-dominated region \citep{Hoang_2025} and used the {\texttt{DustPOL\_py}}\footnote{\href{https://github.com/lengoctram/DustPOL_py} {\texttt{DustPOL\_py}}} code to model thermal dust polarization in \ngc. \texttt{DustPOL\_py} can compute both starlight and emission polarization by including the RAT, MRAT, and RAT-D mechanisms \citep{Lee_2020, Tram_2021}.  

 In this work, we have considered the dust type to be Astrodust, a composite model (mixture of carbon and silicates) of dust grains \citep{Draine_2021}, and their shapes to be oblate \citep[grain aspect ratio = 1.4, e.g.,][]{Cho_2007, Draine_2021}, with the internal dust mass density of $\rho= 3$ g~\cmq. We adopt an MRN grain size distribution that follows a power-law with an index of 3.5 \citep{Mathis_1977}. The $a_{\rm max}$ values are varied from 0.25 $\mu$m to 2.0 $\mu$m to account for grain growth in dense environments.
 
\subsubsection{Ideal polarization model with uniform B-fields}
\label{ideal_model}
To explicitly study the effects of dust physics on the dust polarization, we first consider the ideal model in which the B-fields are uniform and lie in the POS. The polarization degree of this {\it ideal} model is denoted as $P_{\mathrm{mod,\, id}}$, and if the RAT-D effect is included, then it is denoted as $P_{\mathrm{mod,\,id,\,RATD-ON/OFF}}$. The model depends on the input parameters, such as dust grain type and shape, grain size distribution, radiation ($U$) or temperature, gas density, $\gamma$, $\bar{\lambda}$, and $a_{\rm{max}}$. The details of the equations used to compute polarization emission based on these parameters are given in \cite{Lee_2020, Tram_2021}. The $U$ is estimated from $T_{\rm d}$, along with $n_{\rm H} = 2n_{\mathrm{H_2}}$, taken from the ALMA observations. Different anisotropy degree $\gamma$ values are considered for different regions in \ngc, given in Table \ref{tab:pars}. The $\bar{\lambda}$ is taken to be 1.0 $\mu$m. We model the MM1, MM2, MM1\_ext, MM2\_ext, and MM3 regions separately (see Figure \ref{Cdens_map}(a)). Since the other structures (marked in Figure \ref{P_Intensity}) do not have enough pixels in the $S$, $P$, and $T_{\rm d}$ maps, we model the rest of the region in \ngc~altogether, termed as the ``Outer'' region. The values of the aforementioned input parameters for different regions are summarized in Table \ref{tab:pars}. 

The results of ideal thermal dust polarization, $P_{\mathrm{mod,\, id}}$, for different regions are shown in Figure \ref{dustpol_diff_amax}. For comparison, the observed polarization fraction, $P_{\rm obs}$, is also shown in each figure. Only the results for $a_{\rm max}$ = 0.50 $\mu$m are shown here for each region, while the corresponding plots for other $a_{\rm max}$ values are presented in Appendix \ref{Pmodel_appendix}. 
The black solid line in the plots shows the running median of $P_{\rm obs}$, with bins uniformly spaced on the logarithmic column-density scale. To ensure statistical reliability, only bins containing at least 10$-$20 pixels are considered, depending on the total number of pixels in each region. 
From the figures, it can be seen that although the ideal model reproduces the observed polarization trends, the degree of polarization from the model is significantly higher than the observed values. Moreover, $P_{\rm mod,~id}$ increases with increasing $a_{\rm max}$ because of the broader aligned grain size distribution. This shows that the case of a uniform B-field perfectly lying on the POS is not sufficient to produce the observed $P$.

\begin{table*}
\label{tab:pars}
\centering
\caption{Parameters for dust emission polarization modeling. Note:$^*$ The values of $S$ quoted in the brackets are the median values in each region. The quoted parameter values are after a S/N cut of $I/\delta I >$ 5 and $P/\delta P >$ 3.}
\begin{tabular}{lcccccccc}
\hline\hline
Parameter & MM1 & MM2 & MM1\_ext & MM2\_ext & MM3 & Outer \\
\hline
$N_{\mathrm{H_2}} (\cms)$ & 2.67e24$-$2.78e25 & 1.63e24$-$7.61e24 & 1.46e23$-$2.78e25 & 2.90e23$-$7.61e24 & 1.36e23$-$2.36e24 & 7.41e22$-$9.66e24 \\
$T_{\rm d}$ (K)$^*$    & 80$-$131 (101)  & 88$-$138 (100)  & 59.5$-$155 (103) & 61$-$138 (92) & 54.5$-$128 (98)  & 30$-$172 (44) \\
$n_{\mathrm{H_2}} (\cmq)$  & 4.7e7$-$1.5e9 & 2.8e7$-$4.8e8 & 9.0e5$-$1.5e9 & 2.2e6$-$4.8e8 & 9.6e5$-$1.5e8 & 1.0e6$-$8.2e8 \\
$\gamma$ & 0.3  & 1.0  & 0.3  & 1.0 & 0.1 & 0.3 \\
$a_{\rm{max}}$ ($\mu$m) & 0.35$-$2.0  & 0.35$-$2.0  & 0.35$-$2.0  & 0.35$-$2.0 & 0.25$-$1.0 & 0.50$-$2.0   \\
$\eta$   & 0.46  & 0.62  & 0.92 & 0.95 & 0.21  & 0.59  \\
$P_{\mathrm{obs}}$ (\%)   & 0.04$-$1.6  & 0.06$-$1.2 & 0.04$-$7.0 & 0.06$-$5.2 & 0.17$-$8.0 & 0.16$-$10.5  \\
$S$ (degree)$^*$  &  4.4$-$52 (17) & 5.5$-$62 (22)  & 0.4$-$52 (10) & 1$-$62 (10) & 0.1$-$25 (6) & 0.1$-$51 (6)  \\
\hline
\end{tabular}
\end{table*}

\subsubsection{Impact of RAT-D effect}
Although we have taken $a_{\mathrm{max}}$ from 0.25 $\mu$m to 2.0 $\mu$m to account for significant grain growth expected for dense regions like hot cores \citep{Hirashita_2013}, in the regions where dust temperature (radiation intensity) is sufficiently high for RAT-D to be effective, the maximum size of grains is expected to be limited by $a_{\mathrm{disr}}$. The analytical calculations of $a_{\mathrm{disr}}$ show that temperatures $\sim150$ K are capable of disrupting the grains down to 0.03 $\mu$m size even at densities of $\sim10^6$ \cmq. However, there is a constraint on the maximum size of grains that can be disrupted, which depends on the fact that the average RAT efficiency can not exceed its maximum value. Therefore, only grains with $a_{\mathrm{disr}}$ $<$ $\bar{\lambda}/1.8$ can be disrupted, which gives an upper limit on $a_{\mathrm{disr}}$ \citep{Hoang.2021}. In this work, we find this cut-off on $a_{\mathrm{disr}}$ to be $<$ 0.55 $\mu$m. 

In \ngc, the maximum temperature in substructures is $\sim130-150$ K while peak density is $\sim10^9$ \cmq~(see Table \ref{tab:pars} for details). Therefore, to analyze the effect of RAT-D in \ngc, we turned on RAT-D in \texttt{DustPOL\_py} modeling. In the case of MM1, no significant RAT-D effect was found for $S_{\mathrm{max}}$ = 10$^5$ erg \cmq. Although the temperature in MM1 is high, the gas density in this region is also very high, and $a_{\mathrm{disr}} \propto \sqrt{n_{\rm H}}$ (see Eq.~\ref{eq:adisr}). As a result, $a_{\mathrm{disr}}$ becomes larger at higher densities and does not satisfy the grain-size threshold for disruption ($a_{\mathrm{disr}} < \bar{\lambda}/1.8 = 0.55\, \mu$m) in most of the pixels within MM1. A map of $a_{\mathrm{disr}}$ for \ngc~is shown in Figure \ref{align_disr_maps}(c) in Appendix \ref{grain_size_maps}.  For other cases, we turn on RAT-D for $a_{\mathrm{max}} \geq$ 0.50 $\mu$m, while for MM1, RAT-D is not considered. We find that RAT-D somewhat reduces the polarization degree, especially at relatively low density ($N_{\mathrm{H_2}}\, \lesssim$7 $\times 10^{23}$ \cms). However, even with RAT-D, the ideal model ($P_{\mathrm{mod,\, id,\, RATD-ON}}$) remains significantly higher than $P_{\mathrm{obs}}$ (see the golden dots in Figure \ref{dustpol_diff_amax}). Furthermore, the effect of RAT-D at higher densities is found to be minimal due to the aforementioned reasons, suggesting the presence of additional factors responsible for the observed low polarization in \ngc.

\subsubsection{Realistic polarization model with B-field tangling/inclination}
\label{model_w_B-tang}
As discussed in Section \ref{ideal_model}, B-fields are assumed to be on the POS in the ideal model, but in reality, they can be inclined with respect to the LOS and thus affect the overall polarization fraction integrated along the LOS. Additionally, the B-field tangling along the LOS and in the POS within the beam can overall reduce the average polarization. \cite{Hoang_2024} formulated the general expression for the dust polarization model given by 
\begin{eqnarray}
    P_{\rm mod} \approx P_{\mathrm{mod,\, id}} \sin^{2}\gamma_{\rm B} F_{\rm turb},\label{eq:pmodel}
\end{eqnarray}
where $F_{\rm turb}$ is the depolarization coefficient caused by B-field tangling and $\gamma_{\rm B}$ is the B-field inclination angle (angle between the B-field and the LOS) that incorporates the effect of projection on polarization.

Due to the anti-correlation of $F_{\rm turb}$ with the polarization angle dispersion $S$ \citep{Hoang_2024}, one can write Equation (\ref{eq:pmodel}) as 
\begin{equation}
    P_{\mathrm{mod}} = \Phi\,P_{\mathrm{mod,\, id}}\,\left(\frac{S}{1\degree}\right)^{-\eta}, 
\end{equation}
where $\Phi$  is a coefficient that accounts for the B-field's inclination angle and the relationship between $F_{\rm turb}$ and $S$, and $\eta > 0$ quantifies the effect of B-field tangling along the LOS and in the POS \citep{Ngoc_2024}.  The $\eta$ can be calculated from the power-law fit over the observed $P-S$ relation. The value of $\Phi$ is not known, but we can initially fix $\Phi$ using the low-end values of $S\approx 1^{\circ}$ and taking $P_{\mathrm{mod}} \approx P_{\mathrm{mod,\, id}}$, yielding $\Phi\approx 1$. Therefore, we first fix $\Phi$ = 1.0 and vary $a_{\mathrm{max}}$ to analyze the effect of grain growth, RAT alignment, RAT-D, and B-field tangling. 

At sufficiently short wavelengths, dust emission from dense regions of a star-forming region can become optically thick. In this regime, polarized emission from deeper layers can be selectively absorbed by aligned dust grains in the outer layers through dichroic extinction \citep{Hildebrand_2000}. As the optical depth increases, the contribution from dichroic extinction can become significant compared to the polarized thermal emission, potentially rotating the observed polarization direction by $90^\circ$ relative to the optically thin case. This transition can also modify and reduce the observed polarization fraction \citep{Hildebrand_2000}. Assuming that $P_{\mathrm{mod,\, id}}$ is the polarization degree in the optically thin case, then the polarization degree including the effect of optical depth ($\tau$) can be calculated as \citep{Hildebrand_2000}

\begin{equation}
\label{eqn:Ptau}
    P_\tau = \frac{e^{-\tau} \sinh (P_{\rm{mod,\, id}})}{[1-e^{-\tau} \cosh(P_{\rm{mod,\, id}})]}.
\end{equation}
The details for estimation of optical depth in \ngc~are given in Appendix \ref{optical_depth}. When the optical depth effect is considered, the modeled polarization is denoted by $P_{\mathrm{mod,\, \tau}}$, 

\begin{equation}
\label{eqn:Pmod_tau}
    P_{\mathrm{mod,\, \tau}} = \Phi\,P_{\mathrm{\tau}}\,\left(\frac{S}{1\degree}\right)^{-\eta}. 
\end{equation}

\begin{figure*}
    \centering
    \includegraphics[width=8.5cm]{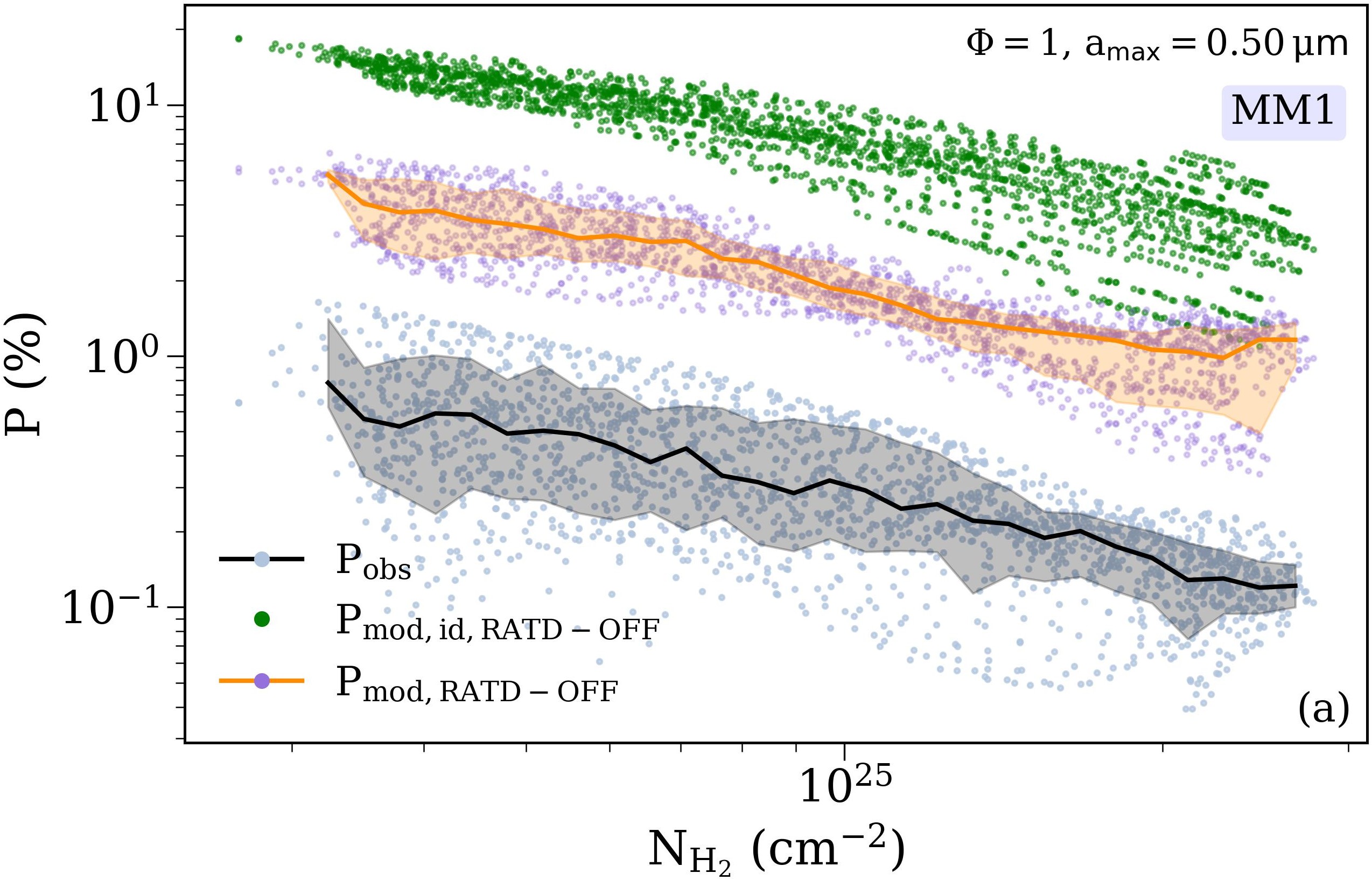} 
    \includegraphics[width=8.5cm]{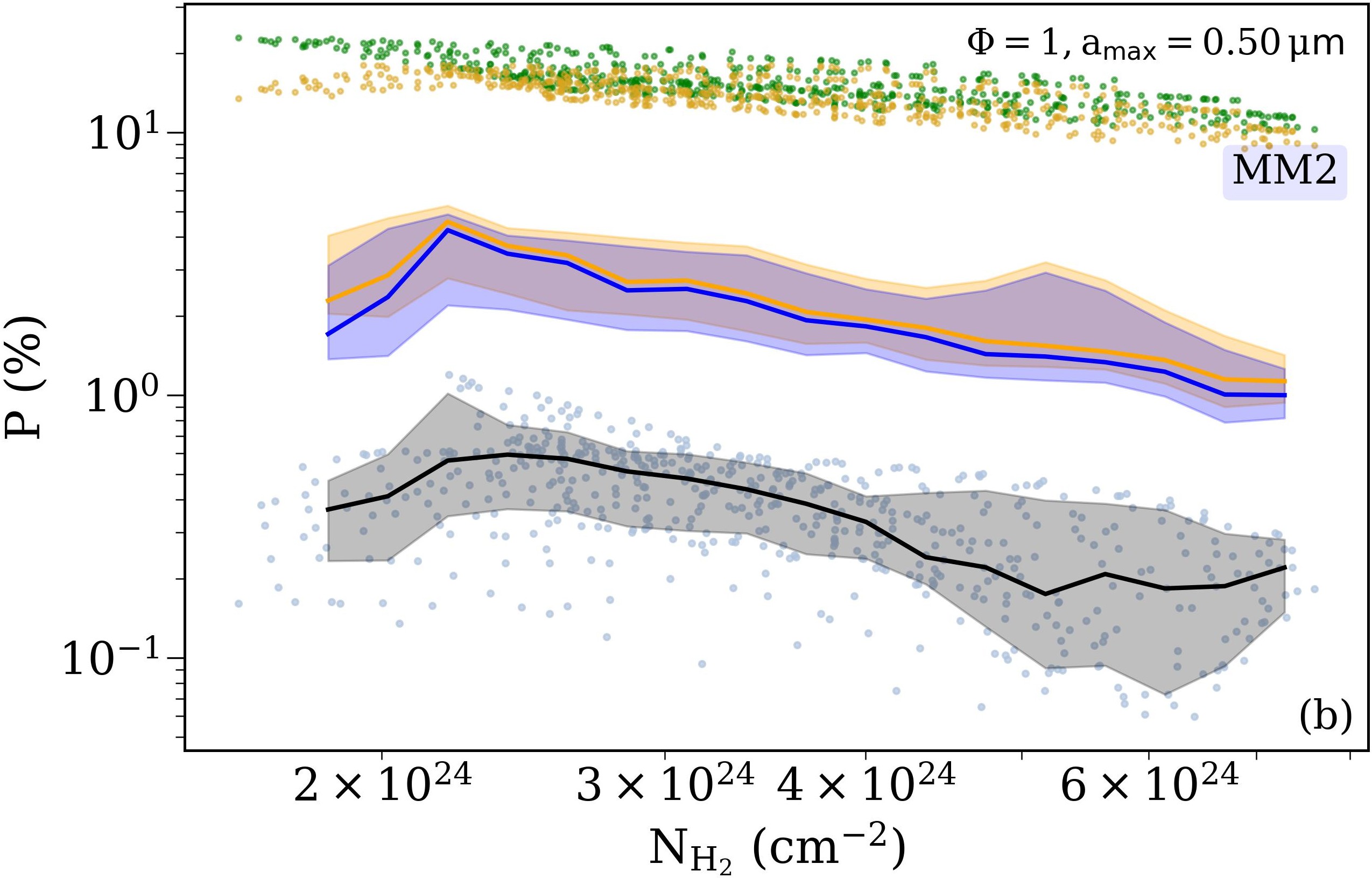}
    \includegraphics[width=8.5cm]{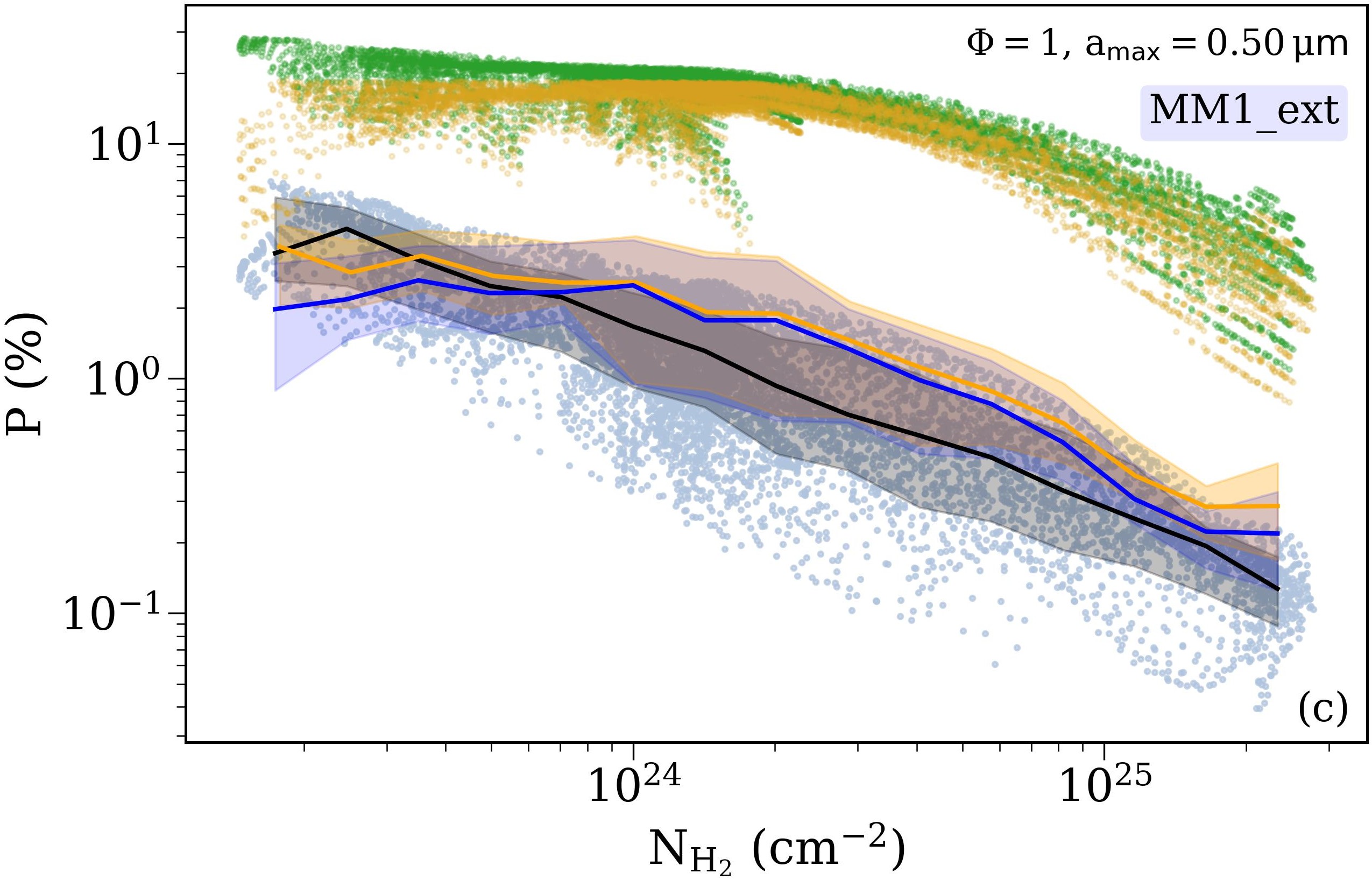}
    \includegraphics[width=8.5cm]{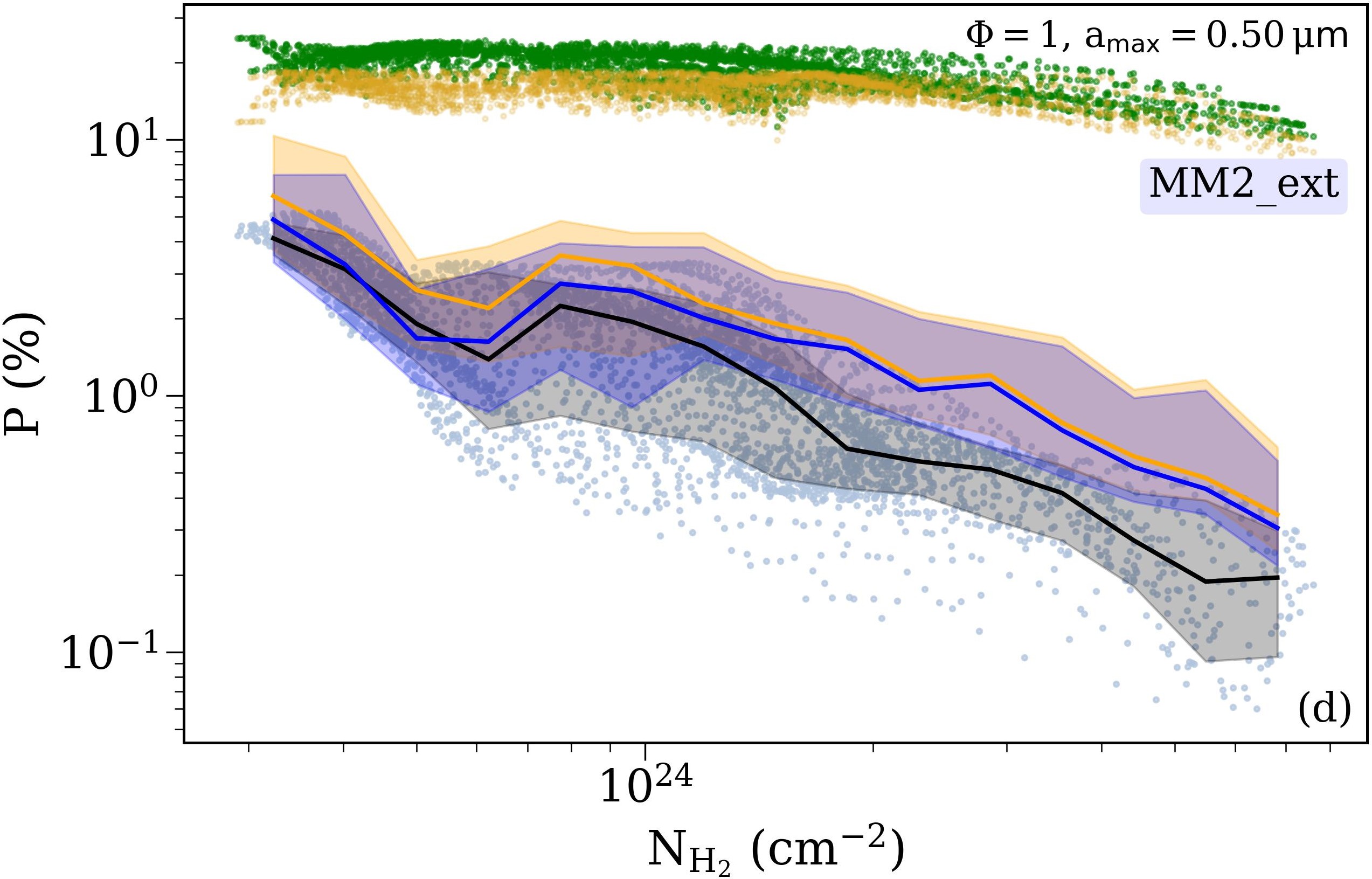}
    \includegraphics[width=8.5cm]{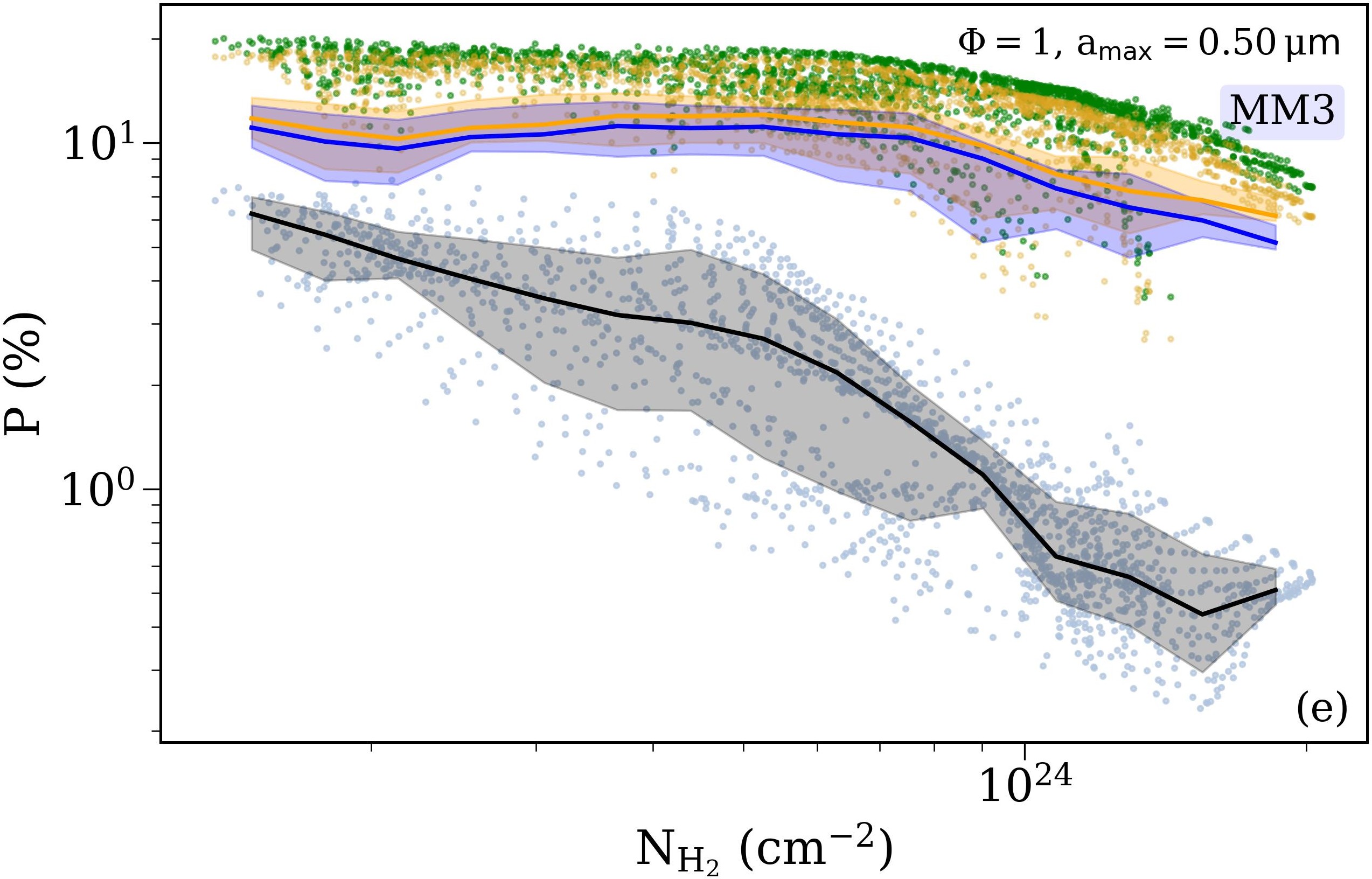}
    \includegraphics[width=8.5cm]{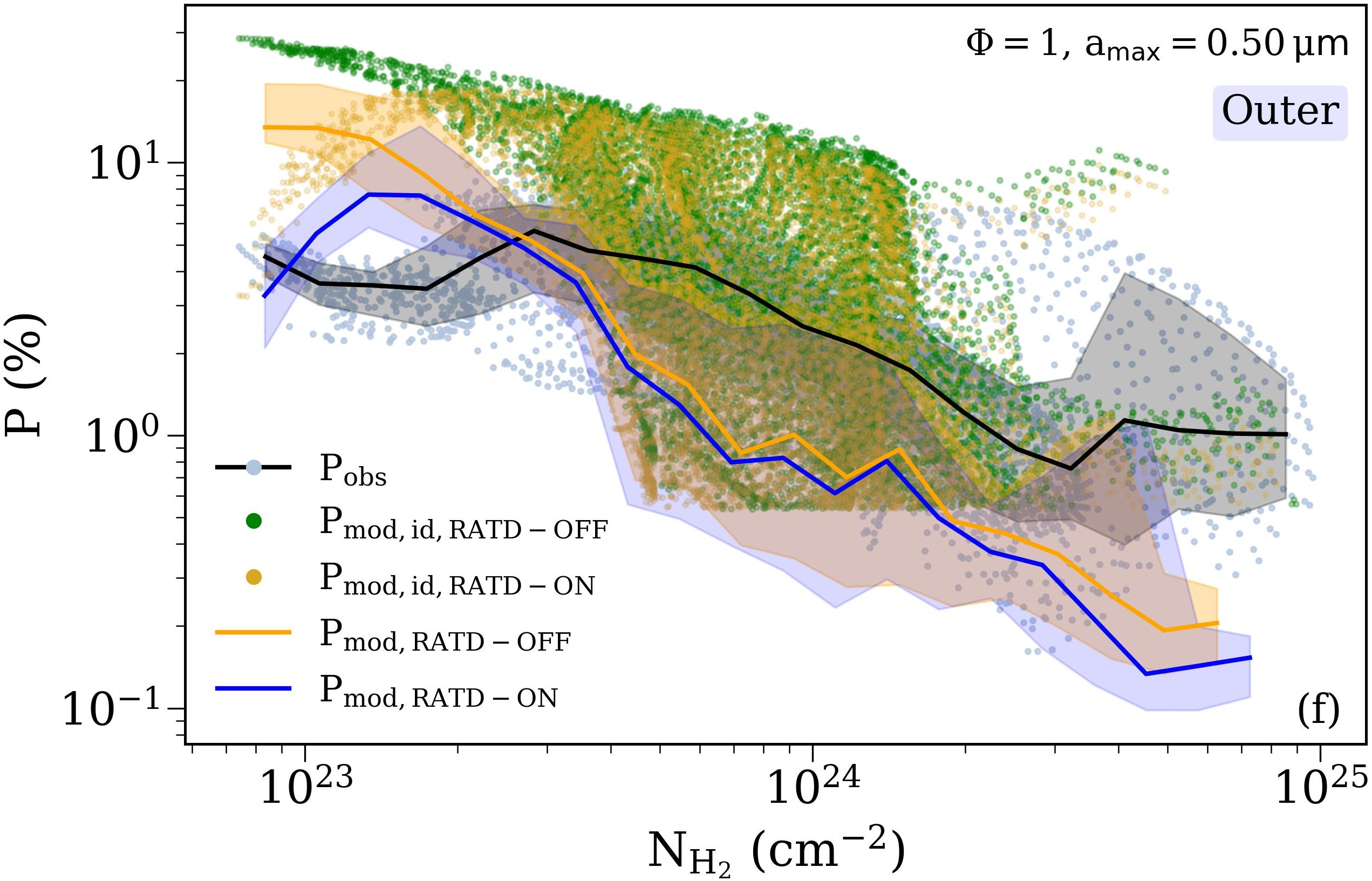} 
 \caption{Results of \texttt{DustPOL\_py} modeled polarization fraction ($P$) versus $N_{\rm H_2}$ for (a) MM1, (b) MM2, (c) MM1\_ext, (d) MM2\_ext, (e) MM3, and (f) the Outer region, assuming $\Phi=1$ and $a_{\rm max}=0.50\,\mu\mathrm{m}$. Light-blue points show the observed polarization fractions, with the black curve and gray shaded region representing the running median and the 16th--84th percentile range, respectively. For MM1 (panel a), where RAT-D is not considered, the green points denote the ideal polarization model, while the purple points correspond to the realistic polarization model (i.e., including B-field tangling and inclination effects). The running median and the 16th--84th percentile range of the realistic model are shown by the orange curve and orange shaded region, respectively. For the remaining regions (panels b-f), green and gold points show the ideal models with RAT-D OFF and ON, respectively. The orange and blue curves represent the running medians of the corresponding realistic modeled polarization fractions, and the shaded regions mark their 16th--84th percentile ranges. RAT-D is applied only for models with $a_{\rm max}\geq0.50\,\mu\mathrm{m}$. The plots for other $a_{\rm max}$ values are shown in Appendix \ref{Pmodel_appendix}. 
 }
    \label{dustpol_diff_amax}
\end{figure*}

The resulting polarization models are shown in Figure \ref{dustpol_diff_amax} for $a_{\mathrm{max}}$ = 0.50 $\mu$m with RATD-ON/OFF, except for MM1, where RAT-D is not considered. For other $a_{\mathrm{max}}$ values, the plots are shown in Appendix \ref{Pmodel_appendix}.  
It can be seen from the figures that the realistic polarization model not only better reproduces the observed polarization trends but also reduces the modeled polarization fractions, bringing them closer to the observed $P$. For almost all the regions, $P_{\rm mod}$ becomes flatter or deviates more from $P_{\rm obs}$ at higher column densities as $a_{\mathrm{max}}$ increases, for a fixed value of $\Phi$. 
Based on the Spearman correlation coefficients ($\gtrsim0.8$) and root mean square errors (RMSEs) of the running median of modeled polarization, we constrained the $a_{\mathrm{max}}$ values for each region. Using these statistics, we find $a_{\mathrm{max}}$ = 0.35 $\mu$m and 0.50 $\mu$m for MM1; 0.35 $\mu$m, 0.50 $\mu$m, and 1.0 $\mu$m for MM2; 0.35 $\mu$m and 0.50 $\mu$m for MM1\_ext; 0.35 $\mu$m, 0.50 $\mu$m, and 1.0 $\mu$m for MM2\_ext; 0.25 $\mu$m for MM3; and 1.5 $\mu$m and 2.0 $\mu$m for the Outer region, best reproduces the observed polarization fractions. Then, fixing these $a_{\mathrm{max}}$ values for every region, we vary the $\Phi$ parameter to visualize the effect of depolarization due to B-field inclination. We take $\Phi$ in the range of 0.3 to 0.7, and $\Phi_{\mathrm{norm}}$, which is obtained by normalizing $P_{\mathrm{mod}}$ with the observed values of $P$ in each region. 

\begin{figure*}
    \centering
    \includegraphics[width=8.9cm]{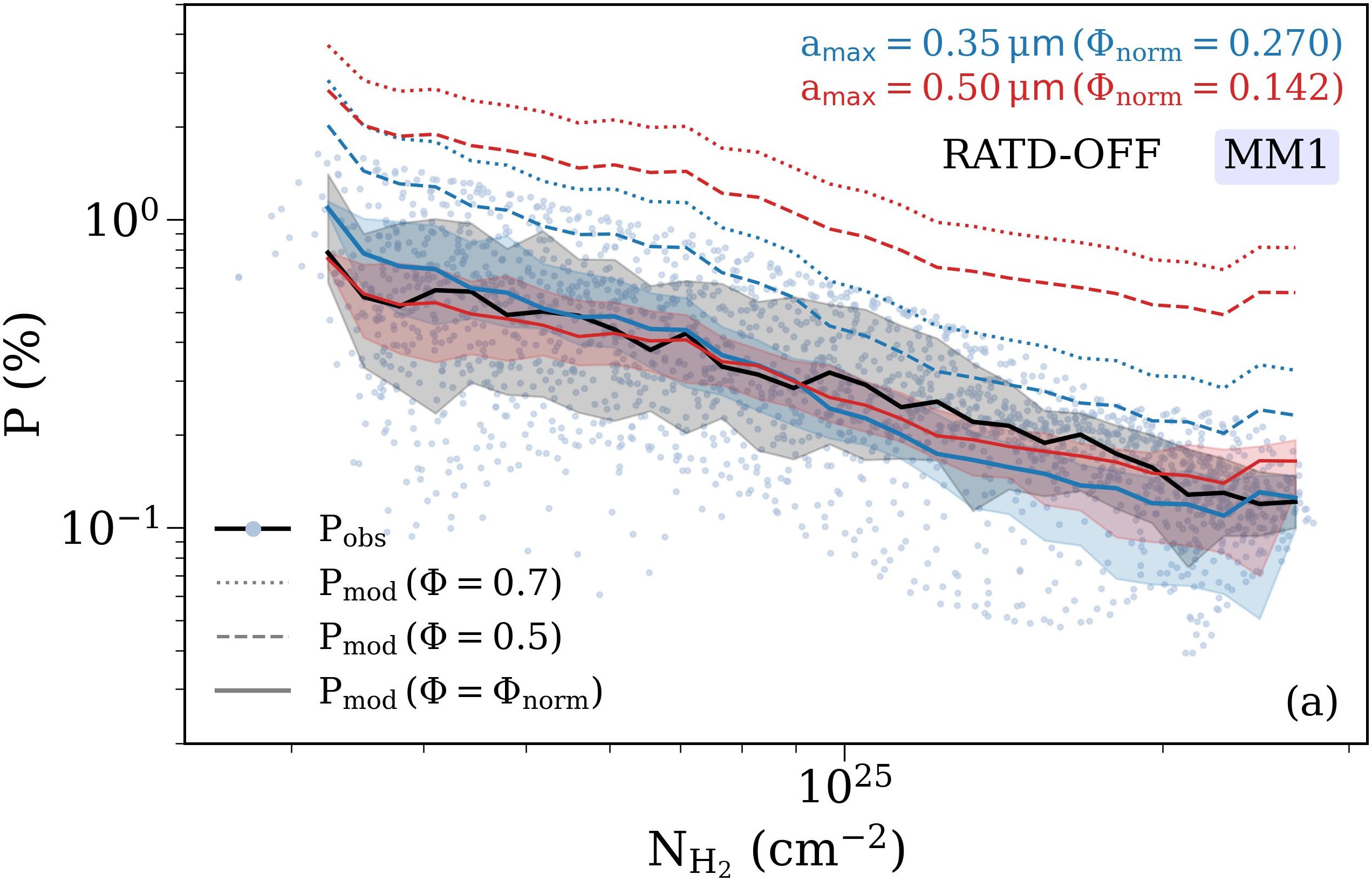}
     \includegraphics[width=8.9cm]{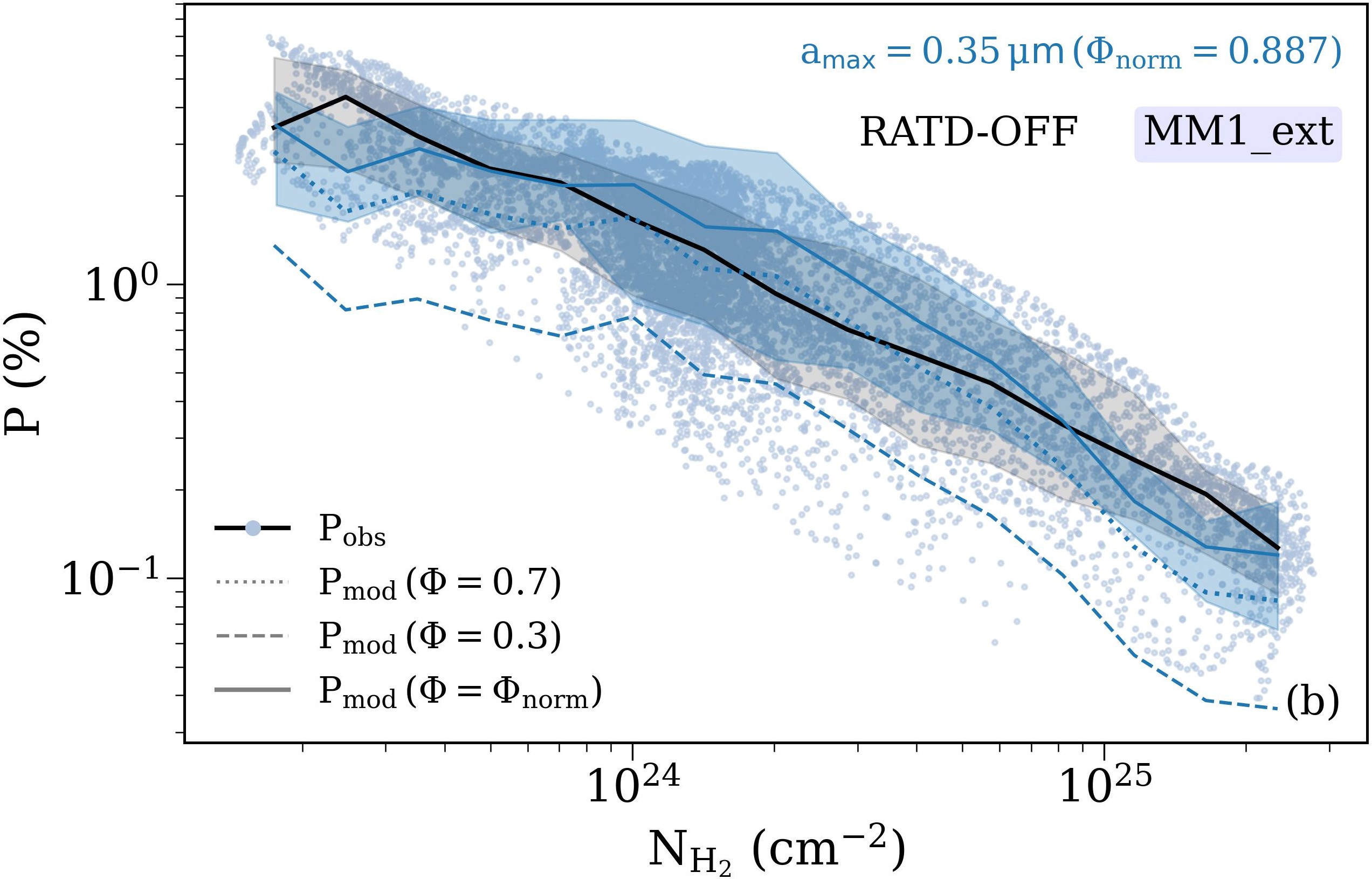}
 \caption{Model results for (a) MM1, using $a_{\mathrm{max}}$ = 0.35 $\mu$m (blue curves) and 0.50 $\mu$m (red curves) and (b) MM1\_ext, using $a_{\mathrm{max}}$ = 0.35 $\mu$m, for different values of $\Phi$. The light-blue dots and black curve are the same as in Figure \ref{dustpol_diff_amax}.}
    \label{MM1_model_combined}
\end{figure*}

\begin{figure*}
    \centering
    \includegraphics[width=5.9cm]{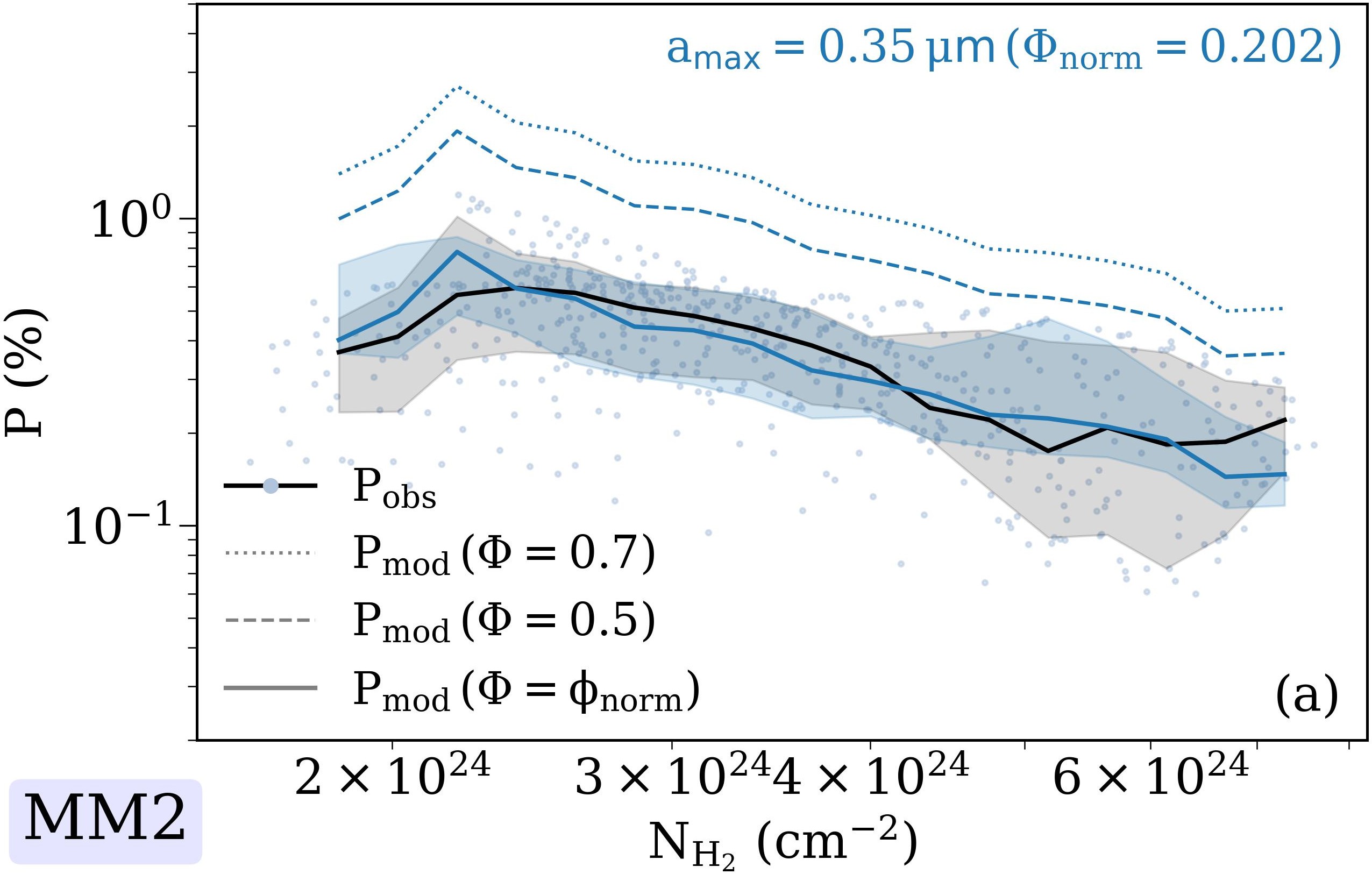} 
    \includegraphics[width=5.9cm]{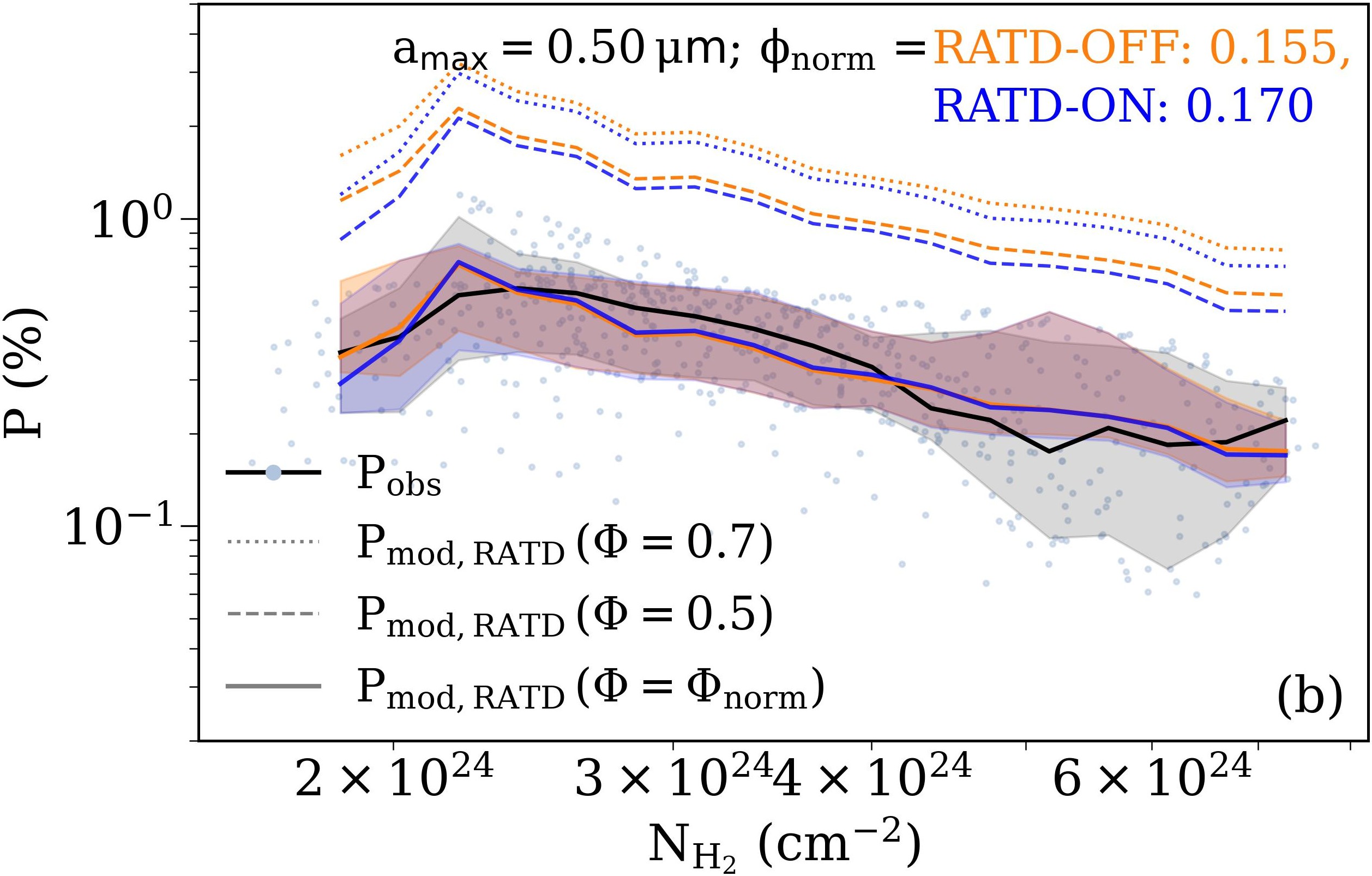}
    \includegraphics[width=5.9cm]{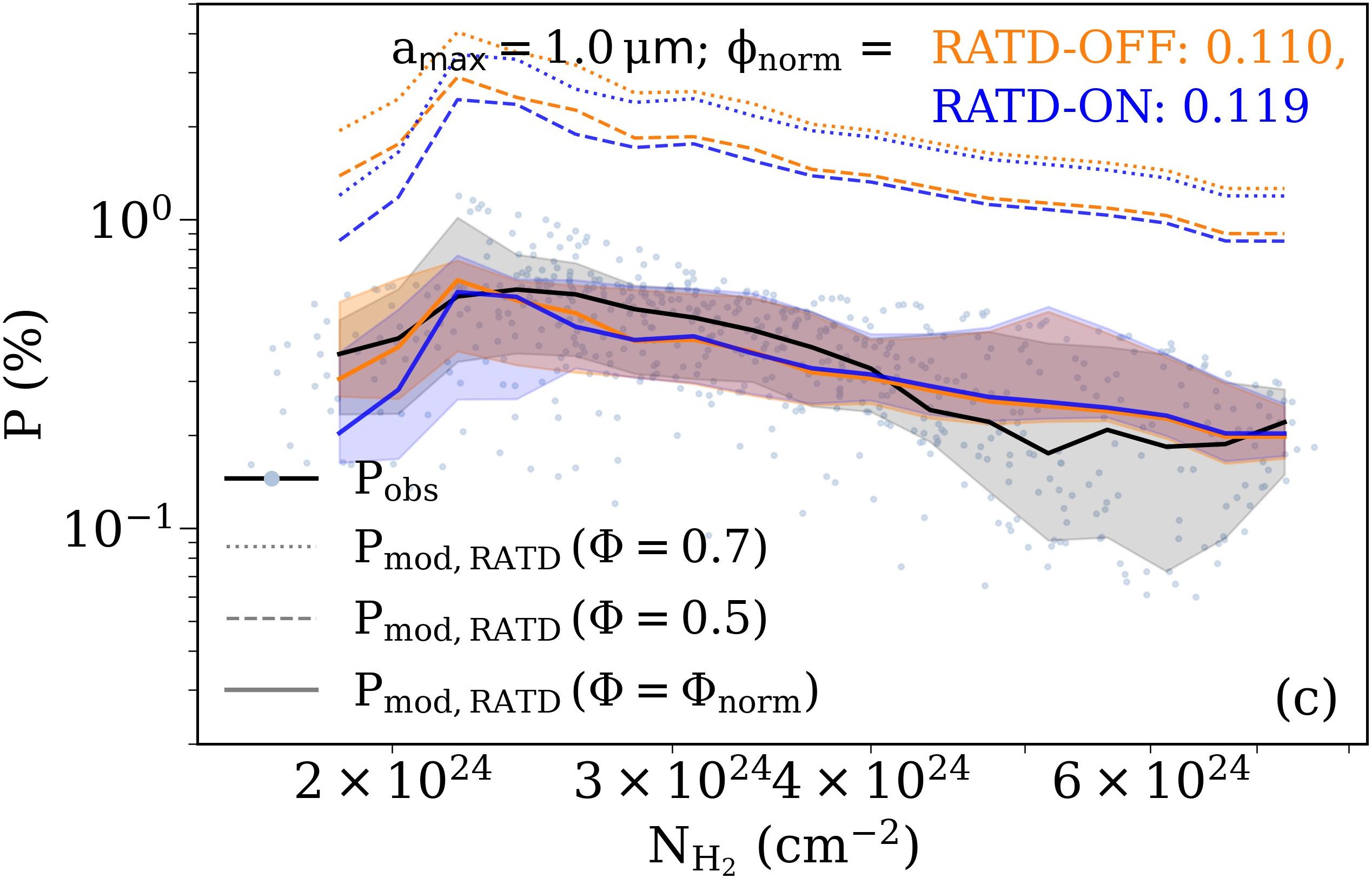}
 \caption{Model results for MM2, using $a_{\mathrm{max}}$: (a) 0.35 $\mu$m, (b) 0.50 $\mu$m, and (c) 1.0 $\mu$m, for different values of $\Phi$. The light-blue dots and black curve are the same as in Figure \ref{dustpol_diff_amax}. In the middle and right panels, the orange and blue curves represent the RATD-OFF and RATD-ON models, respectively.}
    \label{MM2_model_combined}
\end{figure*}

\begin{figure*}
    \centering
    \includegraphics[width=5.9cm]{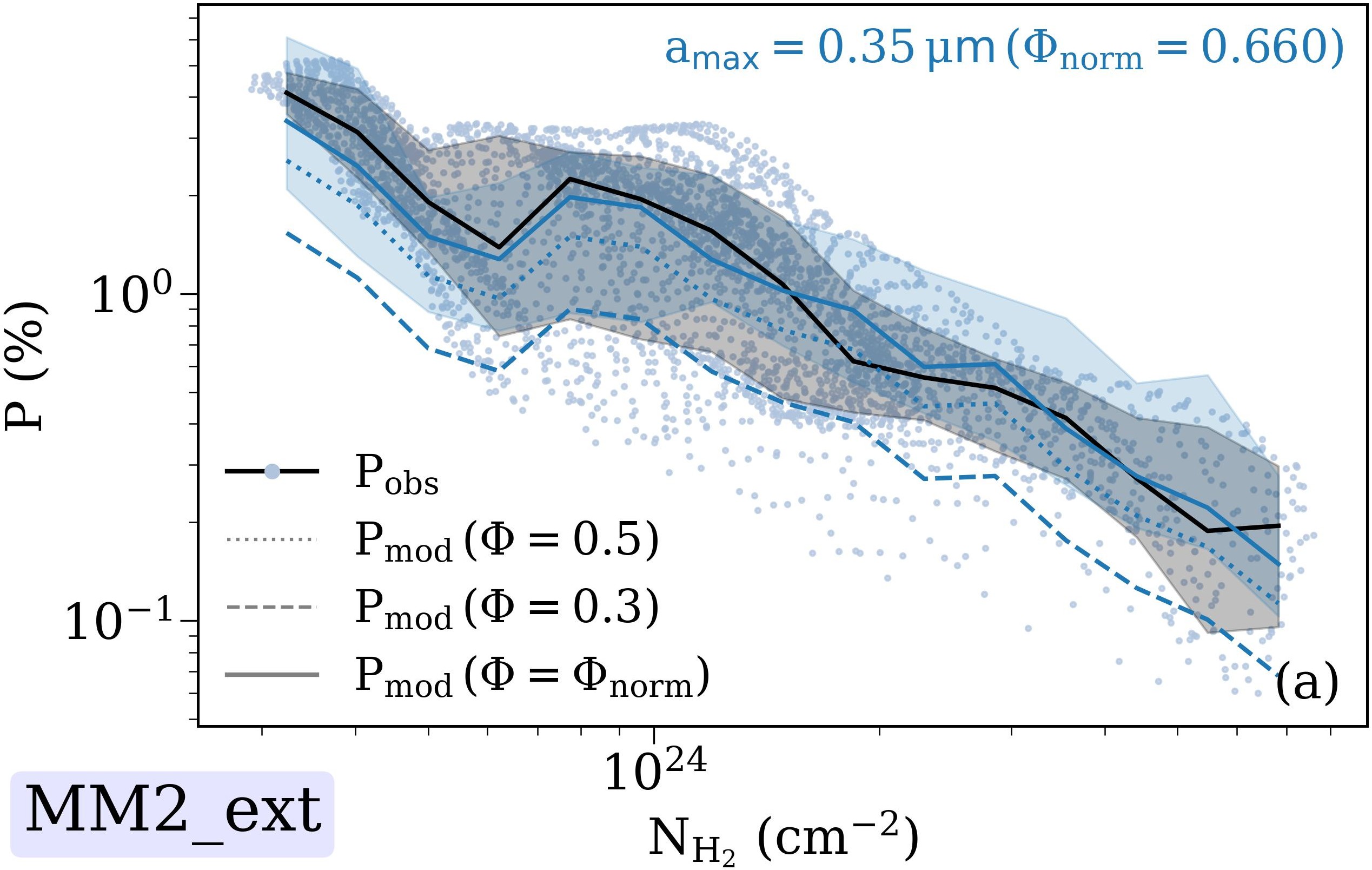}
    \includegraphics[width=5.9cm]{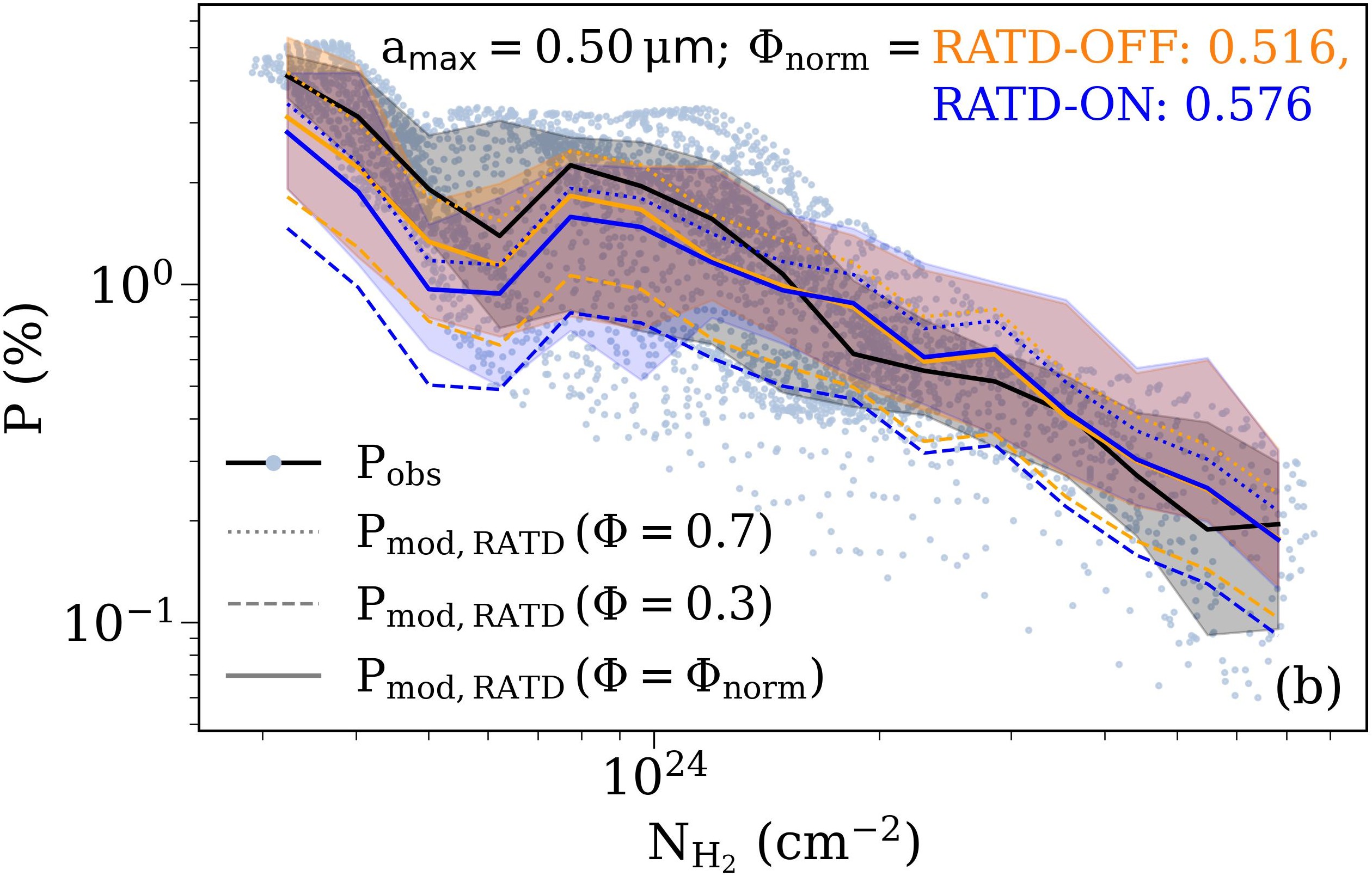} 
    \includegraphics[width=5.9cm]{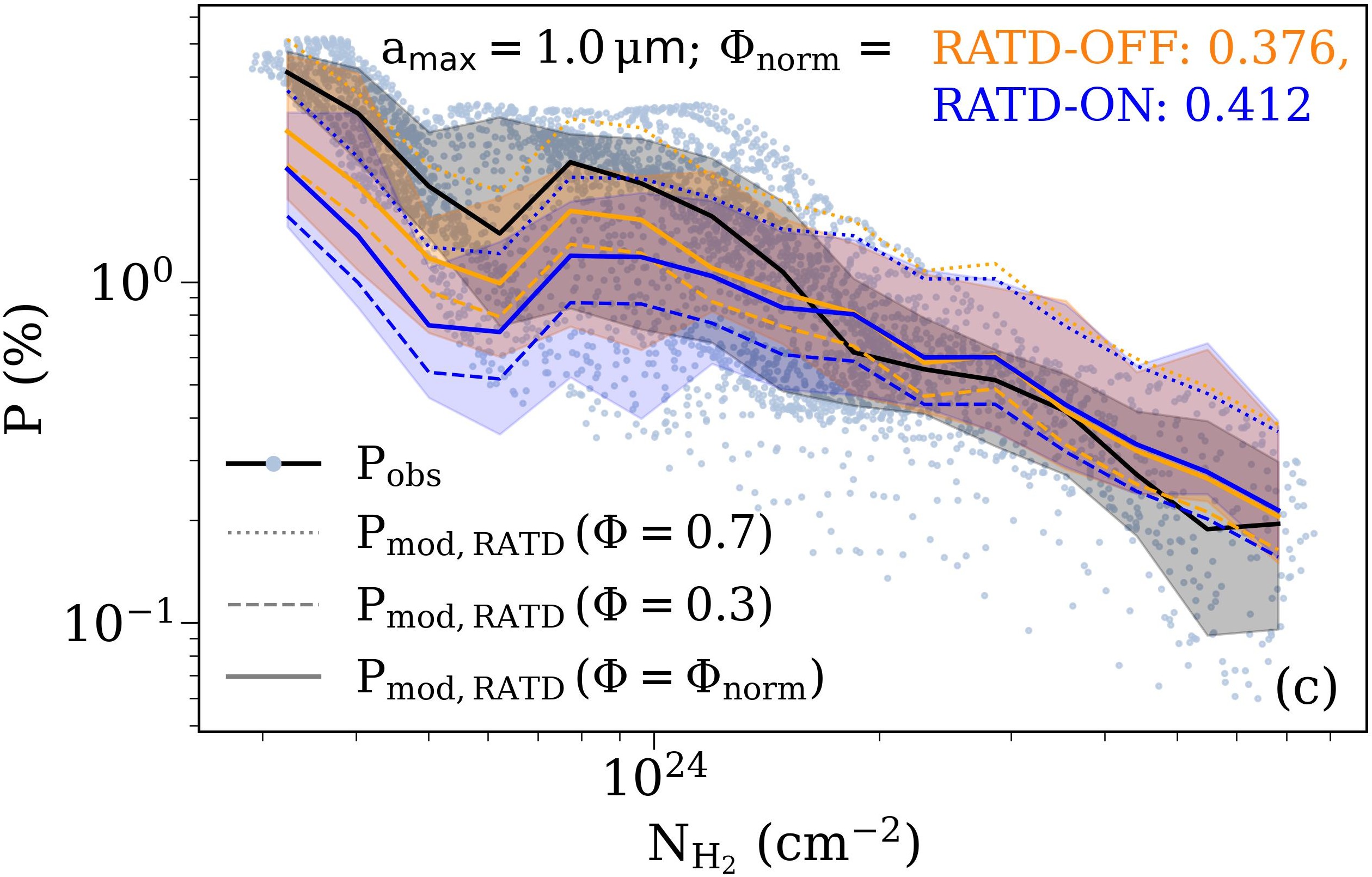}
 \caption{Model results for MM2\_ext, using $a_{\mathrm{max}}$: (a) 0.35 $\mu$m, (b) 0.50 $\mu$m, and (c) 1.0 $\mu$m, for different values of $\Phi$. The colored curves are the same as in Figure \ref{MM2_model_combined}.}
    \label{MM2_larger_model}
\end{figure*}

\begin{figure}
    \centering
    \includegraphics[width=8.5cm]{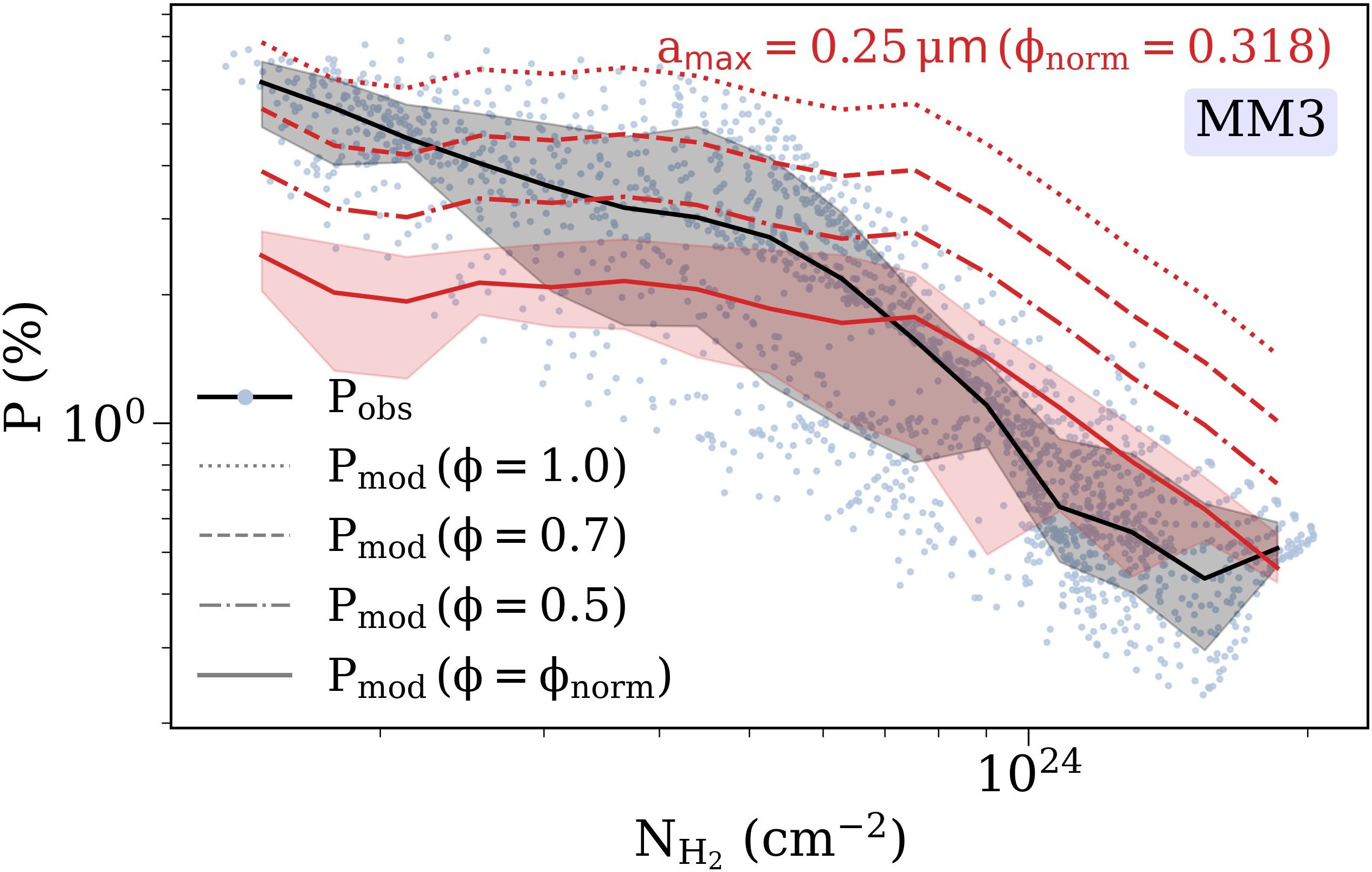}
 \caption{Model results for MM3 using $a_{\mathrm{max}}$ = 0.25 $\mu$m for different values of $\Phi$. The light-blue dots and black curve are the same as in Figure \ref{dustpol_diff_amax}.}
    \label{MM3_large_model}
\end{figure}

\begin{figure*}
    \centering
    \includegraphics[width=8.5cm]{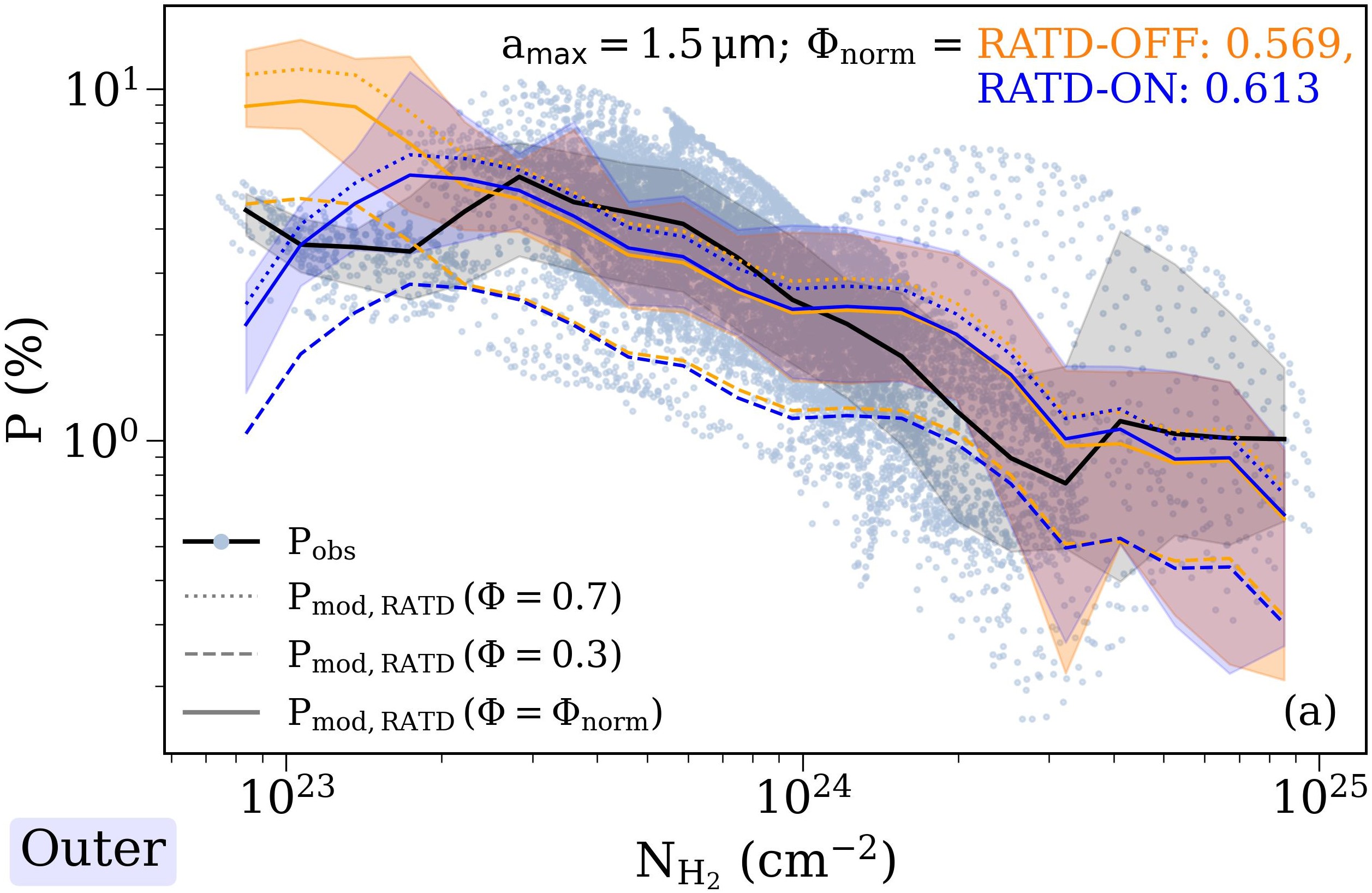} 
    \includegraphics[width=8.5cm]{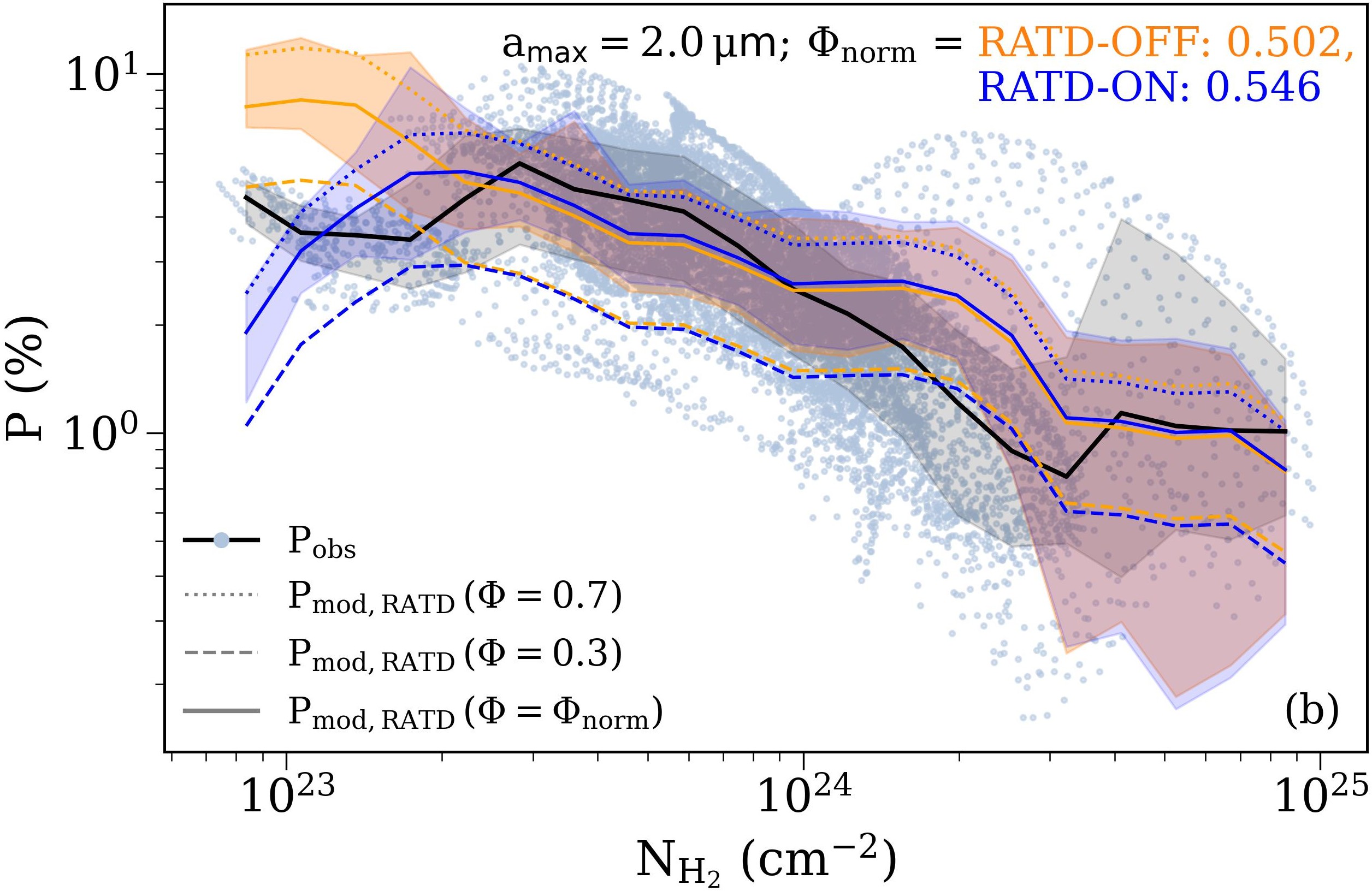}
 \caption{Model results for Outer region, using $a_{\mathrm{max}}$: (a) 1.5 $\mu$m and (b) 2.0 $\mu$m, for different values of $\Phi$. The colored curves are the same as in Figure \ref{MM2_model_combined}.}
    \label{Outer_model_combined}
\end{figure*}

\begin{figure*}
    \centering
    \includegraphics[width=8.9cm]{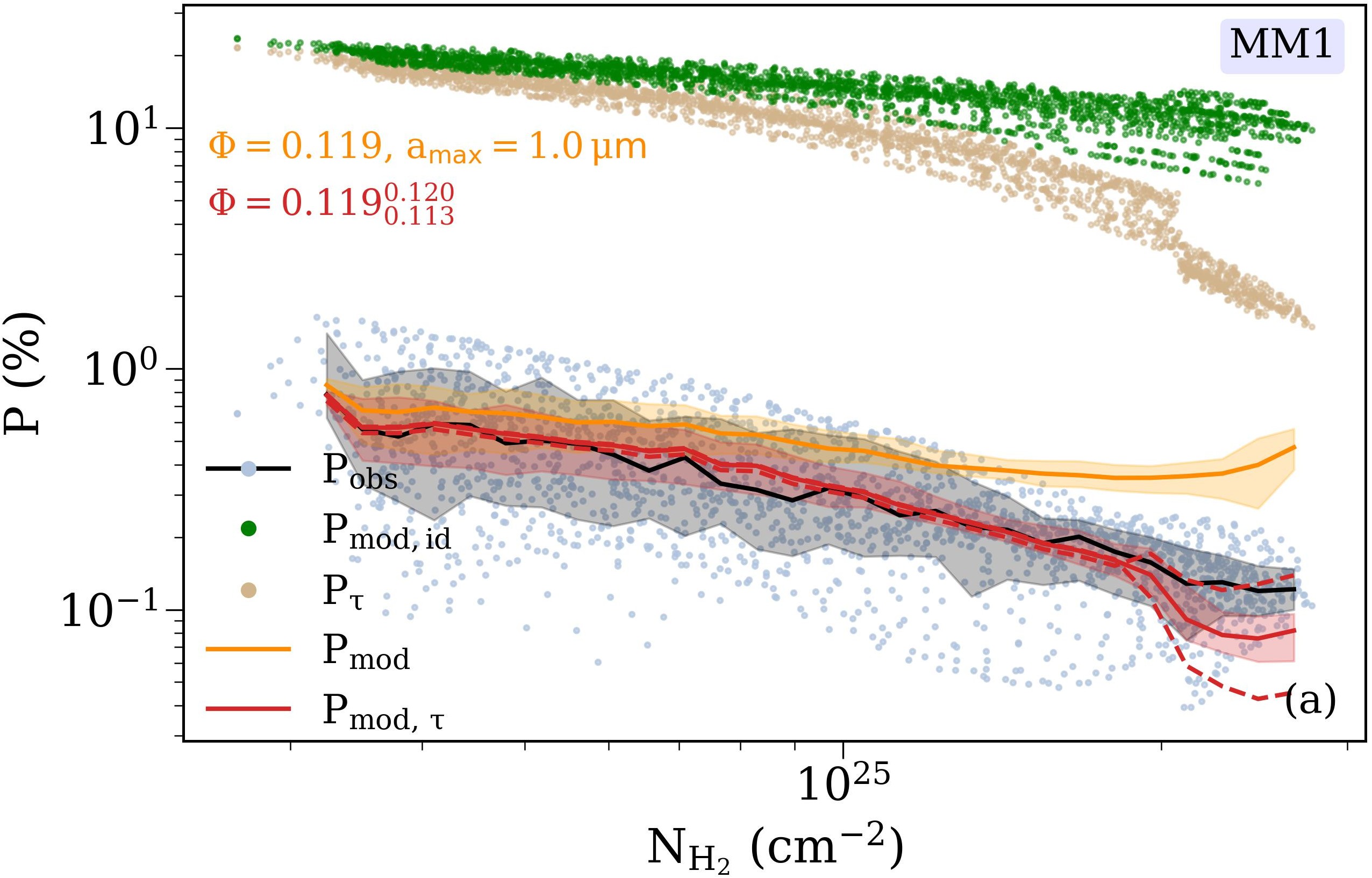}
     \includegraphics[width=8.9cm]{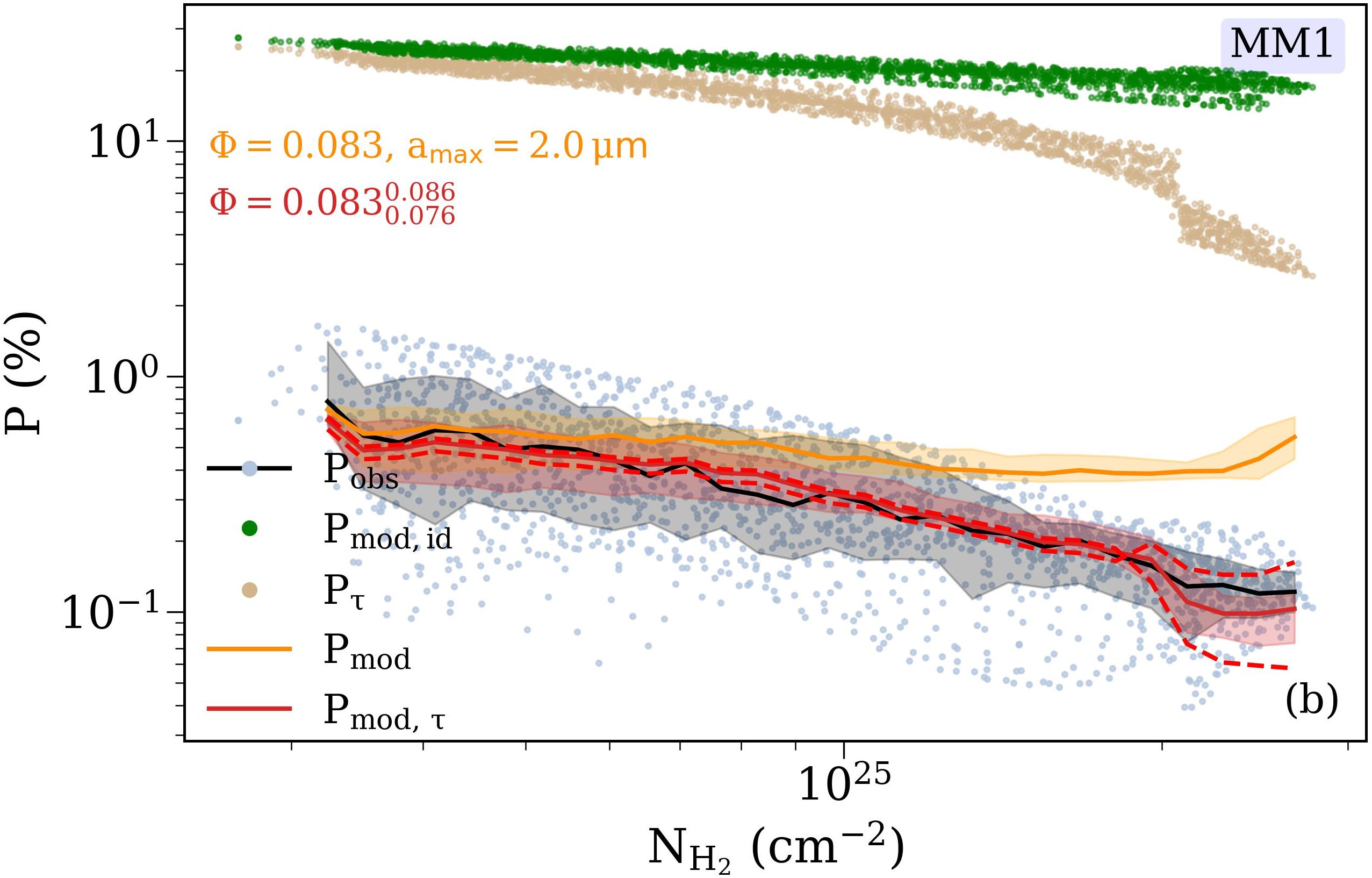}
 \caption{Model results for MM1 including the optical depth ($\tau$) effect and using equations \ref{eqn:Ptau} and \ref{eqn:Pmod_tau} for (a) $a_{\mathrm{max}}$ = 1.0 $\mu$m and (b) 2.0 $\mu$m. The green, light-blue dots, and black solid curve are the same as in Figure \ref{dustpol_diff_amax}. The tan-colored dots show the ideal polarization model that includes optical depth (i.e., equations \ref{eqn:Ptau}). The orange and red curves represent the median realistic polarization models without and with optical depth, respectively. The corresponding shaded regions indicate the 16th–84th percentile ranges around the median models, while the dashed-red curves correspond to an uncertainty of 30\% in the optical depth (see Appendix \ref{optical_depth}).}
    \label{MM1_model_tau}
\end{figure*}

The comparisons of modeled polarization degree for fixed $a_{\mathrm{max}}$ and varying $\Phi$ values with the $P_{\mathrm{obs}}$ are shown in Figure \ref{MM1_model_combined} for MM1  and MM1\_ext, \ref{MM2_model_combined} for MM2, \ref{MM2_larger_model} for MM2\_ext, \ref{MM3_large_model} for MM3, and \ref{Outer_model_combined} for Outer. The corresponding Spearman correlation coefficients and the RMSE values of the comparisons are given in Table \ref{tab:results}. From the figures, it is clearly evident that the realistic dust polarization model with the lowest RMSE B-field inclination factor and the aforementioned best-matched $a_{\mathrm{max}}$ values (with RATD-ON/OFF) reproduces both the observed polarization trend and the polarization degree very well, except for MM3. In MM3, the differences between the observed and modeled polarization degree are higher, specifically in the lower-density part ($\lesssim 7 \times 10^{23}\, \cms$). The result here suggests that the model with $a_{\mathrm{max}}$ = 0.35 and 0.50 $\mu$m best reproduces the observed polarization in MM1, with $\Phi$ decreasing from 0.270 to 0.142. However, considering the optical depth effect in MM1 and using equations \ref{eqn:Ptau} and \ref{eqn:Pmod_tau}, we find that modeled polarization with $a_{\mathrm{max}}$ = 1.0 and 2.0 $\mu$m is also applicable (see Figure \ref{MM1_model_tau}). From Figure \ref{MM1_model_tau}, it can be seen that including optical depth in the polarization model predominantly affects the polarization fractions in the densest regions of MM1, while the effect is less pronounced in the optically thin outer regions.  
From Table \ref{tab:results}, it is evident that the high-density regions (MM1 and MM2) exhibit lower $\Phi$ values, whereas the relatively low-density regions (MM1\_ext, MM2\_ext, MM3, and Outer) show higher $\Phi$ values, which hints towards the change in B-field inclination with density (see discussion in Section \ref{diss_6}).

\section{Discussion}
\label{sec:discuss}
Across the entire NGC6334I region, the polarization fraction decreases with increasing intensity and column density. This trend can partly arise from fluctuations in the B-field orientation within the telescope beam and along the LOS. The regions in \ngc~with polarization angle dispersion of $\sim20\degree-50\degree$ indicate significant fluctuations in the B-field structure and suggest that tangled B-fields contribute to the depolarization observed in \ngc. 
However, the behavior of the $S \times P$ parameter indicates that B-field tangling alone cannot fully explain the observed depolarization. 
Changes in dust grain alignment efficiency or grain size distribution can also play an important role in depolarization in \ngc. In this section, we present the insights obtained from the correlation plots of $P-I$, $P-T_{\rm d}$, and $P-N_{\mathrm{H_2}}$, along with the \texttt{DustPOL\_py} modeling of thermal dust polarization in \ngc. Along with this, we discuss the impact of other factors, such as optical depth and flux loss, as well as the associated limitations and caveats of our analysis.

\subsection{RAT Alignment by Protostellar Radiation in \ngc}
\label{diss_1}

Our results show that although the polarization fraction decreases with increasing density in the sub-regions of \ngc, i.e., MM1, MM2, and MM3, the polarization fraction is higher in some regions exposed to enhanced radiation, as evident from the $P$ vs.\ $T_{\rm d}$ plots in Figure \ref{corr_full} and for the MM1 protocluster in Figure \ref{corr_MM1}. This behavior is consistent with expectations from RAT-A theory, in which grain alignment efficiency depends primarily on the local radiation field, density, and dust grain properties. The high polarization in relatively low-density outer regions can also arise from alignment caused by interstellar radiation flux. However, in high-density regions at moderate temperatures, collisional disalignment of dust grains becomes more prominent than the temperature effect, leading to reduced polarization. 

Also, from analytical calculations of the minimum size of aligned grains, it is clear that high dust temperatures can enable grain alignment down to sizes of $\sim0.015$ $\mu$m (with $a_{\mathrm{align}}$ = 0.032 $\mu$m at a given $n_{\rm{H_2}}$ = 10$^7$ \cmq).  This suggests that radiative torques remain dynamically relevant even in deeply embedded but warm environments. However, there are some regions with small $a_{\mathrm{align}}$ (i.e. $<$ 0.1 $\mu$m) and densities of $\sim10^7$ to 10$^{7.5}$~\cmq, but still show relatively low polarization (i.e. $P\, <$ 1\%, see Figure \ref{a_align_dis}(a) right-panel). Such a low polarization degree can be attributed to a high polarization angle dispersion (see Figure \ref{align_w_S} in Appendix \ref{S_diff}), indicative of a turbulent B-field. 
This observational evidence suggests that the polarization is affected not only by collisional disalignment of grains but also by B-field tangling. Another justification for this effect comes from the modeling of thermal dust polarization in \ngc. 
In every case, we found that although the ideal polarization model (i.e., uniform B-field perfectly on the POS) shows a declining trend with density, it remains significantly higher than the observed polarization. 
This means that the other factors (B-field tangling, RAT-D, and optical depth) are important here to explain the depolarization.

\subsection{Depolarization due to B-field Tangling and Inclination}
\label{diss_2}
In turbulent environments such as NGC6334I, strong protostellar outflows and gravitational accretion gas flows can distort the B-field morphology, producing rapid changes in field orientation across all spatial scales \citep[see][]{Tahani_2023}. In fact, \cite{Cortes_2024} found that the gravitational energy is dominant over the B-field, thermal, kinetic, and outflow energies in \ngc, with the outflow energy being an order of magnitude higher than the B-field energy. Moreover, the cores are found to be super-\alf~and supersonic based on C$^{33}$S line emission \citep{Cortes_2024}. Therefore, it is expected that gravity and the presence of multiple outflows are likely to inject turbulence and perturb the B-field geometry on the scales of the whole \ngc~to individual cores. 
After incorporating the effects of B-field tangling in the POS and along the LOS, as well as the B-field inclination with respect to the LOS, the modeled polarization reduced significantly and became consistent with the observed values. 
The variation in the $\Phi$ value suggests that a uniform B-field inclination angle across the entire \ngc~region is not possible. 
The observed hourglass morphology in MM1 and the flip of lines near MM2 and MM4, if believed to be truly in 3D space, suggest a significant change in B-field inclination in dense regions.

The best-matched modeled polarization for MM1, without accounting for optical depth effects, is obtained for $a_{\mathrm{max}} \lesssim 0.50~\mu$m (see Table \ref{tab:results}). A comparison of observed and modeled polarization fraction in MM1 for $a_{\mathrm{max}}$ = 0.50 $\mu$m is shown in Figure \ref{P_model_comp_MM1_MM2}(a) in Appendix \ref{Pmodel_appendix}. In MM2, it can be seen from Figure \ref{MM2_model_combined} that RAT-D has a minimal effect on the dust polarization, i.e., on the grain size distribution. 
All $a_{\mathrm{max}}$ values (i.e., from 0.35 $\mu$m to 1.0 $\mu$m) in the modeling seem favorable in MM2. 
The best-matched observed and modeled polarization fraction in MM2 for $a_{\mathrm{max}}$ = 0.50 $\mu$m is shown in Figure \ref{P_model_comp_MM1_MM2}(b) in Appendix \ref{Pmodel_appendix}. 
For MM1\_ext, $P_{\mathrm{mod,\,RATD-OFF}}$ with $a_{\mathrm{max}}$ = 0.35 $\mu$m and $\Phi$ = 1.0 is the most probable model. For MM2\_ext, like MM2, $P_{\mathrm{mod}}$ with $a_{\mathrm{max}}$ = 0.35--1.0 $\mu$m are following the $P_{\mathrm{obs}}$ trend  with $N_{\mathrm{H_2}}$. 
For MM3, none of the modeled polarization degrees, corresponding to any choice of $a_{\mathrm{max}}$ under either the RATD-OFF or RATD-ON scenario, fully reproduces the observed polarization. The best agreement is obtained for $a_{\mathrm{max}} = 0.25~\mu$m, although significant deviations from the observations persist at lower densities. Given the presence of a UC \hii region in MM3, along with other evolved \hii regions in the vicinity \citep[see Figure \ref{S_map} and also][]{Tahani_2023}, contamination of the continuum emission by free-free emission may affect the derived polarization properties. This difference might also be due to the varying inclination factor with density or flux loss in ALMA observations (see Section \ref{Caveats}). 

Our modeling results also reveal a degeneracy between $a_{\mathrm{max}}$ and $\Phi$ in reproducing the observed polarization fractions, particularly in MM1 and MM2. As $a_{\mathrm{max}}$ increases, lower values of $\Phi$ are required to reproduce the observed polarization. Therefore, if slightly lower $\Phi$ values (e.g., $\sim$0.11) are considered, grains of size $\sim$1–2 $\mu$m could be present in MM2. This would suggest dust grain growth in this region. Another important point is that the B-field is significantly perturbed in \ngc, and the resulting LOS polarization smearing may also result from the superposition of multiple pre-/protostellar cores along the LOS, particularly in MM1, which is a protocluster region. Only very high angular resolution observations will be able to disentangle the effects of grain growth/destruction from those of complex B-field geometry. Constraining the size of aligned grains in MM1 is complicated by the effects of the outburst and dust optical depth, as discussed in the next section. 

\subsection{Small Grains in MM1: Inefficient Grain Growth or RAT-D by Outburst or optical depth effect?}
\label{Grain_growth}
Grains are expected to grow in dense molecular clouds due to gas accretion and grain-grain collisions. For the dense cores of density $n_{\rm H}\sim 10^{7}$ \cmq, micron-sized grains are expected to be present \citep{Hirashita_2013, Lombart_2026}. Observations by ALMA toward the envelopes of YSOs also reveal the presence of $\sim 10~\mu$m grains (e.g., \citealt{Kwon_2009}). Our detailed modeling here reveals that the best-fit model requires rather small aligned grains of $a_{\mathrm{max}}\sim 0.35-0.50~\mu$m in the dense MM1 region. This could be due to inefficient grain growth in hot cores or some fragmentation mechanism that breaks grown micron-sized grains into small ones, or the optical depth effect. The insufficient grain growth appears inconsistent with current theory and observations of large grains in extremely dense environments such as hot cores \citep{Hirashita_2013}. 

\begin{figure}
    \centering
    \includegraphics[width=8.5cm]{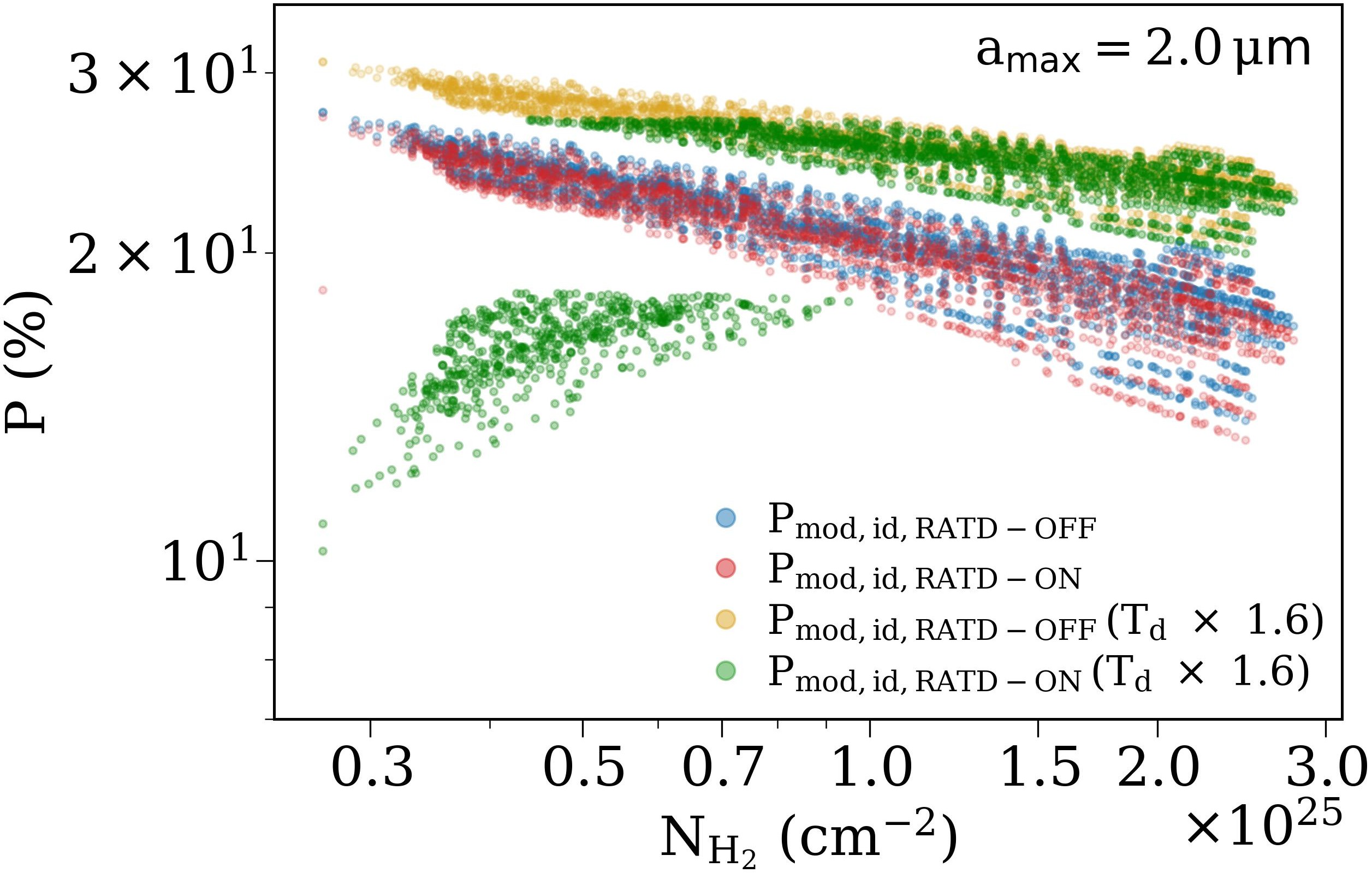}   
 \caption{Modeled polarization for MM1 with increased dust temperature ($T_{\rm d} \times 16^{1/6}$) and $a_{\rm{max}}$ = 2.0 $\mu$m, showing the effect of RAT-D in MM1 due to outburst. }
    \label{effect_outburst}
\end{figure}

An accretion outburst was detected from the embedded protocluster MM1 in \ngc~by \cite{Brogan_2016, Hunter_2017}, which increased the total luminosity of the region by a factor of $\sim16$, to be $L_{\rm bol}^{\rm burst}\sim 47,600$ \lsun~ from the pre-burst luminosity of $\sim 2900$ \lsun~ \citep[][]{Hunter_2021}. The surrounding dust is likely heated by infrared radiation from the outburst, which would have raised its temperature \citep[][]{Hunter_2017, Jayender_2026}. Using Eq. (\ref{eq:adisr}), one can estimate the grain disruption size as 
\begin{eqnarray}
a_{\rm disr}^{\rm burst}\sim 0.34 \left(\frac{L_{\rm bol}^{\rm burst}}{47000~L_{\odot}}\right)^{-1/2}\left(\frac{R}{10^{3} ~{\rm au}}\right) \mu m,
\end{eqnarray}
where $U = L_{\rm bol}/(4\pi R^{2} c ~u_{\rm ISRF})$ with $u_{\rm ISRF} = 8.64\times 10^{-13}$~erg~\cmq~(see \citealt{Hoang.2021}) and $R$ is the distance to the central protostar, assuming the typical parameters of $\gamma=0.1$ and $S_{\rm max}=10^{7}$~erg~\cmq. 
Meanwhile, the disruption size for the pre-burst is $a_{\rm disr}^{\rm pre}\sim 1.38 \mu$m. This implies that micron-sized grains grown pre-outburst can be disrupted into sub-micron grains via the RAT-D mechanism during outburst. To investigate this fact, we increased the dust temperature in MM1 by a factor of $\sim 1.6$, corresponding to the increase in the bolometric luminosity (i.e., $T_{\rm d} \times 16^{1/6}$; see equation \ref{eqn:align}), assuming that the dust must have been heated during the outburst before cooling down. Using this elevated $T_{\rm d}$, we modeled the polarized emission to examine the impact of RAT-D in MM1. We take $a_{\rm{max}}$ = 2.0 $\mu$m, and the other parameters are the same as given in Table \ref{tab:pars}. Figure \ref{effect_outburst} presents a comparison of $P_{\mathrm{mod,\, RATD\text{-}ON/OFF}}$ before and after increasing $T_{\rm d}$ in MM1. The results indicate that enhanced radiative heating during the outburst allows grain disruption to become effective at lower densities and consequently reduces the polarization degree. However, grains in the densest parts of MM1 are likely to survive disruption owing to the shielding by the high gas density.

We also estimated the rotational disruption timescale, $t_{\rm disr}$ (see Appendix~\ref{timescales}), to range from $\sim4$ years to $\sim108$ days for $a_{\rm disr}=0.05$--$2.0~\mu$m, assuming $T_{\rm d}\approx80$ K, corresponding to the minimum dust temperature in MM1. For the maximum dust temperature of $\sim130$ K in MM1, the disruption timescale decreases further to $\sim80$--$6$ days for the same grain-size range. Therefore, in the hot cores of MM1, the disruption timescale is much shorter than the outburst duration, which is reported to be $>6$ years in \ngc~\citep[see][and references therein]{Hunter_2021}. Hence, dust grains in relatively low-density regions of MM1 
could have been disrupted very early during the outburst. In comparison to the disruption timescale, the grain-growth timescale in MM1 is much longer (see Appendix \ref{timescales} for calculation). For grains to grow to sizes of $\sim1~\mu$m, assuming a median density of $n_{\rm H}\sim6\times10^{8}~{\rm cm^{-3}}$ in MM1, a timescale of at least $10^{3}$ years is required, while the growth of $2~\mu$m-sized grains would require an even longer timescale. An alternative mechanism is grain shattering by shocks from protostellar outflows. However, the volume impacted by outflows in comparison to that of RAT-D by outburst radiation is uncertain. Also, the grain-shattering timescale is, in general, much longer than the disruption timescale \citep{Hirashita_2013}.

On the other hand, increasing optical depth can also reduce the net polarization integrated along the LOS. This occurs because polarization produced by dichroic extinction becomes significant and can partially offset the polarized thermal dust emission. When optical depth ($\tau$) is included in the dust polarization modeling of MM1 (see section \ref{model_w_B-tang}), the $a_{\rm{max}}$ = 1.0$-$2.0 $\mu$m became more probable to match the observed polarization (see Figure \ref{MM1_model_tau}). This suggests that dust grain growth is also possible in MM1 if the optical thickness of the region ($\tau>1.0$) is considered in the polarization model. To further constrain our grain-growth results, accurate measurements of the dust opacity index from multi-frequency observations probing the same spatial scales are needed, as grain growth is expected to produce a lower dust opacity index (e.g., $\lesssim1.0$) in dense cores.



\subsection{Effect of magnetic relaxation on RAT Alignment: MRAT}
As discussed in Section \ref{sec:mod}, the grains consisting of iron clusters (i.e., superparamagnetic inclusions-SPIs) in these dense regions can achieve perfect alignment \citep{Hoang_Laz_2016, Hoang_2025}. To study this effect on grain alignment in detail, we calculate the parameter $\delta_{\mathrm{mag}}$, defined as the ratio of the gas-collision damping timescale $\tau_{\rm gas}$ to the magnetic relaxation time $\tau_{\rm mag,\,sp}$, given by 

\begin{equation}
\delta_{\rm mag,\, sp}
= \frac{\tau_{\rm gas}}{\tau_{\rm mag,\, sp}}
= 56\, a_{-5}^{-1}
\frac{N_{\rm cl}\,\phi_{\rm sp,\, -2}\,\hat{p}^{2}\,B_{3}^{2}\,k_{\rm sp}(\omega)}
{\hat{\rho}\, n_{4}\, T_{{\rm gas},\, 1}^{1/2}}
\frac{1}{T_{d,\, 1}}, 
\end{equation}
where $B_3 = B_{\mathrm{tot}}/(10^3 \mu$G), $n4 = n_{\rm H}/(10^4\, \cmq)$, $T_{\mathrm{gas,\, 1}}$ = $T_{\mathrm{gas}}/(10\, K)$, $T_{\mathrm{d,\, 1}}$ = $T_{\mathrm{d}}/(10\, K)$, $\hat{p} = p/5.5$ is the coefficient for magnetic moment of Fe atom, $N_{\mathrm{cl}}$ is the number of Fe atoms per iron cluster, $\phi_{\rm{sp},\,-2}$ =  $\phi_{\mathrm{sp}}/0.01$ is the volume filling factor of iron clusters, and $k_{\mathrm{sp}}\, (\omega)$ is the function of grain rotation frequency $\omega$. For our estimates, we choose $\phi_{\rm{sp}}=0.1$, corresponding to 10$\%$ of iron abundance in the form of iron clusters--the typical value in the THEMIS dust model \citep{Jones_2017}. We take $p$ = 5.5 \citep[for silicate dust,][]{Draine_1996_conf}, $k_{\mathrm{sp}}\, (\omega)$ $\approx 1$, and calculated  $B_{\mathrm{tot}} \approx 1.3 B_{\mathrm{POS}}$. Using the B-field strength of the POS component in the range of $\sim1$ mG to $\sim 11$ mG in \ngc~ measured by \cite{Cortes_2024} and $N_{\rm cl}=10^3$ \citep{Hoang_Laz_2016}, we estimate the role of magnetic relaxation on RAT alignment to test the MRAT mechanism in \ngc. We adopt a fiducial value of $N_{\rm cl}=10^3$, which lies within the commonly considered range of $N_{\rm cl}\sim20$--$10^5$ for superparamagnetic iron inclusions in interstellar dust grains \citep[e.g.,][]{Jones_1967, Bradley_1994, Giang.2024}.

For MM1 ($B_{\mathrm{POS}}\approx 7{-}11$ mG, $\bar{n}_{\rm H}\approx 9\times10^8\,\mathrm{cm^{-3}}$, 
and $\bar{T_{\rm d}}\approx 101$ K), $\delta_{\rm mag,\, sp}$ is found to be in the range of $\sim16.1{-}39.6$. 
For MM2 ($B_{\mathrm{POS}}\approx 2{-}4$ mG, $\bar{n}_{\rm H}\approx 3.0\times10^8\,\mathrm{cm^{-3}}$, 
and $\bar{T}_{\rm d}\approx 105$ K), we obtain $\delta_{\rm mag,\, sp}\sim3.7{-}14.8$. 
For MM3 ($B_{\mathrm{POS}}\approx 1{-}3$ mG, $\bar{n}_{\rm H}\approx 4.4\times10^7\,\mathrm{cm^{-3}}$, 
and $\bar{T}_{\rm d}\approx 94$ K), $\delta_{\rm mag,\, sp}$ spans $\sim7.5{-}67.2$.  For the whole \ngc, taking $B_{\mathrm{POS}}\approx 0.5{-}11$ mG, $\bar{n}_{\rm H}\approx 1.3\times10^8\,\mathrm{cm^{-3}}$, 
and $\bar{T}_{\rm d}\approx 77$ K, the $\delta_{\rm mag,\, sp}$ lies in the range $\sim 1-425$ ($\sim12.7$ for $\langle{B_{\mathrm{POS}}}\rangle \approx1.9$ mG). Through numerical simulations, \cite{Hoang_Laz_2016} showed that super-paramagnetic grains can generally achieve perfect alignment when magnetic relaxation is much stronger than gas collisional damping, i.e., when $\delta_{\rm mag,\, sp} > 10$. 

In MM1, $\delta_{\rm mag,\, sp}$ clearly exceeds this threshold. Although the lower limits of $\delta_{\rm mag,\, sp}$ in MM2 and MM3 are smaller than this value, their upper limits are significantly higher than the threshold. Furthermore, adopting $N_{\rm cl}=10^4$ for iron nanoparticles in larger grains increases $\delta_{\rm mag,\, sp}$ by a factor of 10, placing even the lower limits above the threshold. By taking into account the B-field strength of the LOS component measured through Zeeman observations towards the MM3 UC\hii region, $\langle B_{\mathrm{LOS}}\rangle \approx$ 3.4 mG \citep{Hunter_2018}, \cite{Cortes_2024} derived the average total B-field strength of $\sim 4 \pm 1$ mG in \ngc. Considering $B_{\mathrm{tot}} \sim 4 \pm 1$ mG, $\delta_{\rm mag,\, sp}$ across the entire \ngc~would be around $33\pm 17 $. 
Hence, for grains with SPIs in this collisionally dominated regime, large grains can be assumed to be perfectly aligned due to the MRAT mechanism \citep{Hoang_2025}. For this reason, our choice of $f_{\rm max}=1$ in the polarization modeling with \texttt{DustPOL\_py} is appropriate.

\startlongtable
\begin{deluxetable*}{cccccc}
\tablecaption{Results of dust polarization modeling of \ngc. The lowest RMSE values and their corresponding inclination factors are highlighted in bold. Where two values appear in a cell, they correspond to $P_{\mathrm{mod,\, RATD-OFF}}$ and $P_{\mathrm{mod,\, RATD-ON}}$, respectively.}  
\label{tab:results}
\tablehead{
\colhead{Region} &
\colhead{Model} &
\colhead{$a_{\max}$ ($\mu$m)} &
\colhead{$\Phi$} &
\colhead{Spearman ($r_s$)} &
\colhead{RMSE}
}
\startdata
MM1 & $P_{\mathrm{mod,\, RATD-OFF}}$ & 0.35 & 1.0 & 0.981 & 1.239 \\
    &  &  & 0.7 &  & 0.757 \\
    &  &  & 0.5 &  & 0.437 \\
    &  &  & \textbf{0.270} &  & \textbf{0.092} \\
    &  & 0.50 & 1.0 & 0.978 & 2.163 \\
    &  &  & 0.7 &  & 1.401 \\
    &  &  & 0.5 &  & 0.893 \\
    &  &  & \textbf{0.142} &  & \textbf{0.037} \\
    & $P_{\mathrm{mod,~\tau}}$ & 1.0 &
      $\Phi_{\rm norm} = 0.119\tablenotemark{a}$ [0.113, 0.120] &
      0.990 & 0.037 [0.028, 0.045] \\
    &  & 2.0 &
      $\Phi_{\rm norm} = 0.083$ [0.076, 0.086] &
      0.990 & 0.043 [0.062, 0.047] \\
\noalign{\vskip 4pt}
MM2 & $P_{\mathrm{mod,\, RATD-OFF}}$ & 0.35 & 1.0 & 0.931 & 1.589 \\
    &  &  & 0.7 &  & 1.000 \\
    &  &  & 0.5 &  & 0.607 \\
    &  &  & \textbf{0.202} &  & \textbf{0.069} \\
    & $P_{\mathrm{mod,\, RATD-OFF}}$, & 0.50 & 1.0 &
      0.931, 0.944 & 2.081, 1.863 \\
    & $P_{\mathrm{mod,\, RATD-ON}}$ & & 0.7 &
      & 1.342, 1.190 \\
    &  & & 0.5 & & 0.850, 0.742 \\
    &  & & \textbf{0.155, 0.170} &
      & \textbf{0.057, 0.059} \\
    & $P_{\mathrm{mod,\, RATD-OFF}}$, & 1.0 & 1.0 &
      0.944, 0.865 & 2.927, 2.559 \\
    & $P_{\mathrm{mod,\, RATD-ON}}$ & & 0.7 &
      & 1.934, 1.678 \\
    &  & & 0.5 & & 1.273, 1.091 \\
    &  & & \textbf{0.110, 0.119} &
      & \textbf{0.057, 0.076} \\
\noalign{\vskip 4pt}
MM1\_ext & $P_{\mathrm{mod,\, RATD-OFF}}$ & 0.35 &
      \textbf{1.0} & 0.975 & \textbf{0.557} \\
    &  &  & 0.7 &  & 0.789 \\
    &  &  & 0.3 &  & 1.402 \\
    &  &  & 0.887 &  & 0.625 \\
    & $P_{\mathrm{mod,\, RATD-OFF}}$, & 0.50 &
      \textbf{1.0} & 0.982, 0.896 & \textbf{0.622, 0.795} \\
    & $P_{\mathrm{mod,\, RATD-ON}}$ & & 0.7 &
      & 0.663, 0.990 \\
    &  & & 0.3 & & 1.306, 1.490 \\
    &  & & 0.550, 0.644 & & 0.946, 1.070 \\
\noalign{\vskip 4pt}
MM2\_ext & $P_{\mathrm{mod,\, RATD-OFF}}$ & 0.35 &
      1.0 & 0.989 & 0.536 \\
    &  &  & 0.5 &  & 0.641 \\
    &  &  & 0.3 &  & 1.081 \\
    &  &  & \textbf{0.660} &  & \textbf{0.309} \\
    & $P_{\mathrm{mod,\, RATD-OFF}}$, & 0.50 &
      1.0 & 0.993, 0.986 & 0.925, 0.485 \\
    & $P_{\mathrm{mod,\, RATD-ON}}$ & & \textbf{0.7} &
      & \textbf{0.230, 0.392} \\
    &  & & 0.3 & & 0.968, 1.134 \\
    &  & & 0.516, 0.576 & & 0.428, 0.594 \\
    & $P_{\mathrm{mod,\, RATD-OFF}}$, & 1.0 &
      1.0 & 0.993, 0.954 & 1.541, 0.775 \\
    & $P_{\mathrm{mod,\, RATD-ON}}$ & & \textbf{0.7} &
      & \textbf{0.600, 0.437} \\
    &  & & 0.3 & & 0.798, 1.085 \\
    &  & & 0.376, 0.412 & & 0.600, 0.851 \\
\noalign{\vskip 4pt}
MM3 & $P_{\mathrm{mod,\, RATD-OFF}}$ & 0.25 &
      1.0 & 0.896 & 2.682 \\
    &  &  & 0.7 &  & 1.357 \\
    &  &  & \textbf{0.5} &  & \textbf{1.128} \\
    &  &  & 0.318 &  & 1.686 \\
\noalign{\vskip 4pt}
Outer & $P_{\mathrm{mod,\, RATD-OFF}}$, & 1.5 &
      1.0 & 0.860, 0.817 & 5.359, 2.389 \\
      & $P_{\mathrm{mod,\, RATD-ON}}$ & &
      0.7 & & 3.100, 1.108 \\
      & & & 0.3 & & 1.460, 1.718 \\
      & & & 0.569, \textbf{0.613} &
      & 2.215, \textbf{0.924} \\
      & $P_{\mathrm{mod,\, RATD-OFF}}$, & 2.0 &
      1.0 & 0.832, 0.803 & 5.824, 2.904 \\
      & $P_{\mathrm{mod,\, RATD-ON}}$ & &
      0.7 & & 3.364, 1.364 \\
      & & & 0.3 & & 1.334, 1.587 \\
      & & & 0.502, \textbf{0.546} &
      & 1.909, \textbf{0.916} \\
\enddata
\tablenotetext{a}{Only the normalized $\Phi$ values are shown. Values in square brackets indicate the resulting variation in $\Phi$ and RMSE due to a $\pm30\%$ uncertainty in the optical depth (see Section~\ref{optical_depth}).}
\end{deluxetable*}

\subsection{Is Polarization Fraction Influenced by B-field–Local Gravity Alignment?}

The relative orientations of different geometries, such as B-field, local gravity, intensity gradient, and velocity gradient, have become a statistical tool to analyze the impact of gravity and gas flow on the B-field structure and understand the physical conditions of star formation \citep[e.g.][]{Koch_2012a, Wang_2020b, Liu_2023b}. In \ngc, an ordered radial pattern of B-field geometry around the substructures MM1 and MM2 is evident (see Figure \ref{pol_map}(a)), which becomes more disordered towards the dense cores within MM1 and MM2, possibly due to feedback from multiple outflows. 
The alignment measure (AM) parameter is commonly used to quantify the alignment between different orientations and is defined as 
\begin{align}
\mathrm{AM}_{\rm B}^{\mathrm{LG}} = \langle \cos(2\phi_{\rm B}^{\mathrm{LG}}) \rangle,
\end{align}
 where $\phi_{\rm B}^{\mathrm{LG}} = \left| \theta_{\rm B} - \theta_{\mathrm{LG}} \right|$ is the angle difference between B-field (B) and local gravity (LG) orientations and is in the range of $0\degree$ to $90\degree$ \citep{Gonz_2017}. The AM lies between -1 (perpendicular ) and +1 (parallel). The uncertainty in AM is given by 
 {\small
\begin{align}
    \delta \mathrm{AM} =\frac{1}{N'}\sqrt{\left\langle \left(\cos(2\phi_{\rm B}^{\mathrm{LG}}) \right)^2 \right\rangle- \mathrm{AM}^2+ \sum_i^{N'} \left( 2 \sin(2\phi_i)\,\delta\phi_i\right)^2},
    \end{align}}
where $N'$ is the number of data points taken for AM calculation \citep{Liu_2023b} and $\delta\phi_{\rm B}^{\rm{LG}}$ = $\sqrt{\delta \theta_{\rm B}^2 + \delta \theta_{\mathrm{LG}}^2}$. 
    
\begin{figure*}
    \centering
    \includegraphics[width=5.95cm]{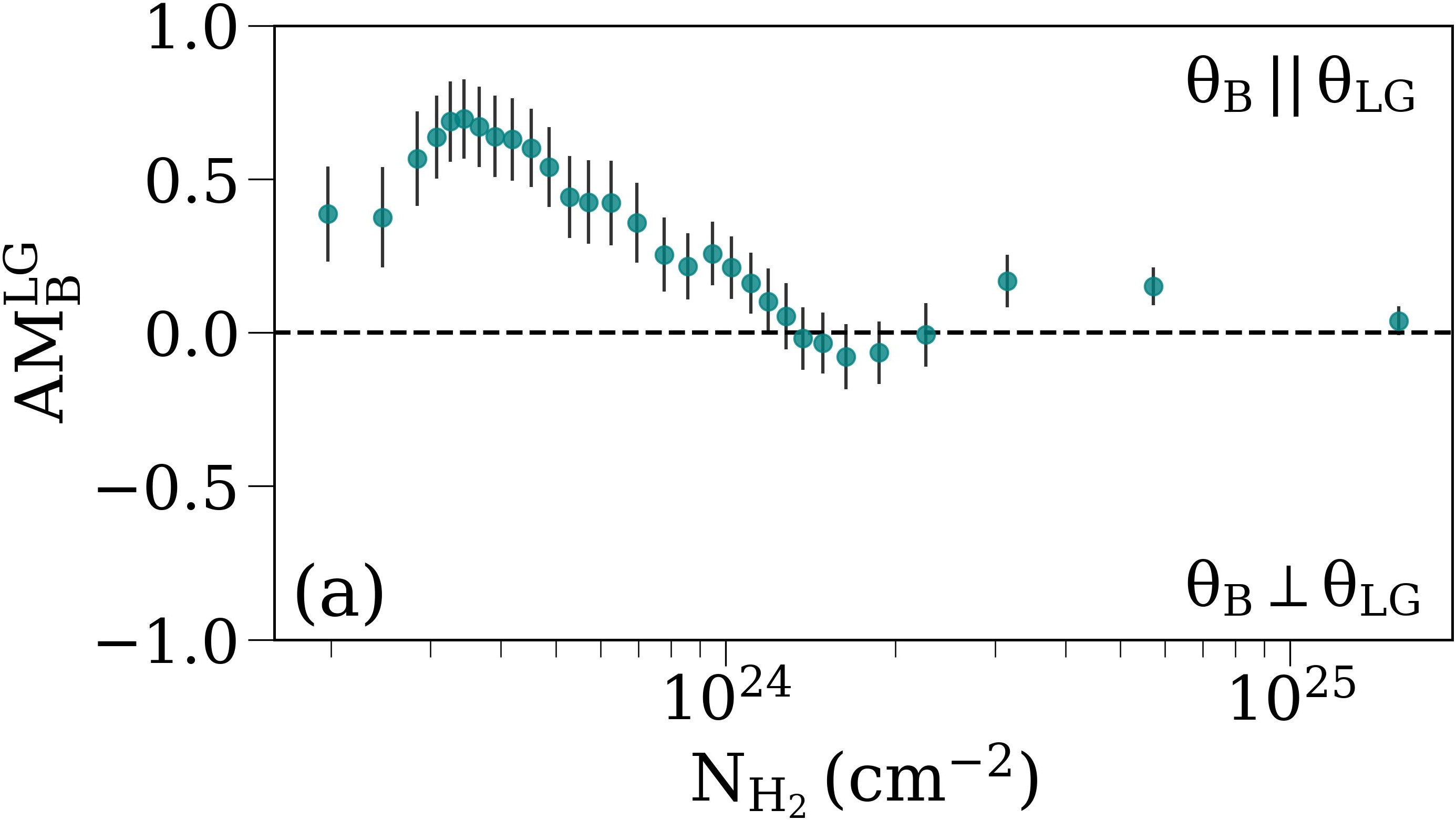} 
    \includegraphics[width=5.95cm]{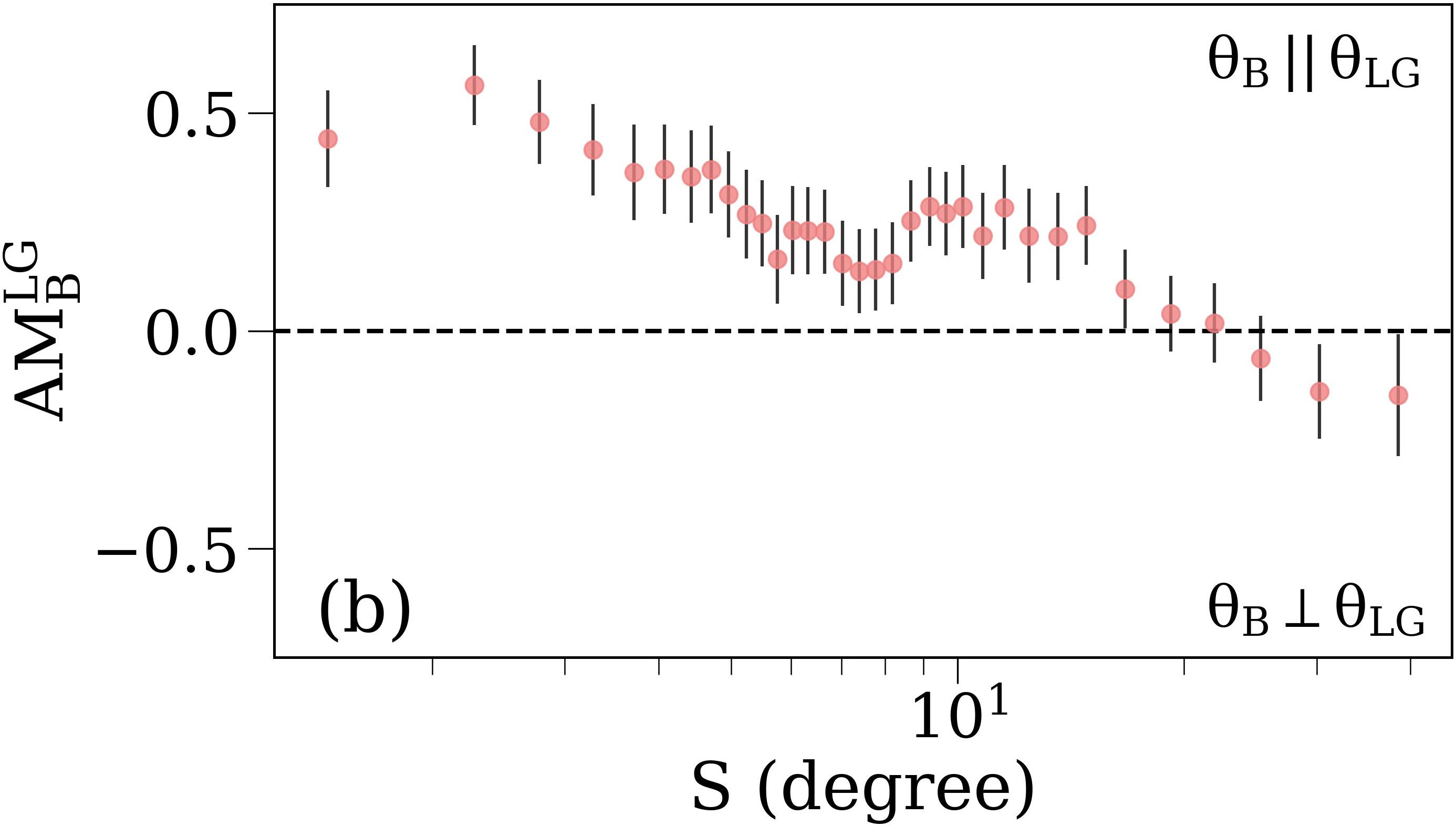} 
    \includegraphics[width=5.95cm]{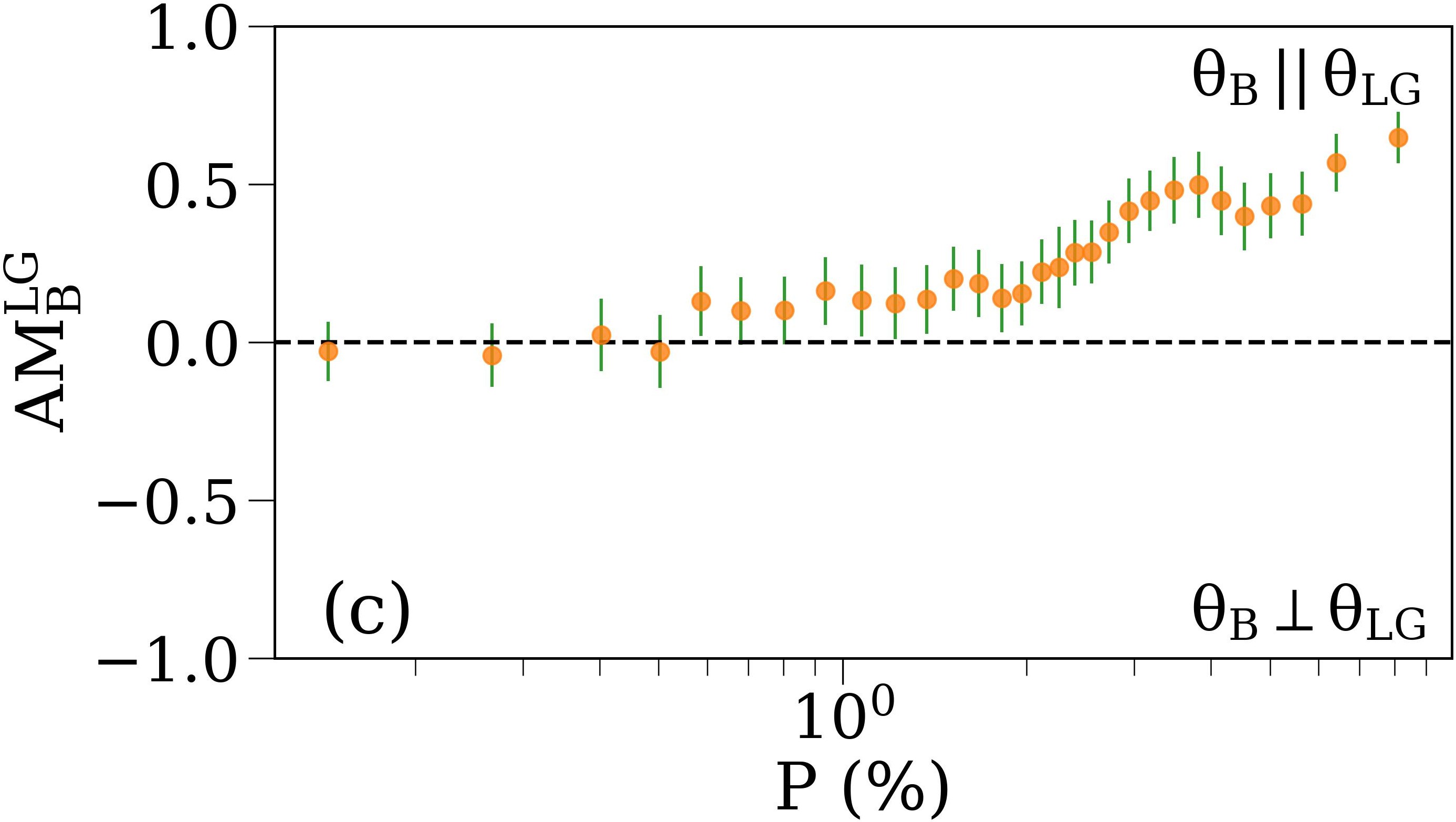}     
 \caption{Relative orientations between B-field ($\theta_{\rm B}$) and local gravity ($\theta_{\mathrm{LG}}$) characterized by AM shown as a function of (a) $N_{\mathrm{H_2}}$, (b) $S$, and (c) $P$.}
    \label{AM}
\end{figure*}

In \ngc, \cite{Liu_2023b} found a more parallel alignment between the B-field and local gravity, with the alignment becoming less parallel toward higher-density regions. However, here we aim to analyze the variation of $\mathrm{AM}_{\rm B}^{\mathrm{LG}}$ with the polarization degree and the polarization angle dispersion. For the local gravity map, we calculated the 2D orientations of LG at each pixel with $I/\mathrm{\delta I} > 5$ from the gas mass map using the gravitational force equation \citep[e.g.,][]{Koch_2012a, Wang_2020b}. The acute angle differences between the B-field and LG were computed at each pixel. We derived $\mathrm{AM}_{\rm B}^{\mathrm{LG}}$ in column density, polarization degree, and dispersion bins containing a uniform number of pixels. The corresponding uncertainty, $\delta \mathrm{AM}$, was estimated by considering only the error in $\phi_{\rm B}$, since uncertainties in local gravity are not constrained by the signal-to-noise ratio of the map. The data points with $\delta\phi_{\rm B}^{\mathrm{LG}} = \delta\theta_{\rm B} > 10\degree$ (as $\delta\theta_{\rm {LG}}$ = 0) are excluded from the calculation. The variation of $\mathrm{AM}_{\rm B}^{\rm{LG}}$ with $N_{\mathrm{H_2}}$, $P$, and $S$ is shown in Figure \ref{AM}. 

Figure \ref{AM}(a) shows a predominantly parallel alignment between the B-field and local gravity across almost all column densities in \ngc, consistent with the results of \cite{Liu_2023b}. However, the alignment becomes progressively less parallel toward higher column densities, which suggests that the B-field is relatively ordered in low-density regions, while it becomes relatively perturbed at high densities, likely due to gravitational pinching, turbulent outflows, or gas inflow in the cores of \ngc. A similar behavior is observed with polarization angle dispersion. The $\mathrm{AM}_{\rm B}^{\mathrm{LG}}$ decreases from $\sim 0.6$ to $\sim 0$ with increasing $S$ (see Figure \ref{AM}(b)), indicating that larger angle differences between the B-field and local gravity are associated with a more disordered B-field geometry.
Consequently, Figure \ref{AM}(c) shows that regions where the B-field and local gravity are more aligned exhibit higher polarization degrees. This is consistent with a more ordered B-field structure, which leads to enhanced polarization, whereas misalignment corresponds to increased disorder and reduced polarization degree. 

Comparing the $S$ map (Figure \ref{S_map}) obtained here with the Alfvén Mach number map of \cite{Cortes_2024} (see their Figure 12), we find a good spatial correlation between the two. This correlation is expected because, under the Davis–Chandrasekhar–Fermi method \citep{Davis_1951}, the Alfvén Mach number is approximately proportional to the polarization-angle dispersion. This suggests that the distortion in the B-field geometry, and thus the dispersion of polarization position angles, is influenced by supersonic turbulent gas flows along with gravity. Therefore, the presence of stellar feedback from the protocluster (MM1) and other regions (MM2, MM3, and MM4) in \ngc~strongly affects the B-field geometry and hence leads to a low polarization fraction, especially in the dense cores. However, multiple layers along the LOS with different polarization properties can also affect the observed polarization properties, which are not considered in polarization modeling here.   
       
\subsection{Evidence for bending of B-fields toward denser regions}
\label{diss_6}
If the B-field is inclined toward the LOS, the measured polarization fraction decreases even when grain alignment is efficient. Therefore, part of the low polarization fraction attributed to inefficient grain alignment could instead arise from geometric depolarization. From Table \ref{tab:results}, it is clear that our assumption of B-field lies predominantly in the POS (i.e., $\Phi=1.0$) is not mostly true. This suggests that any realistic inclination of the B-field leads to an overestimation of the modeled polarization, especially in dense cores where the intrinsic polarization fraction is already low.

The different $\Phi$ values required to match the modeled polarization with the observed polarization indicate that a single $\Phi$ value cannot explain the polarization across outer low-density to inner high-density regions. This result is expected because the B-field geometry in \ngc~significantly changes from outer to inner regions. Recently, \cite{Pravash_2026} also found that $\Phi$ decreases with increasing column density, consistent with our results. In our study, we also find that $\Phi \approx 0.1$–$0.3$ is more appropriate for the high-density MM1 and MM2 regions, whereas for the extended regions—MM1\_ext, MM2\_ext, MM3, and the outer regions, $\Phi \approx 0.5$–$1.0$ better reproduces the observed polarization.

Taking an average $B_{\mathrm{LOS}} \approx 3.4$ mG measured toward the MM3 UC\hii region \citep{Hunter_2018}, and adopting $B_{\mathrm{POS}} \approx 0.5$–$11$ mG with a mean value of $\sim 1.9$ mG \citep{Cortes_2024}, we obtain $\Phi \sim \sin^2(\gamma_B)$ in the range $\sim 0.02$–$0.91$, with a mean value of $\sim 0.24$. This range of $\Phi$ agrees well with our adopted values in the dust emission polarization modeling for different regions. However, the Zeeman effect in OH maser emission is observed only toward the MM3 and CM2 sources in \ngc~\citep{Hunter_2018}, but not in the MM1 and MM2 hot cores, which are the primary regions of interest in this paper. Therefore, although we adopted the best-match $\Phi$ values based on the RMSEs, the effect of changes in the B-field inclination angle on the polarization fraction is difficult to distinguish from other contributing factors. And there is no direct measurement of this factor for the whole \ngc. Hence, this result should be interpreted with some caution.

This highlights the importance of constraining the 3D B-field geometry in star-forming regions through complementary observations, such as dust polarization, Zeeman measurements, and Faraday rotation. Improved characterization of the 3D B-field structure would help break the degeneracy between the inclination factor and the maximum size of aligned grains, thereby providing more robust constraints on dust grain growth. Also, 3D B-fields would help better characterize the field tangling both in the POS and along the LOS. In addition, multiwavelength, high-angular-resolution observations of dust optical depth and opacity in \ngc\ would provide independent constraints on grain growth and complement those derived from polarization modeling.

\subsection{Caveats of the study}
\label{Caveats}
In this paper, we have used ALMA Band 6 (1.2 mm) 12m array data that has a maximum recoverable scale of around 6\arcsec, whereas the typical size of \ngc~is around 10\arcsec$-$15\arcsec. Therefore, the emission from extended structures in Stokes $I$ can be filtered out in the interferometric observations. As a result, the polarization fraction would be affected. To quantify the effect of spatial filtering, we used single-dish APEX Telescope Large Area Survey of the Galaxy (ATLASGAL) data at 870 $\mu$m \citep{Schuller_2009} that has a spatial resolution of $\sim 18\arcsec$. To estimate the missing flux, we convolved and regridded the MagMaR data to the ATLASGAL beam and pixel grid (pixel size $\sim 6\arcsec$), and used a spectral index in the range 2.4 to 3.2 \citep{Hunter_2017} to match the frequency differences between MagMaR ($\sim 250.508$ GHz) and ATLASGAL ($\sim 345$ GHz).

The ATLASGAL data of the NGC 6334 region were obtained in August 2007, while an outburst was reported in 2015 in \ngc, which enhanced the dust emission from the region. \cite{Hunter_2017} estimated the excess amount of flux at 870 $\mu$m by matching the preburst SMA data (2008) with the postburst ALMA data (2016) and found that it increased by a factor of $\sim 4$ in MM1, while no significant change was found in MM2, MM3, and MM4. However, they also reported that, when observed with a low-resolution single-dish telescope (beam size $\sim18\arcsec$), the outburst-induced increase in MM1 would correspond to only a $\sim30\%$ increase in the total flux density of \ngc. Since the single-dish APEX observations do not resolve the individual substructures within \ngc, and MM1 is the dominant millimeter source in the region, we scaled the entire ATLASGAL 870 $\mu$m emission from \ngc~by a factor of 1.3 to approximately reproduce the post-outburst dust continuum emission. Under this assumption, we find that the ALMA Band 6 data could miss 24--40\% of the total flux for spectral indices between 3.2 and 2.4, respectively.          

\ngc~is an active star-forming region that has massive YSOs in MM1, MM2, and MM3 (UC\hii), and therefore might have free-free emission. At longer wavelengths (millimeter), dust emission can be contaminated by free-free emission in the evolved \hii regions. As free-free emission is not polarized like dust emission, if not subtracted, the polarization fraction would be underestimated. \cite{Brogan_2016} found that MM3 is affected by free-free emission with a spectral index of $\lesssim$0.5 using 1.3 mm and 3 mm ALMA data, while other regions studied in this paper mostly show an index of $\geq$ 2.0. \cite{Hunter_2021} also found that MM1, MM2, and MM4 are dominated by dust emission, while MM3 shows optically thin free-free emission. Hence, in our analysis, free-free emission would mostly impact the observed polarization fraction in MM3, while in other regions, the effect would be minimal.  

There are some other factors that also impact the observed polarization, such as changes in grain axial ratio with gas density due to anisotropic grain growth \citep{Hoang.2022}. A higher axial ratio would increase the intrinsic polarization fraction, lowering the required maximum grain size \citep{Tram.2025}. Though we used a $^{13}\rm{CH_3OH}$-based rotational temperature map, which is relatively optically thin compared to $\rm{CH_3OH}$, the lines may become optically thick in the highest-density regions, such as the center of MM1 ($N_{\rm{H_2}} \sim 10^{24} - 10^{25}\, \cms$). Thus, the rotational temperatures derived for these regions could be biased by optical depth effects and should be treated with caution. 

\section{Summary}
\label{sec:conc}
In this paper, we investigate dust physics in a high-mass star-forming region, \ngc, using high-resolution thermal dust polarization data at 1.2 mm from the ALMA MagMaR survey. Through observed polarization properties, analytical calculations, and polarization modeling, we studied various physical processes underlying dust polarization, including grain alignment, grain growth and disruption, gas randomization, B-field tangling and inclination, and optical depth effects, in \ngc. The summary of the findings is as follows:

\begin{itemize}
    \item 
The observed polarization degree in the dense regions of \ngc~is very low ($<$ 2\%), while in outer low-density regions, it reaches up to 10\%, suggesting a depolarization trend. 
The dust temperature ranges from $\sim30$ K to 180 K, and the number density ($n_{\mathrm{H_2}}$) ranges from $\sim 10^5$ to $10^9\, \cmq$ in \ngc. The B-field structure shows a pinched/spiral morphology in MM1, and a flip by 90\degree~near MM2 and MM4 is observed. Consequently, the polarization angle dispersion ($S$) values are high in these regions (i.e., 20\degree - 50\degree).

  \item 
The variation of $P$, $S$, and $S \times P$ with $I$, $T_{\mathrm{d}}$, and $N_{\mathrm{H_2}}$ suggests that in MM1, the inefficient dust grain alignment due to gas randomization likely contributes significantly to the observed depolarization. In MM2 and MM3, the depolarization can be explained by a combination of B-field tangling and reduced grain alignment efficiency. 

  \item 
The analytical calculations based on the RAT-A theory suggest that grains up to $\sim 0.015$ $\mu$m in size can be aligned in regions of strong radiation in \ngc, while high-density regions suppress alignment due to stronger randomization. Conversely, the dust grains might also get disrupted in hot cores due to very high temperatures. 
Hence, the size distribution of aligned grains is determined by the competition between temperature-enhanced alignment, density-driven randomization, and radiative disruption.     
  
  \item 
Using thermal dust polarization modeling, we find that the ideal RAT-A model, i.e., a uniform B-field perfectly aligned with the POS, overestimates the observed polarization in all substructures. Including the B-field dispersion, inclination angle, and optical depth effects in the model better estimates the observed $P$. This suggests that the depolarization observed in \ngc\ is driven by a combination of these factors and reduced grain alignment efficiency. 
The AM analysis shows that in regions where the B-field and local gravity are better aligned, the dispersion in the B-field is lower, and the polarization degree is higher. The best-fit inclination factor, $\Phi$, decreases from $\sim0.5$--$1.0$ in the low-density outer regions to $\sim0.1$--$0.3$ in the denser inner regions, indicating that the B-field in NGC6334I transitions from being primarily oriented in the POS to becoming increasingly aligned with the LOS as the density increases. 
This provides quantitative support for the bending of B-field lines in dense regions, potentially driven by gravitational accretion, and motivates future studies of the 3D B-field structure.

\item Our detailed polarization modeling suggests possible grain growth in MM2.  
In MM1, we found a smaller maximum grain size ($\sim0.35-0.50~\mu$m) than in MM2. 
We suggest that the rotational disruption effect induced by the intense outburst event in MM1 provides a plausible explanation for the reduced grain sizes in the relatively low-density regions of MM1. 
Alternatively, the depolarization effect by high optical depth ($\tau\gtrsim1$) in the densest regions of MM1 suggests large micron-sized grains ($\sim1.0-2.0~\mu$m), implying significant grain growth in this region. High-angular-resolution multi-wavelength continuum observations are required to accurately determine the dust opacity index and, in turn, grain sizes, thereby disentangling the two aforementioned scenarios (grain disruption versus high optical depth).

\item This study on \ngc~demonstrates that dust polarization is governed by a combination of grain physical properties (e.g., size), local physical conditions such as gas density, radiation field, and B-field geometry, and optical depth. 
These findings demonstrate that the RAT theory, which intrinsically provides a link between alignment efficiency and the local physical conditions, can successfully reproduce dust polarization observed in massive star-forming regions. However, other effects, such as highly tangled B-fields and multiple substructures along the LOS, also contribute to the observed dust polarization. Synthetic polarization observations by post-processing MHD simulations of massive star-forming regions are needed to disentangle these competing effects. 
\end{itemize}

\section*{Acknowledgements}
We thank the anonymous referee for the useful comments and
suggestions to improve the quality of the paper. V. Rawat and T.H. acknowledge the support from the major research project (No. 2026183200) from the Korea Astronomy and Space Science Institute (KASI) funded by the Ministry of Science and ICT (MSIT). This work is partially supported by the Vietnam National Foundation for Science and Technology Development (NAFOSTED) under grant number 103.99-2024.36. P.S. was partially supported by a Grant-in-Aid for Scientific Research (KAKENHI Number JP26H02066 and JP23H01221) of JSPS. MFL and JMG acknowledge support from the PID2023-146675NB-I00 (MCI-AEI-FEDER, UE) program. This work was also partly supported by the Spanish program Unidad de Excelencia María de Maeztu CEX2020-001058-M, financed by MCIN/AEI/10.13039/501100011033, and by the MaX-CSIC Excellence Award MaX4-SOMMA-IC. Y.C.  was partially supported by a  Grant -in- Aid for  Scientific Research  (KAKENHI number JP24K17103) of the JSPS. L.A.Z. acknowledges financial support from CONACyT-280775, UNAM-PAPIIT IN110618, and IN112323 grants, México. We thank Le Ngoc Tram, N\^{g}an L\^{e}, Nguyen Chau Giang, Bao Truong, and Raiga Kashiwagi for all the discussions related to the DustPOL\_py tool and the results of polarization modeling. We also thank Mengke Zhao for the discussion on the \emph{Volume Density Mapper} tool.

\clearpage

\appendix
\makeatletter
\@addtoreset{figure}{section}
\renewcommand{\thefigure}{\Alph{section}\arabic{figure}}
\makeatother

\section{Total intensity at 1.2 mm dust emission}
\label{Total_intens}

\begin{figure}
    \centering
    \includegraphics[width=8.5cm]{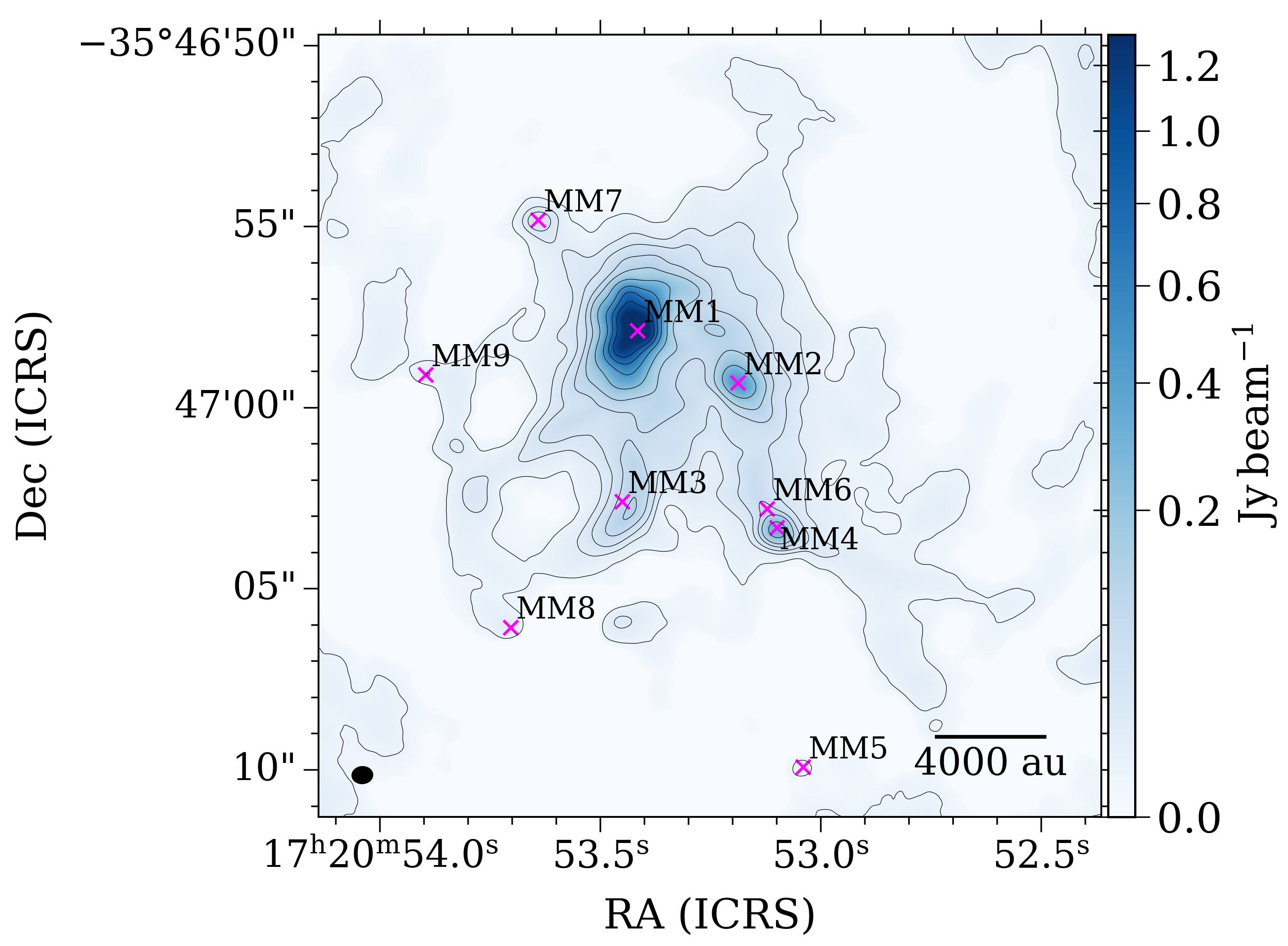}    
 \caption{ 1.2 mm total intensity dust emission map of NGC6334I, over which the contour levels are shown at 0.0043, 0.018, 0.036, 0.072, 0.120, 0.240, 0.480, 0.720, and 0.960 Jy $\mathrm{beam}^{-1}$. }
    \label{intensity}
\end{figure}

Figure \ref{intensity} shows the ALMA 1.2 mm dust continuum emission in \ngc. The total intensity map (Stokes $I$) reveals the millimeter sources identified in \cite{Brogan_2016}, in which the dominant source of dust emission is the MM1 protocluster region. 

\section{Rotational Temperature from $\rm{^{13}CH_3OH}$ and variation with column density}
\label{Trot}
To estimate the gas temperature, we conduct a rotational diagram analysis \citep{Goldsmith_1999} of $\rm{^{13}CH_3OH}$ multiple transition lines obtained from ALMA observations under the assumption of local thermodynamic equilibrium (LTE). The details of lines from CDMS are given in Table\ref{tab:trot}. The population in the upper state level of $\rm{^{13}CH_3OH}$ is given by 

\begin{equation}
    N_\mathrm{u} = \frac{N_{\mathrm{^{13}CH_3OH}}}{Z} g_\mathrm{u} e^{-E_\mathrm{u}/kT_{\mathrm{rot}}},
\end{equation}
where $N_{\rm u}$ is the upper state column density, $N_{\mathrm{^{13}CH_3OH}}$ is the total column density of $\rm{^{13}CH_3OH}$, $g_{\rm u}$ is the statistical weight of the upper state, $E_{\rm u}$ is the upper level energy, and $Z$ is the partition function. All the transition lines were modeled simultaneously using a single excitation temperature, column density, systemic velocity, and line width at each pixel. In pixels where the highest $E_{\rm u}$ = 322 K line is not detected, we only fit the other three lines, and if the $E_{\rm u}$ = 288 K line is also not detected, we do not fit those pixels. The rotational temperature derived here is taken as a proxy for dust temperature (see Figure \ref{Cdens_map}).   

\begin{table*}[htbp]
\label{tab:trot}
\centering
\caption{Details of $^{13}$CH$_3$OH transition lines}
\begin{tabular}{ccccc}
\hline
$E_u$ (K) & Frequency (GHz) & Transition & $A_{ij}\, (10^{-5}\, \rm s^{-1})$ & $g_u$ \\
\hline
114.94 & 256.671817 & $9(0,9)-8(1,7)$ & 4.10 & 19 \\
122.33 & 246.143897 & $5(-4,2)-6(-3,4)$ & 0.73 & 11 \\
287.84 & 256.826572 & $14(3,12)-14(2,13)+-$ & 8.88 & 29 \\
321.79 & 257.421792 & $15(3,13)-15(2,14)+-$ & 8.96 & 31 \\
\hline
\end{tabular}
\label{tab:13ch3oh_lines}
\end{table*}

\begin{figure}
    \centering
    \includegraphics[width=8.5cm]{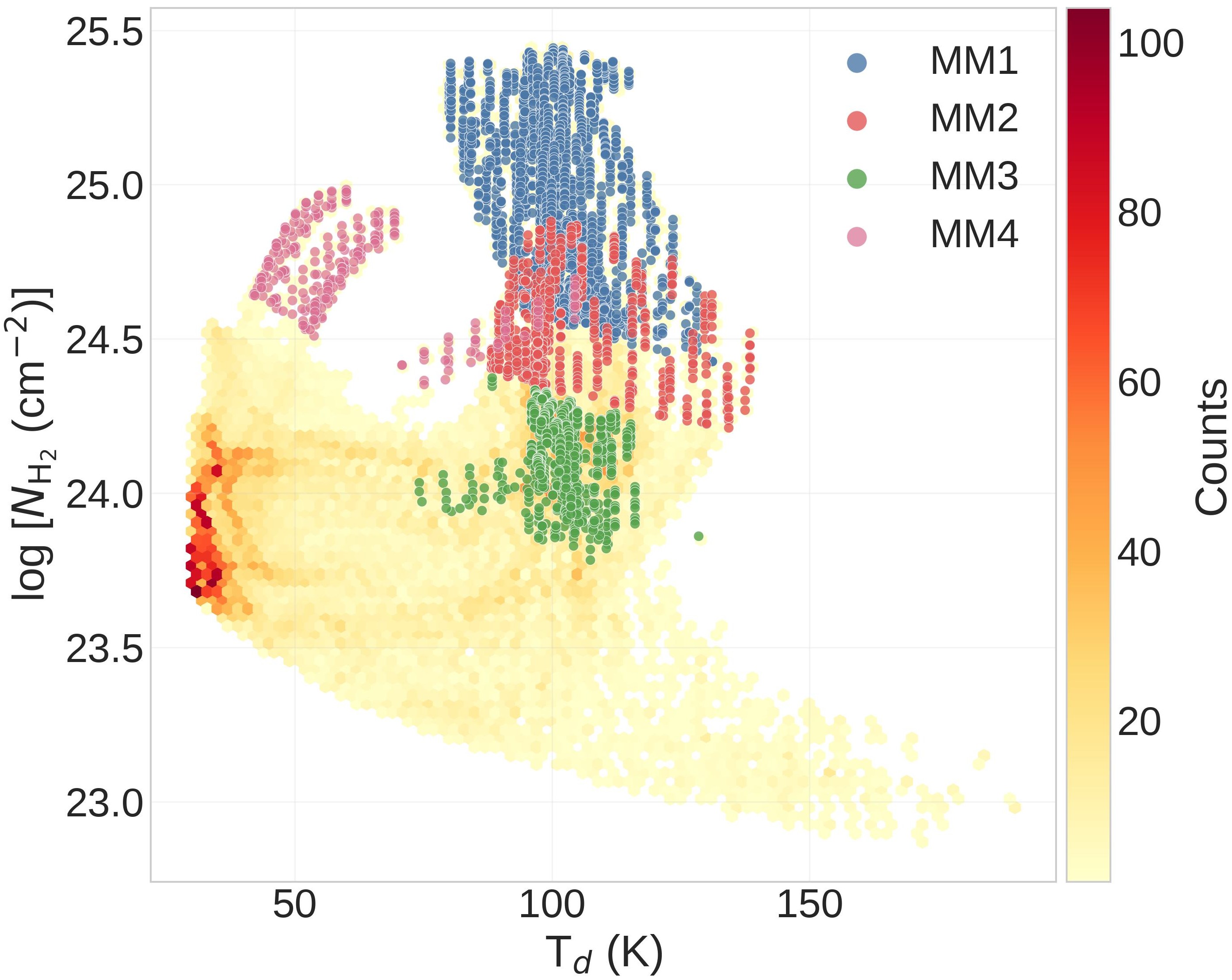}
 \caption{Variation of column density with dust temperature. The colored hexbins denote the whole \ngc, while the colored solid circles show different subregions within \ngc.}
    \label{NH2_vs_Td}
\end{figure}

Figure \ref{NH2_vs_Td} shows the variation of column density with dust temperature inferred from the $\mathrm{^{13}CH_3OH}$ rotational temperature diagram. The distribution of column density as a function of dust temperature shows a broad but systematic behavior. Overall, higher column densities are associated with relatively lower dust temperatures, while the column density gradually decreases toward higher temperatures. The MM1 region shows the highest $N_{\mathrm{H_2}}$ values (log $N_{\mathrm{H_2}} \gtrsim 24.6$) and is concentrated around $T_{\rm d} \sim 90-$110 K, indicating dense and warm material. MM2 spans intermediate column densities, with a wider temperature spread, and does not exhibit a clear monotonic trend, although a slight decrease in $N_{\mathrm{H_2}}$ with $T_{\rm d}$ remains apparent. In contrast, the MM3 region primarily traces lower column densities (log $N_{\mathrm{H_2}} \lesssim 24.2$) and spans a broad temperature range, suggesting more diffuse, heated material. However, it is to be noted that the lower temperature limit in MM3 is due to our assumption of the temperature model, i.e., convergence to $\sim 30$ K at the non-detected pixels. In reality, the temperature in the MM3 UC\hii region can be much higher due to photoionizing radiation from the young massive protostar. Thus, the derived temperature and column densities in MM3 should be treated with caution, given the limitations of the available observations. 

\section{Volume Density Mapper}
\label{VDM}
This method is based on the idea of hierarchical evolution of a molecular cloud under the relative effects of physical factors such as gravity, turbulence, and the B-field, where the density of structures increases as the scale size decreases. Therefore, it evaluates multiple spatial scales (e.g., filaments, clumps, and cores) and distinguishes them from the large-scale diffuse background solely based on the observed column-density map. The MDR applies a \emph{constrained diffusion algorithm}, which identifies and separates structures by iteratively smoothing the image to successive scales of $2^n$ ($n$ = 1, 2, 3,...; where $n$ is the number of pixels). Each component map contains structures whose dominant size falls between $2^n$ and $2^{n+1}$ pixels, and summing all components reconstructs the original image exactly. After the multi-scale decomposition above, the characteristic scale ($l_c$) at each pixel is computed as the intensity-weighted average scale across all component maps. The $l_c$ represents the size of the structure that dominates the column density contribution at that pixel, not how compact the local emission appears. Therefore, instead of a simple linear average of all the widths of the structures, the MDR method uses intensity-weighted averaging over widths within a given scale size \citep[][]{Zhao_2026}. In the diffuse surroundings, the column density at each pixel is dominated by the large-scale envelope of the cloud, so $l_c$ is large there. In the dense regions, the column density is dominated by the compact clump/core, so $l_c$ is smaller. It assumes that each structure's POS width is comparable to its LOS depth. So the apparent smaller $l_c$ in the dense regions and larger $l_c$ in the diffuse outskirts actually reflects the nested three-dimensional (3D) hierarchical structure of the cloud.

\section{Comparison of $S$ with different lag sizes}
\label{S_diff}
\begin{figure*}
    \centering
    \includegraphics[width=17.5cm]{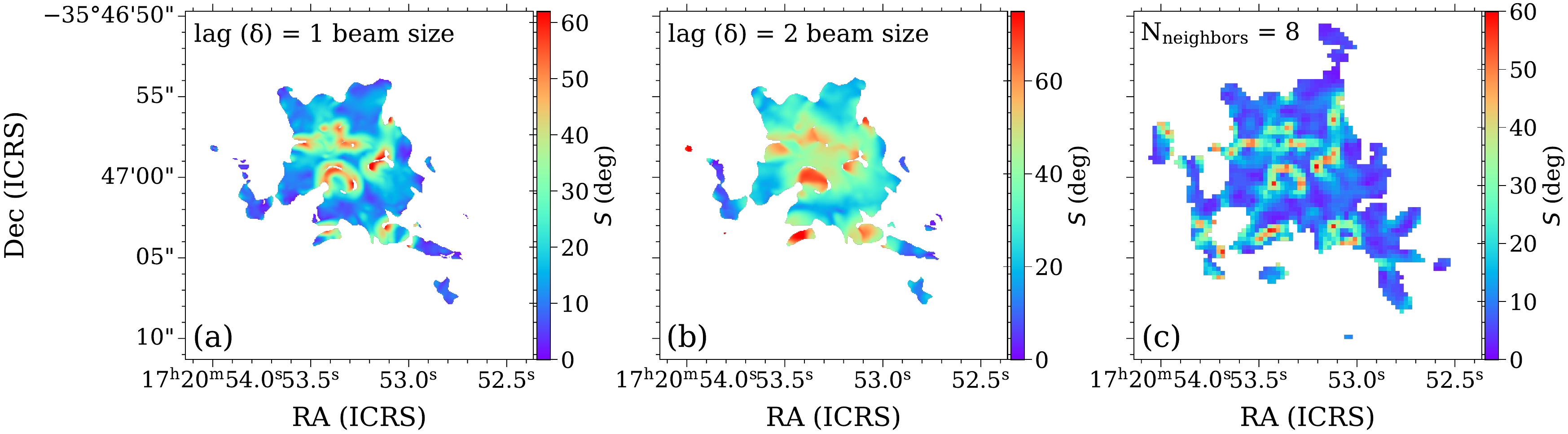} 
 \caption{Polarization angular dispersion ($S)$ maps computed from the $Q$ and $U$ maps of pixel sizes = 0.05\arcsec with lag ($\delta$) = (a) 0.5\arcsec and (b) 1\arcsec. (c) $S$ is calculated by taking 8 nearest neighbors in the re-gridded Q and U maps (pixel size = 0.25\arcsec; see Section \ref{sec:pol}).}
    \label{S_different}
\end{figure*}

\begin{figure*}
    \centering
    \includegraphics[width=17.5cm]{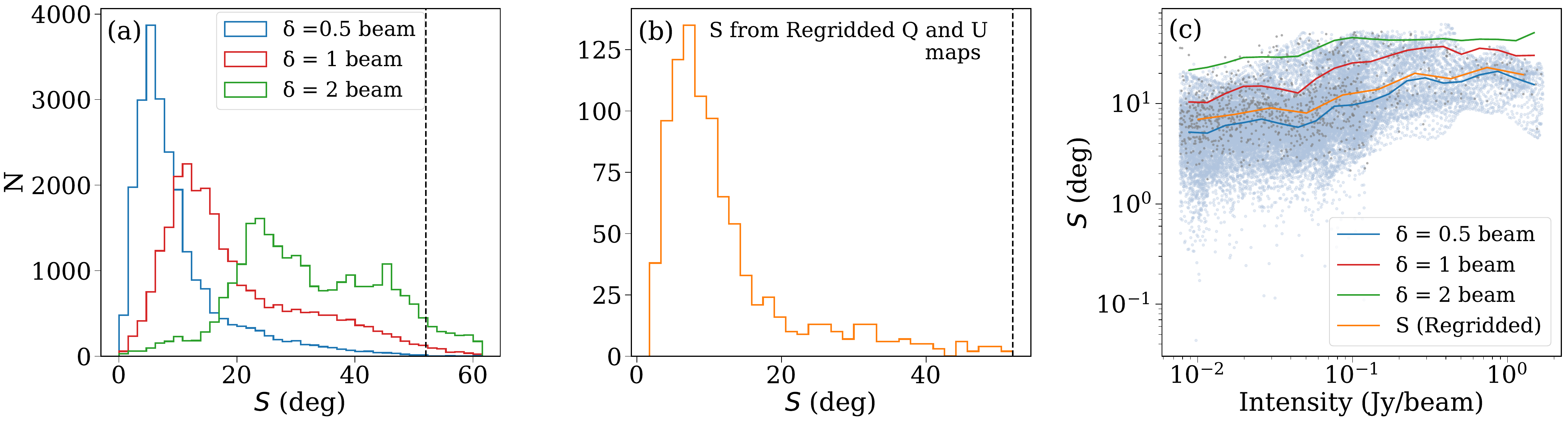} 
 \caption{Histogram distributions of $S$ maps (a) computed with lag ($\delta$) = 0.5, 1, and 2 beam sizes (shown in Figures \ref{S_map}(b) and \ref{S_different}(a) and (b)), (b) computed from the re-gridded $I$, $Q$, and $U$ maps (shown in Figure \ref{S_different}(c). (c) Variation of $S$ with intensity for different $S$ maps. The blue points show $S$ calculated with $\delta = 0.5\,\mathrm{beam}$, while the grey points show $S$ calculated from the re-gridded Stokes maps.}
    \label{S_distribution}
\end{figure*}

Figure \ref{S_different} shows the comparison of different $S$ maps (see Section \ref{sec:pol}). The maps shown in Figure \ref{S_different} (a) and (b) are obtained using equation \ref{eqn:S1}, with lags of one and two times the beam sizes, respectively. While the $S$ (regridded) map shown in Figure \ref{S_different} (c) is computed using equation \ref{eqn:S3} from regridded $I$, $Q$, and $U$ maps (pixel size = 0.25\arcsec). Figures \ref{S_distribution}(a) and (b) show the histogram distribution of the aforementioned $S$ maps, and Figure \ref{S_distribution}(c) shows the variation of $S$ values from these different maps as a function of total intensity. From the figures, it can be seen that the $S$ with $\delta$ = 0.5 beam size closely matches with the $S$ (regridded).  

\begin{figure}
    \centering
    \includegraphics[width=8.5cm]{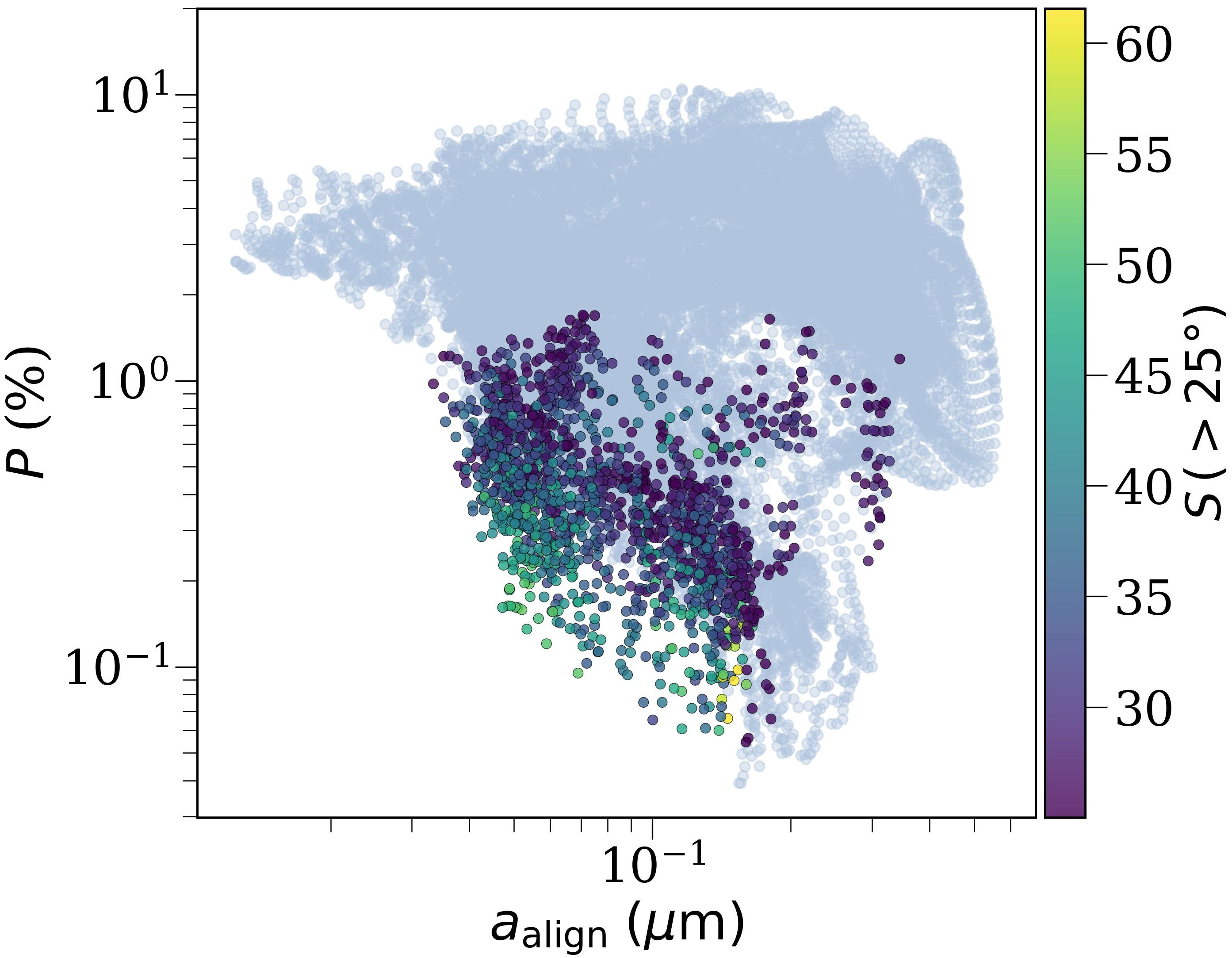} 
 \caption{Variation of polarization degree with $a_{\mathrm{align}}$. The points with $S > 25 \degree$ are highlighted with a color scale.}
    \label{align_w_S}
\end{figure}

Figure \ref{align_w_S} shows the variation of polarization degree with the minimum size of aligned grains, where pixels with polarization angle dispersion greater than 25 degrees are highlighted using a colormap. This shows that a low polarization fraction might also be due to B-field tangling, along with gas randomization and optical depth at high density.

\section{Maps of grain alignment and disruption size}
\label{grain_size_maps}
\begin{figure*}
    \centering
    \includegraphics[width=17.5cm]{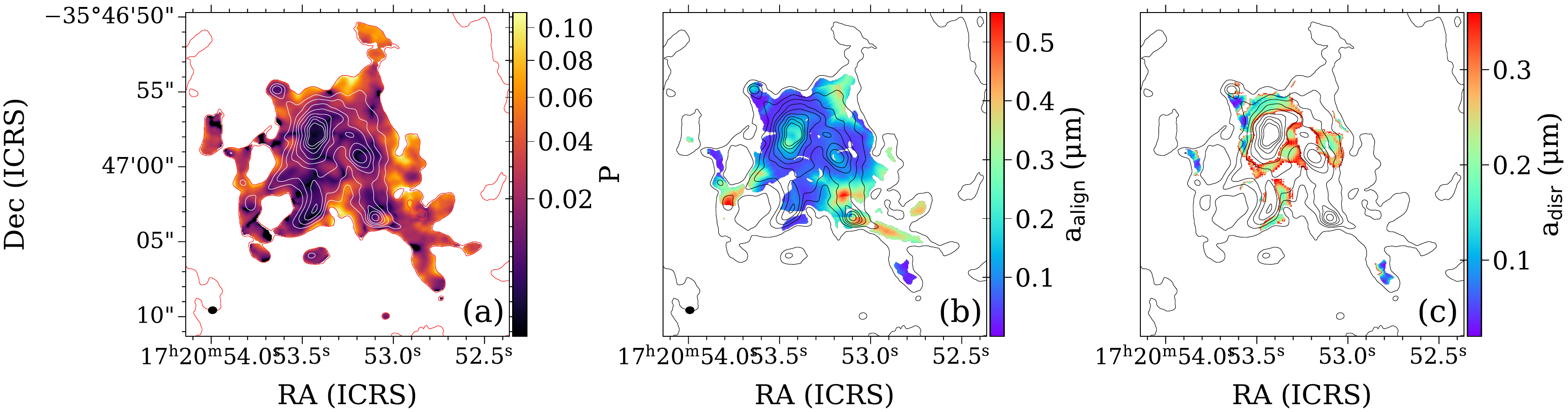} 
 \caption{(a) Polarization fraction same as Figure \ref{pol_map}(b). (b) $a_{\mathrm{align}}$ map. (c) $a_{\mathrm{disr}}$ map where the threshold of $a_{\mathrm{disr}} < \bar{\lambda}/1.8$ holds.}
    \label{align_disr_maps}
\end{figure*}

Figures \ref{align_disr_maps}(b) and (c) show the maps of the minimum size of grains that can be aligned ($a_{\mathrm{align}}$) and the size of grains that can be disrupted ($a_{\mathrm{disr}}$), respectively. From the figure, it is evident that in regions where $a_{\mathrm{align}}$ is lower, i.e., where smaller grains can be aligned due to relatively lower density and stronger radiation, the polarization fraction is higher, as the size distribution of aligned grains becomes broader. In contrast, rotational disruption appears to be effective only in low-density and strong-radiation (i.e., high dust temperature) regions in \ngc, due to the condition $a_{\mathrm{disr}} < \bar{\lambda}/1.8$.

\section{Modeled polarization fractions for different grain sizes}
\label{Pmodel_appendix}
\begin{figure*}
    \centering
    \includegraphics[width=5.9cm]{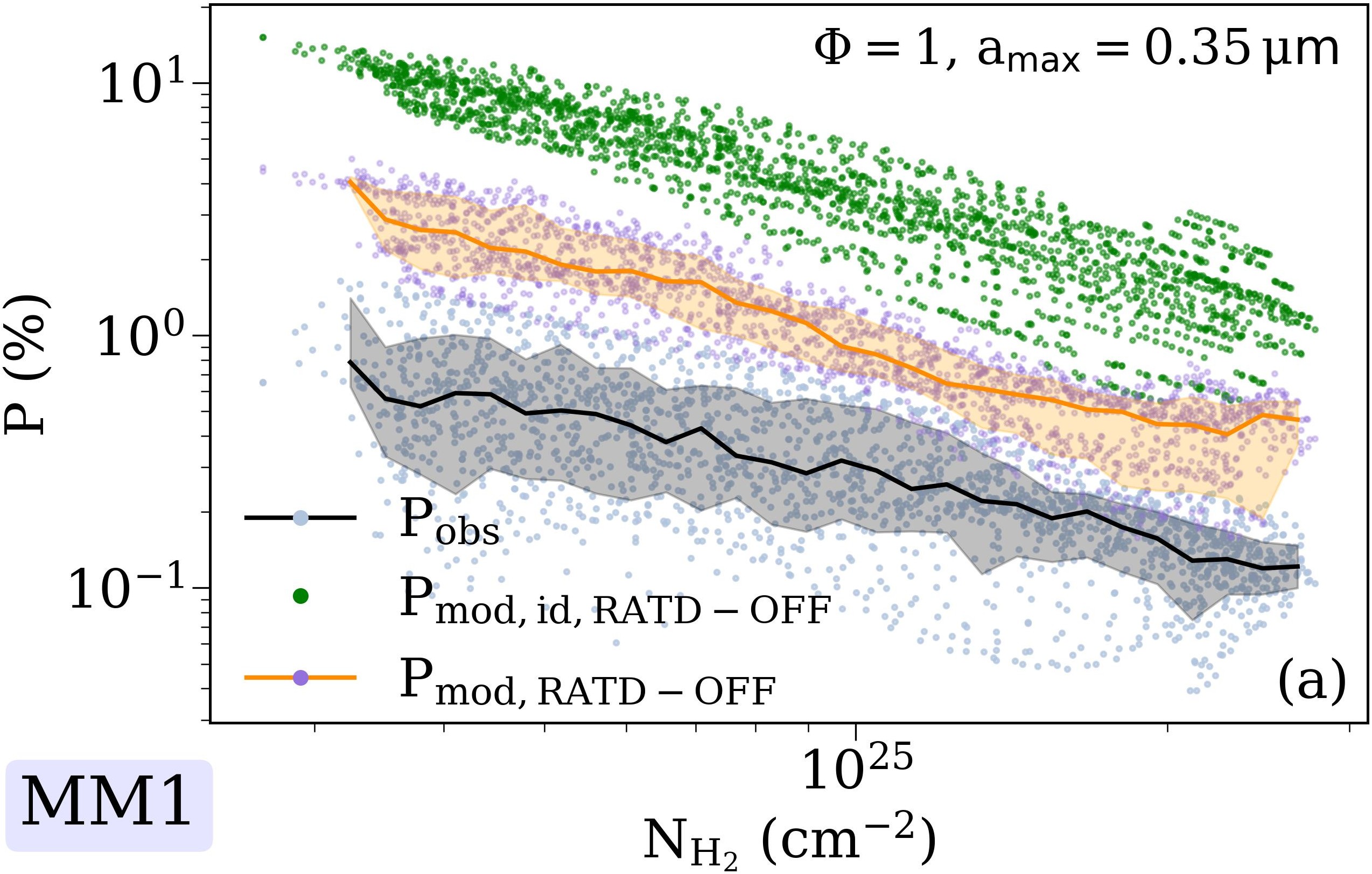} \includegraphics[width=5.9cm]{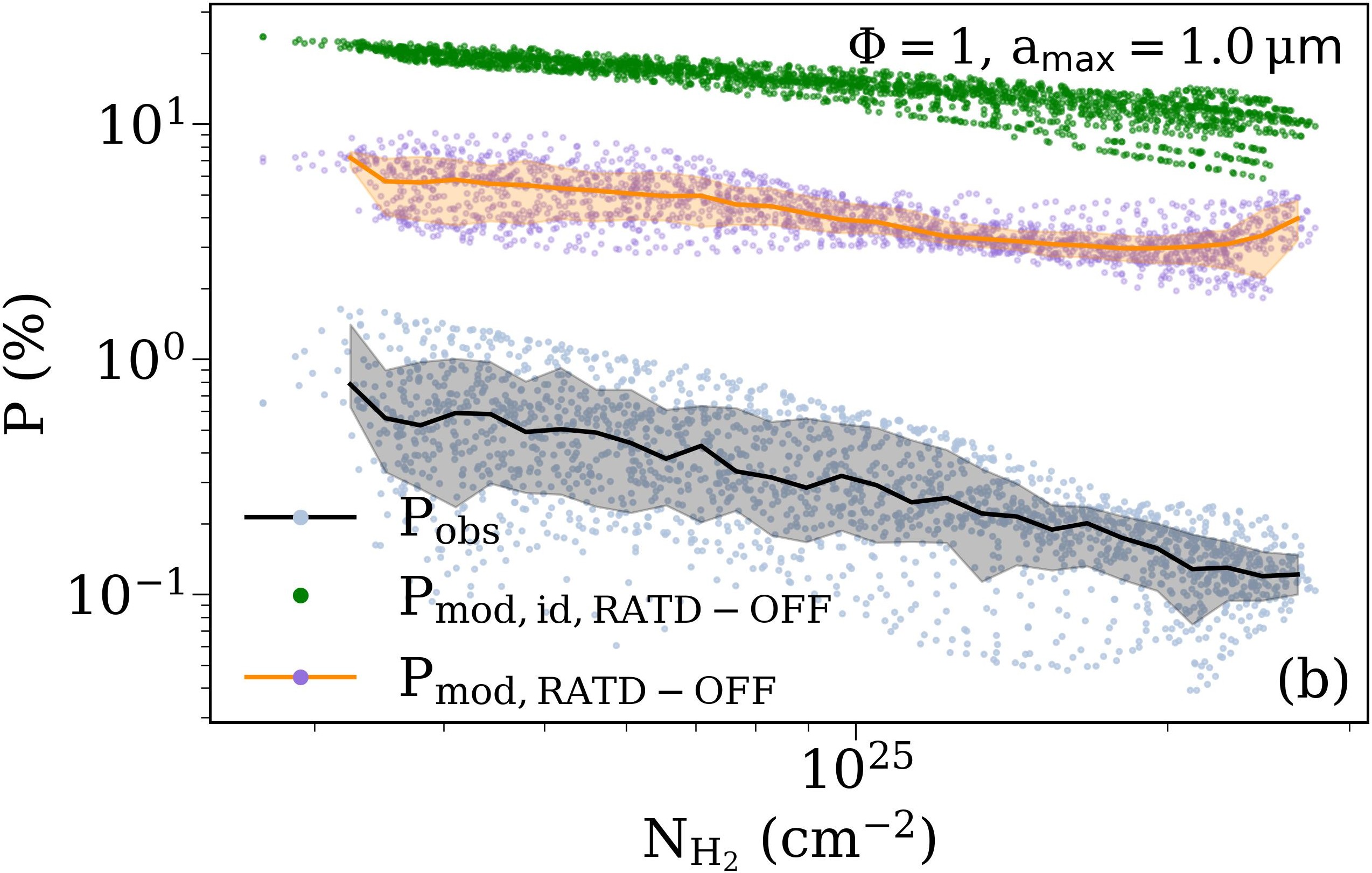} \includegraphics[width=5.9cm]{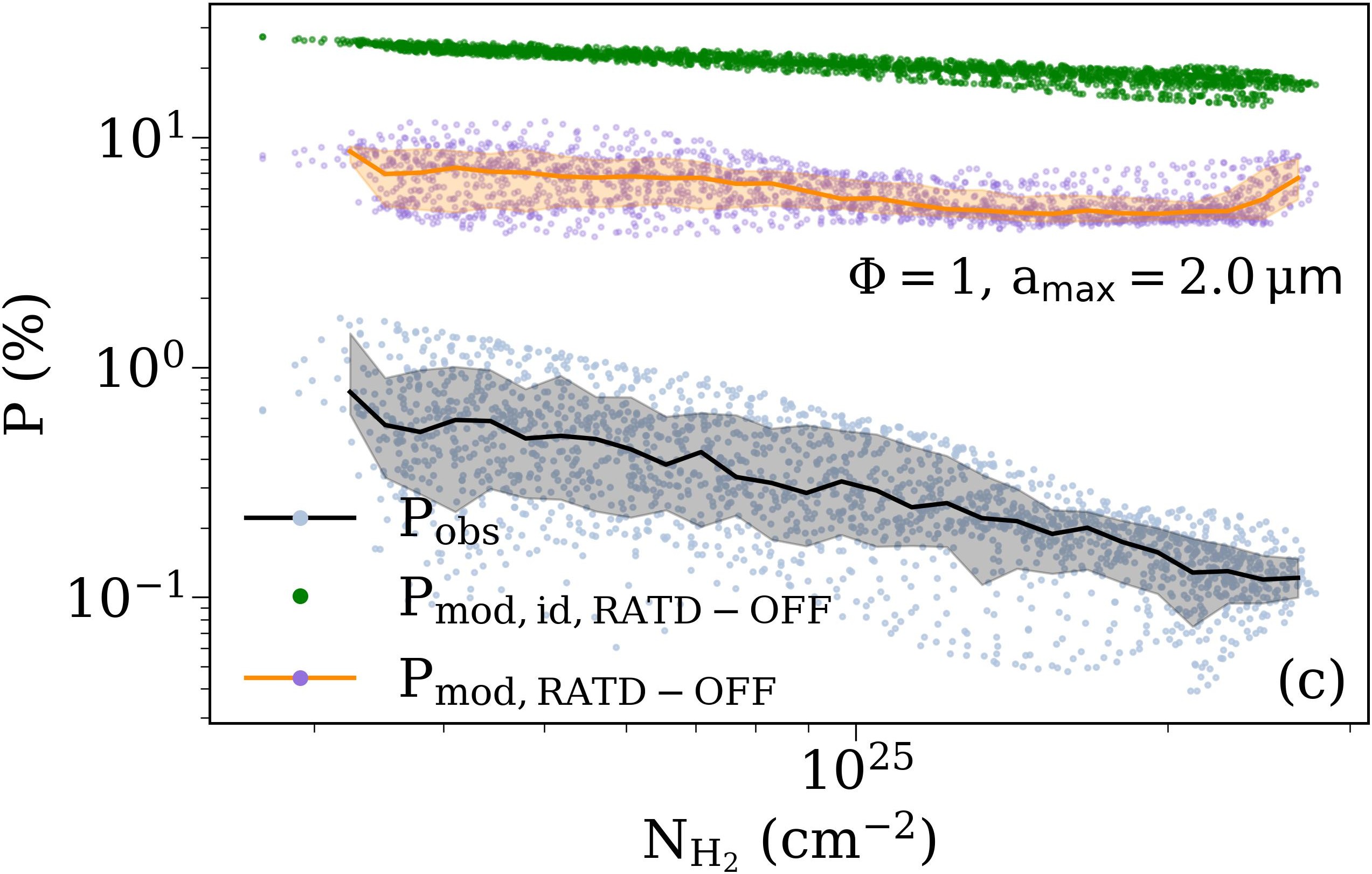} \includegraphics[width=5.9cm]{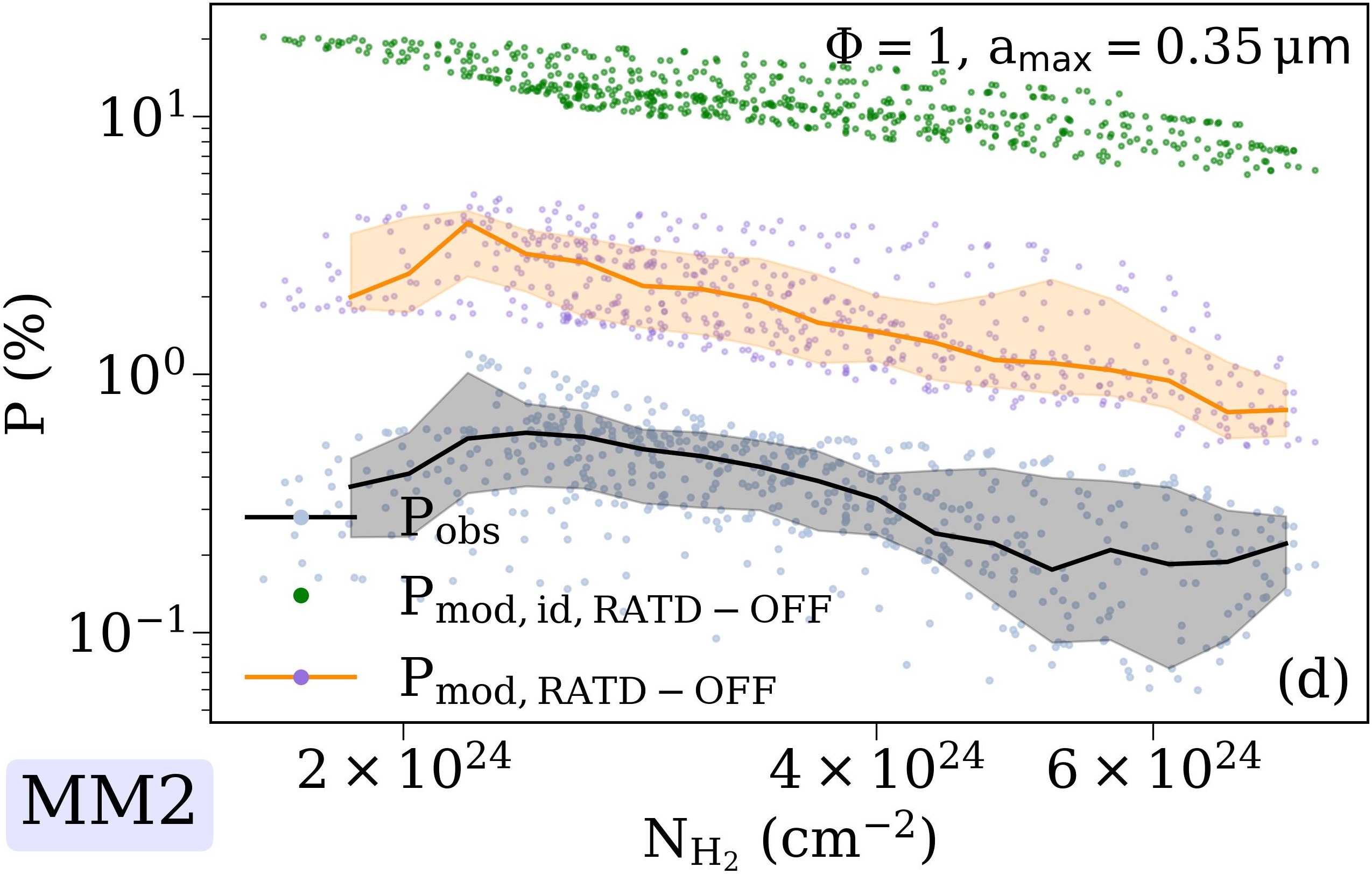} \includegraphics[width=5.9cm]{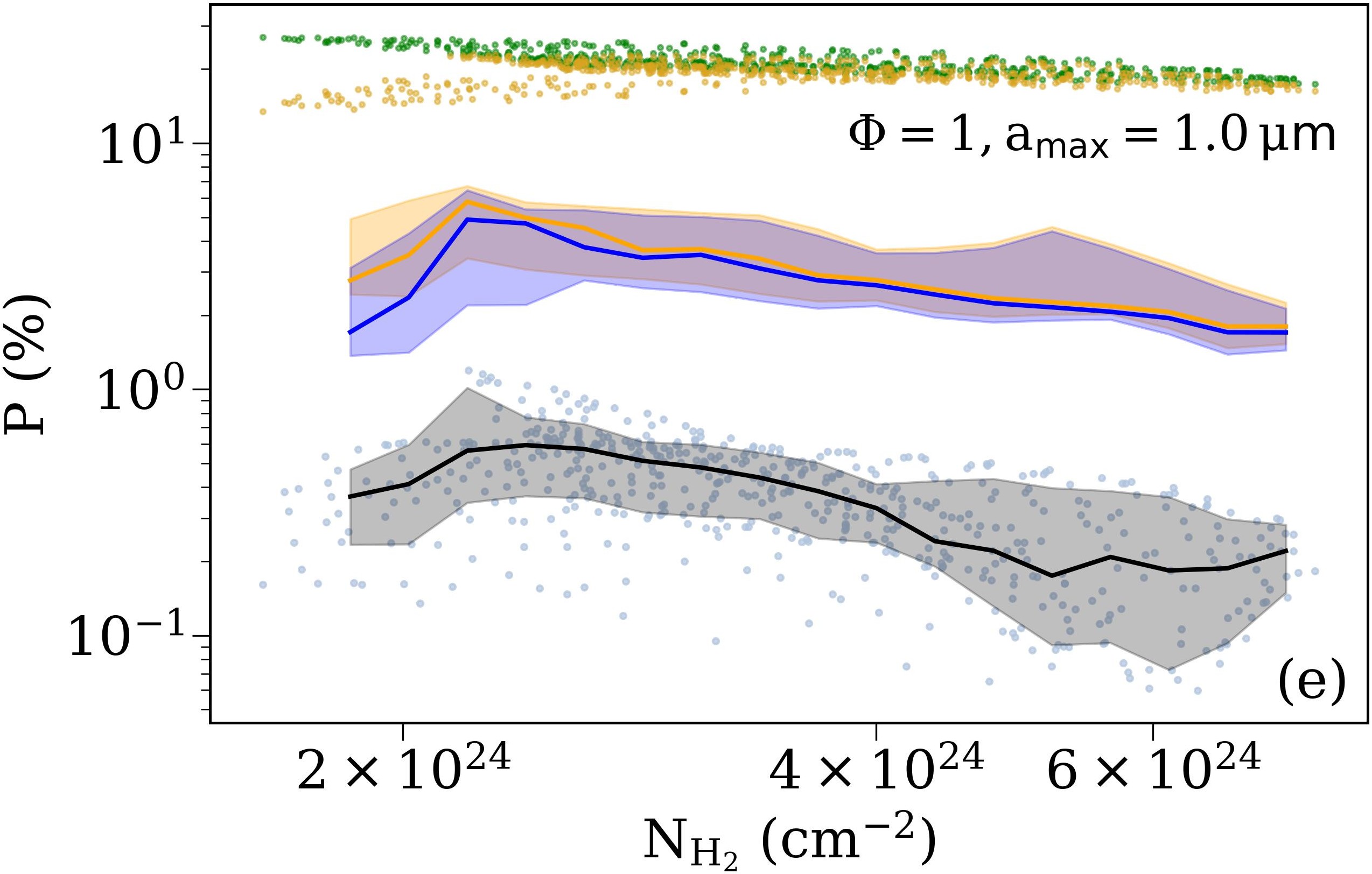} \includegraphics[width=5.9cm]{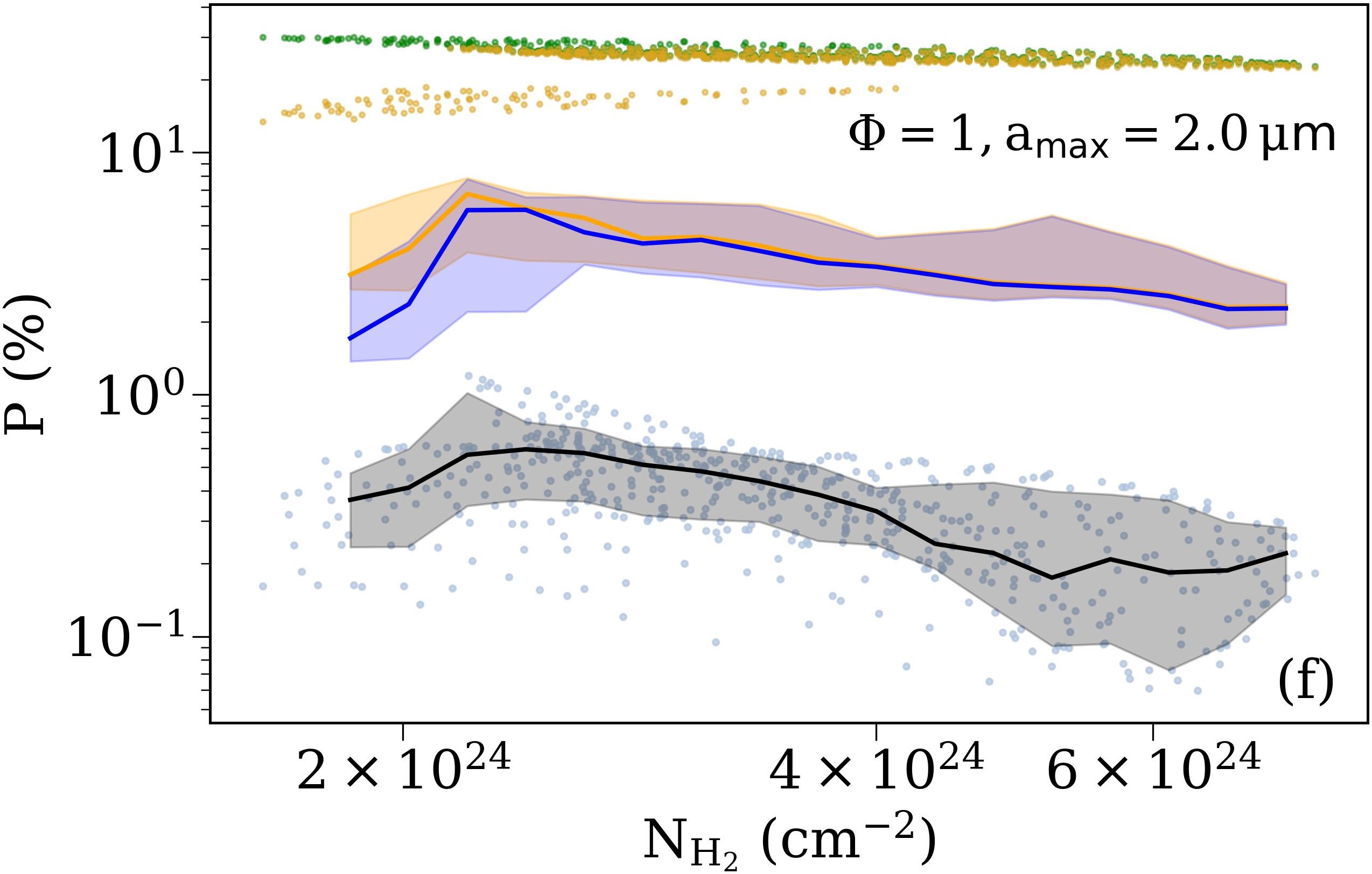} \includegraphics[width=5.9cm]{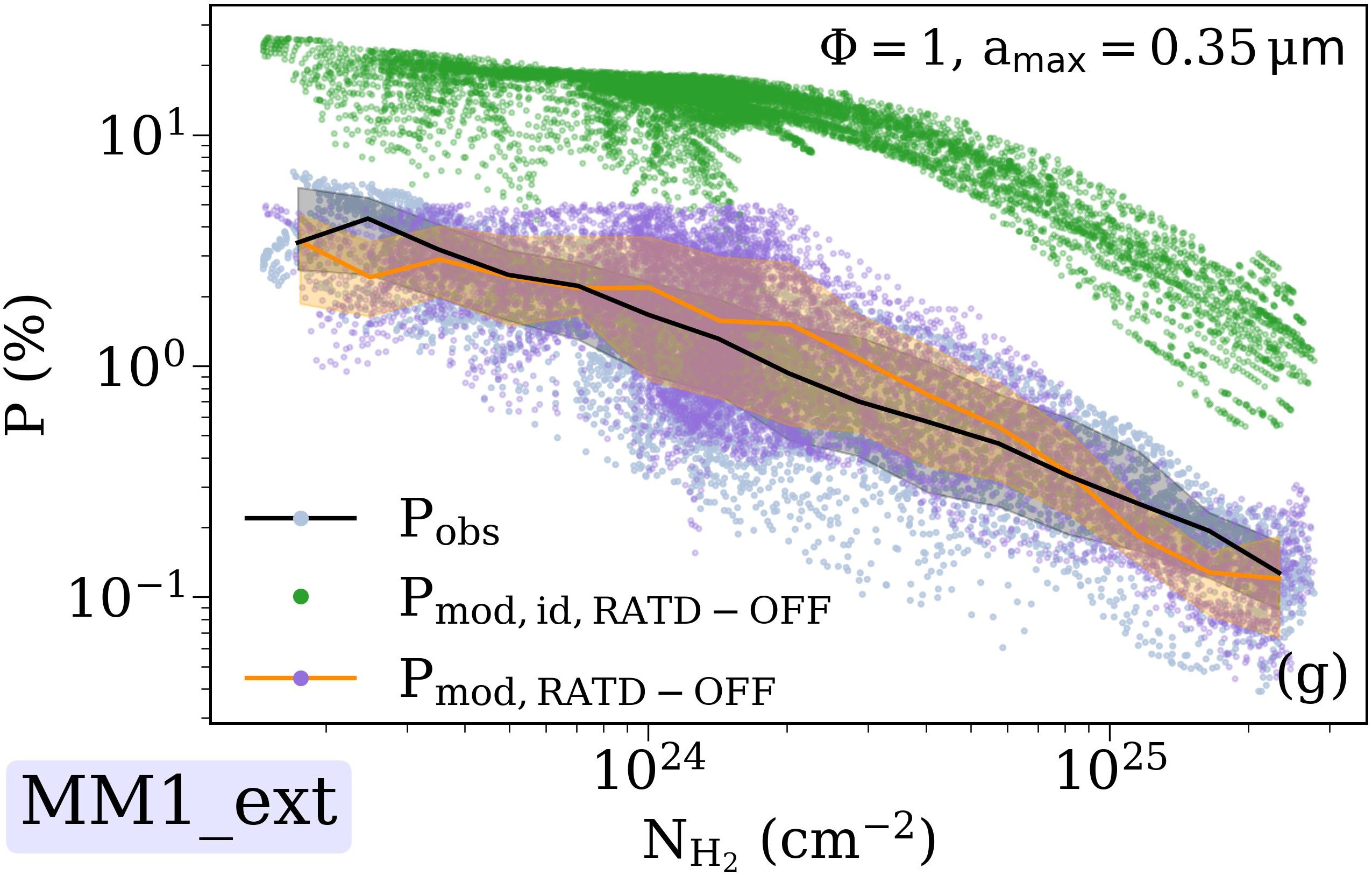} \includegraphics[width=5.9cm]{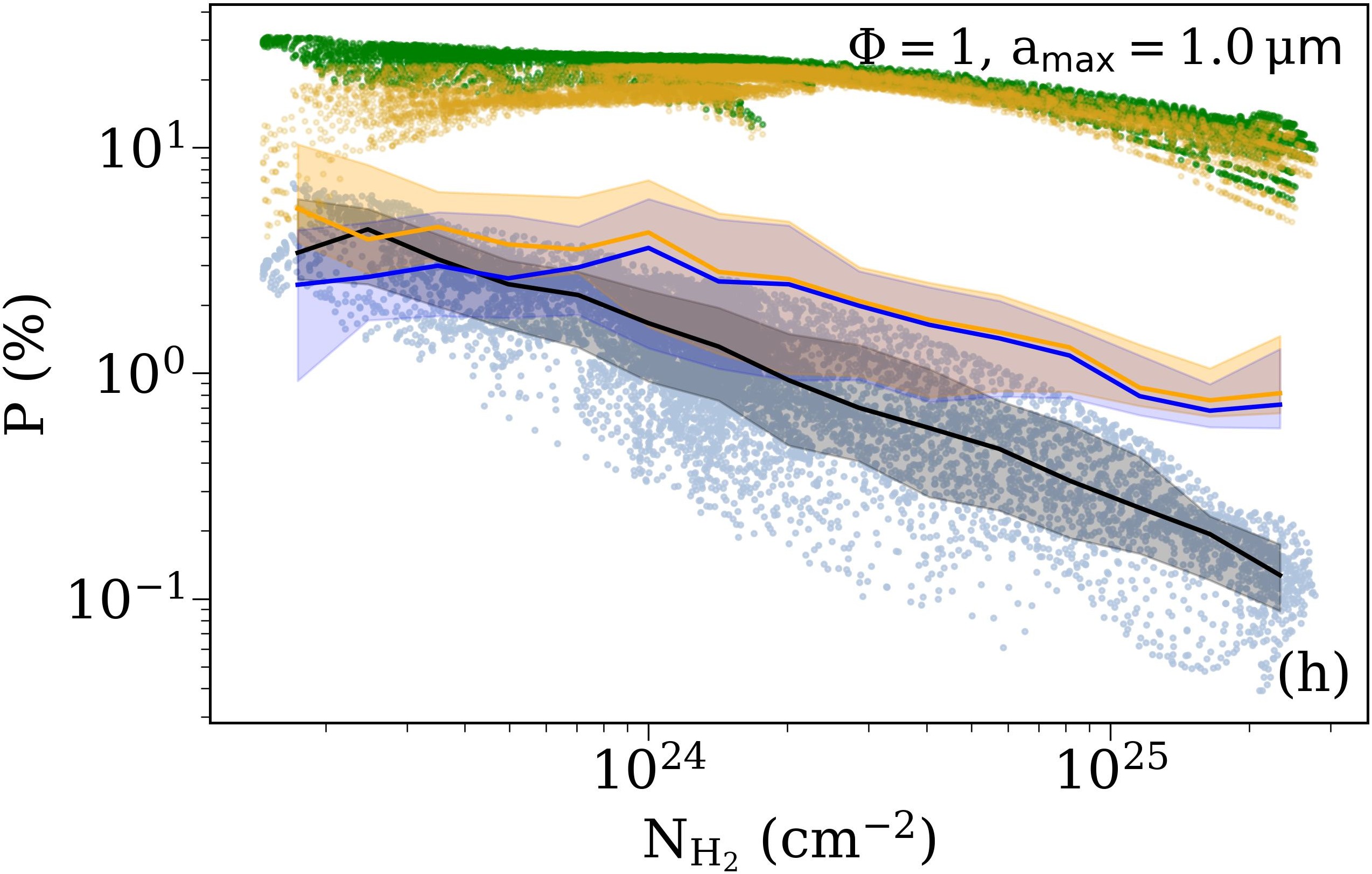} \includegraphics[width=5.9cm]{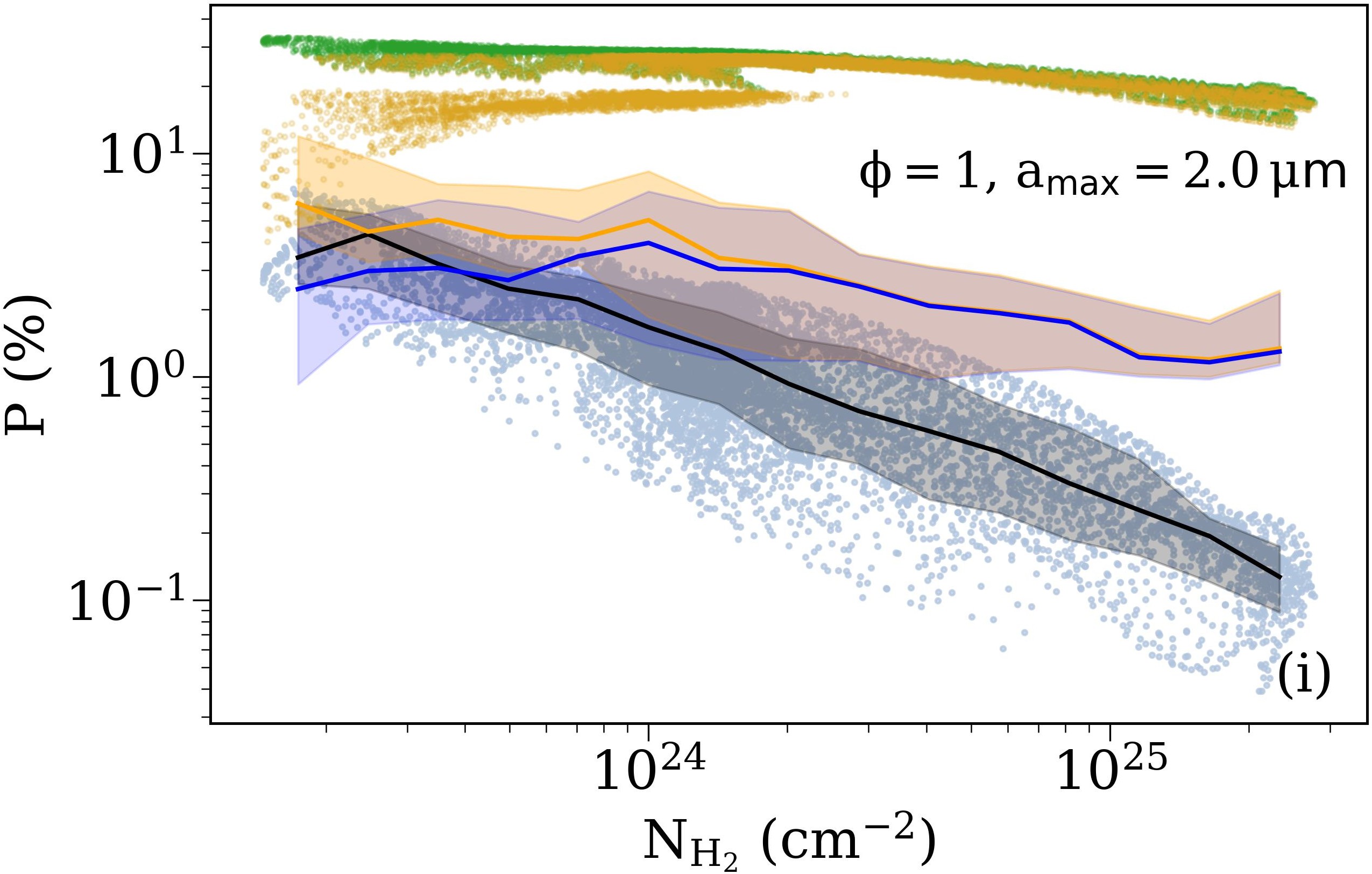} \includegraphics[width=5.9cm]{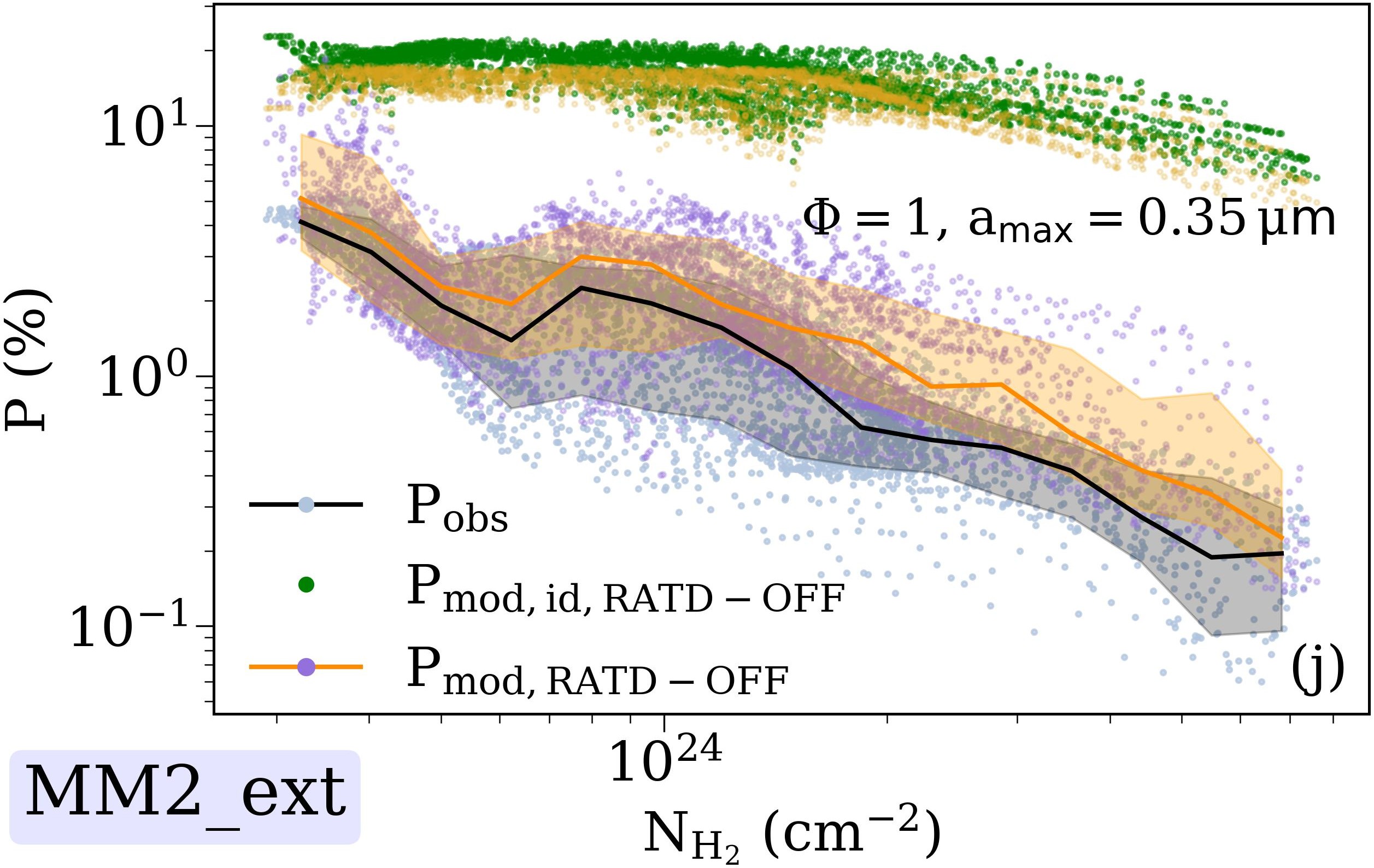} \includegraphics[width=5.9cm]{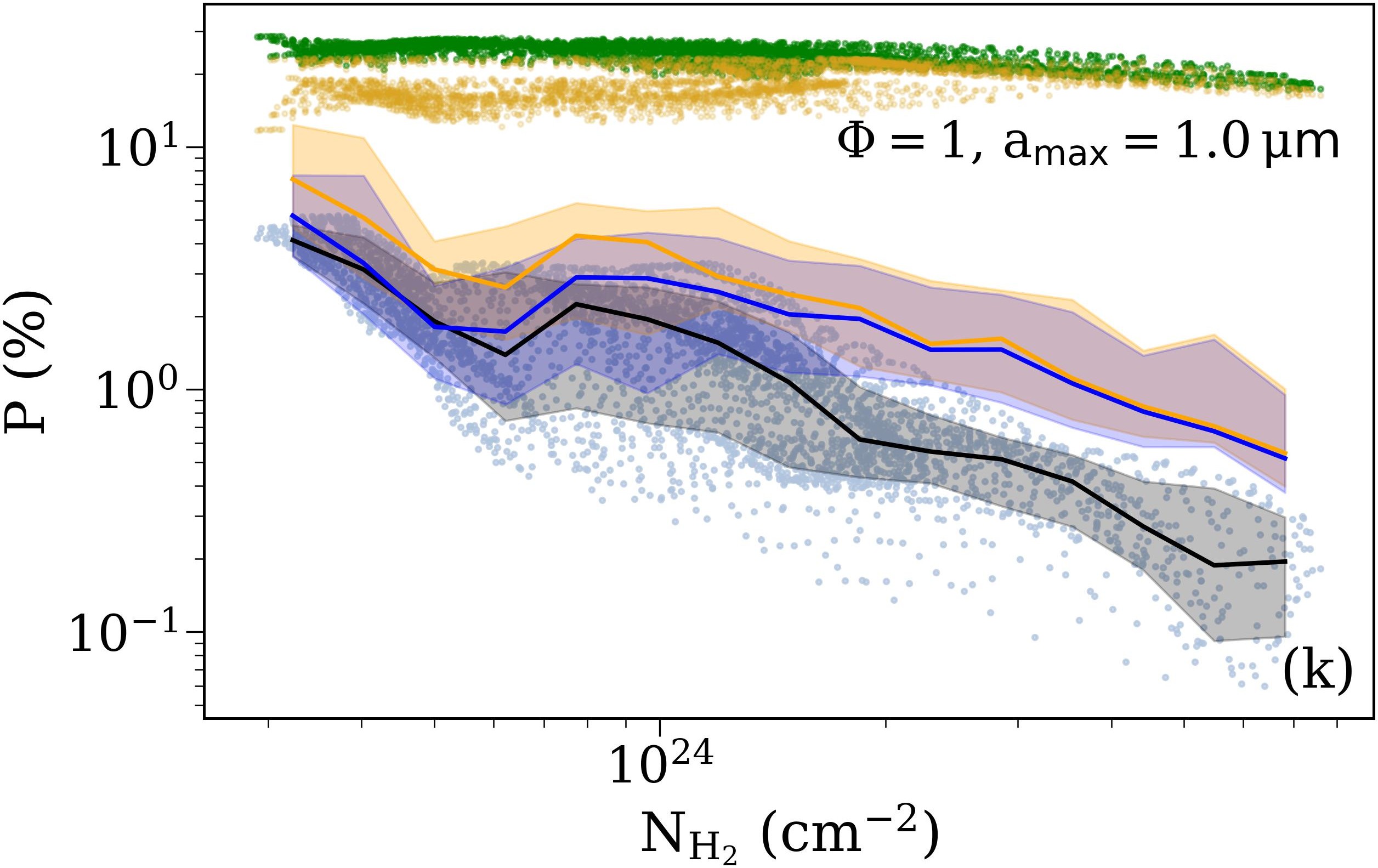} \includegraphics[width=5.9cm]{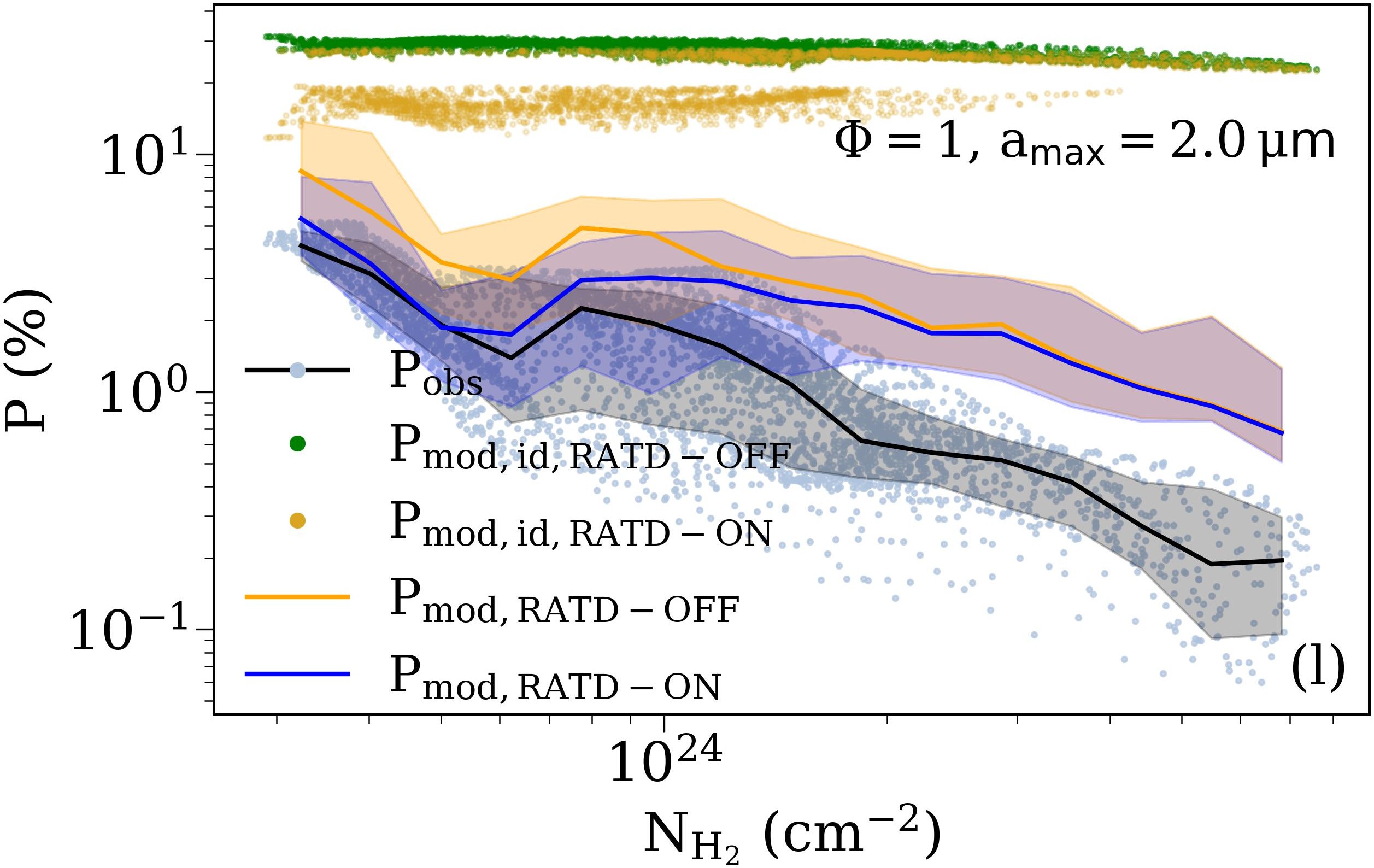}

    \caption{Model results for the MM1, MM2, MM1\_ext, and MM2\_ext regions, using different $a_{\mathrm{max}}$ values with fixed $\Phi=1.0$.}
    \label{Model_other_figures}
\end{figure*}

\begin{figure*}
    \centering
    \includegraphics[width=5.9cm]{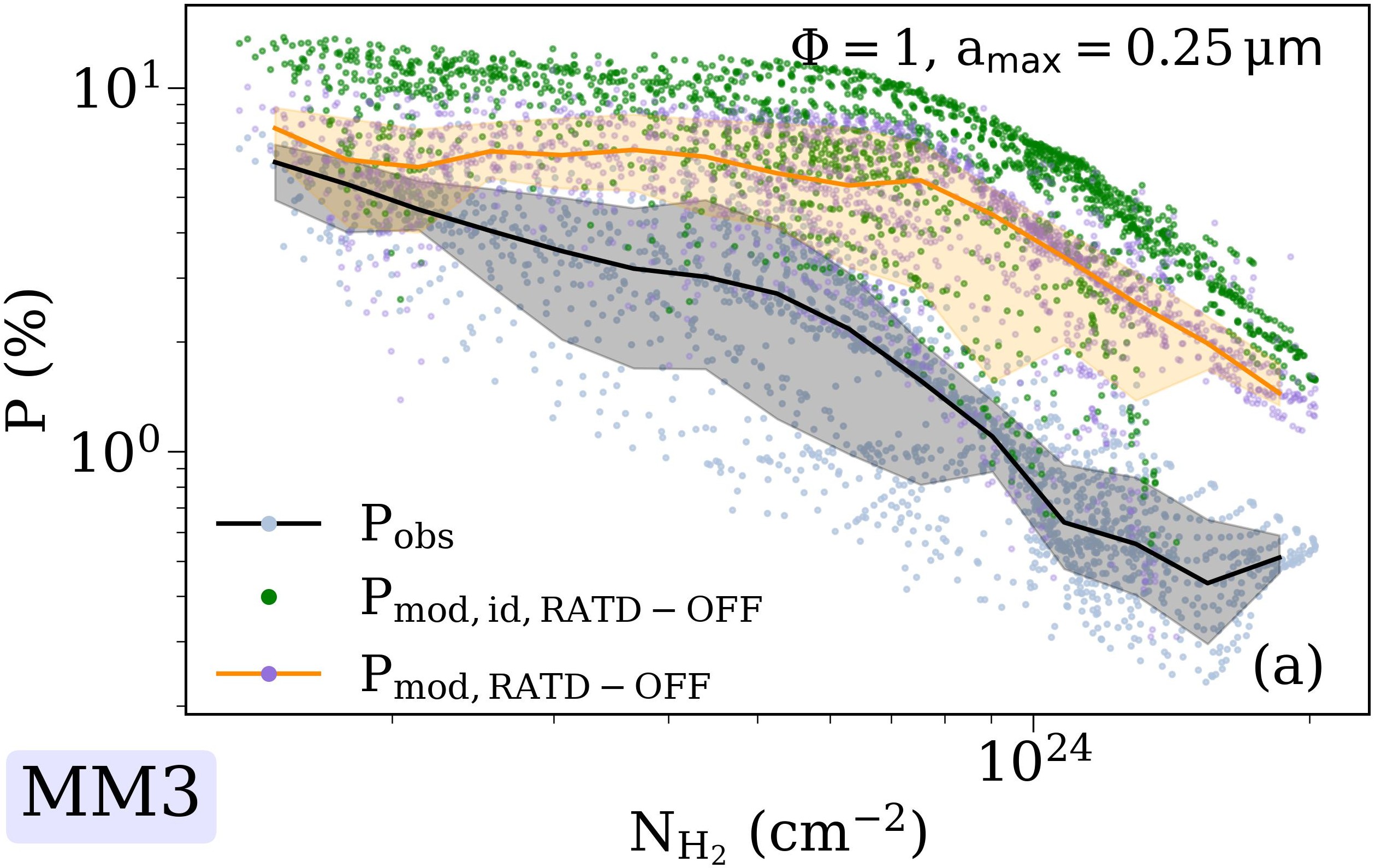}
    \includegraphics[width=5.9cm]{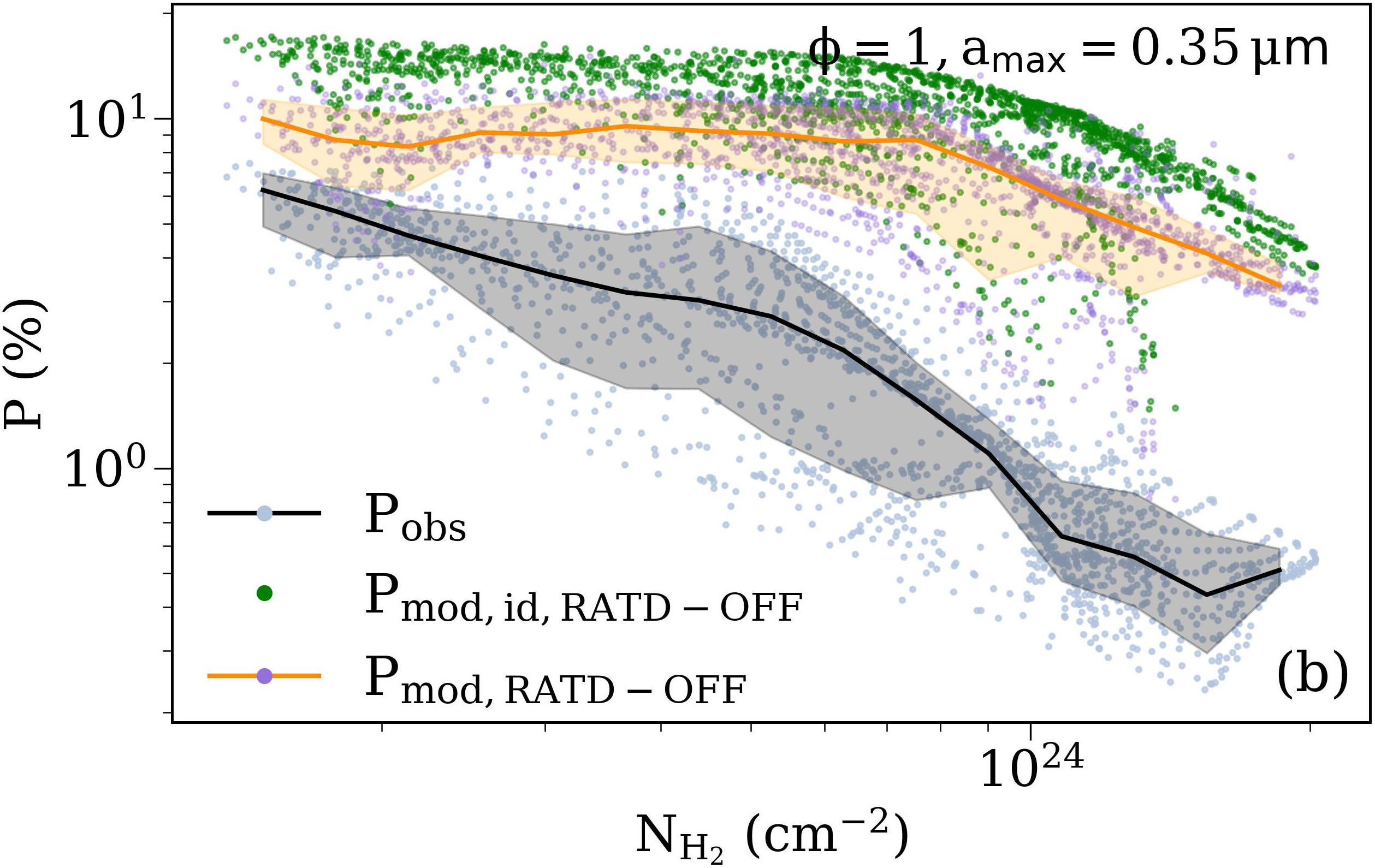}
    \includegraphics[width=5.9cm]{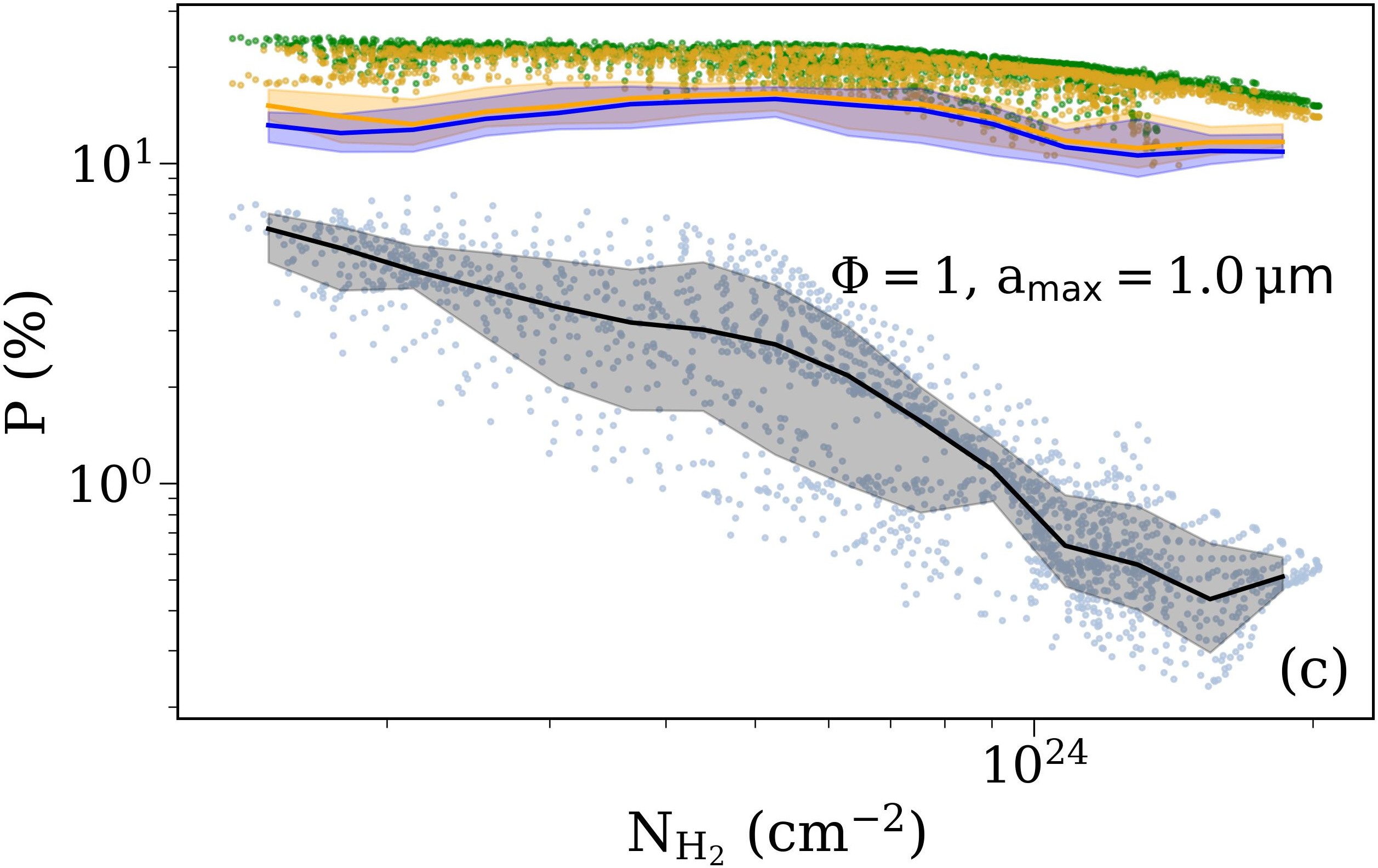}
    \includegraphics[width=5.9cm]{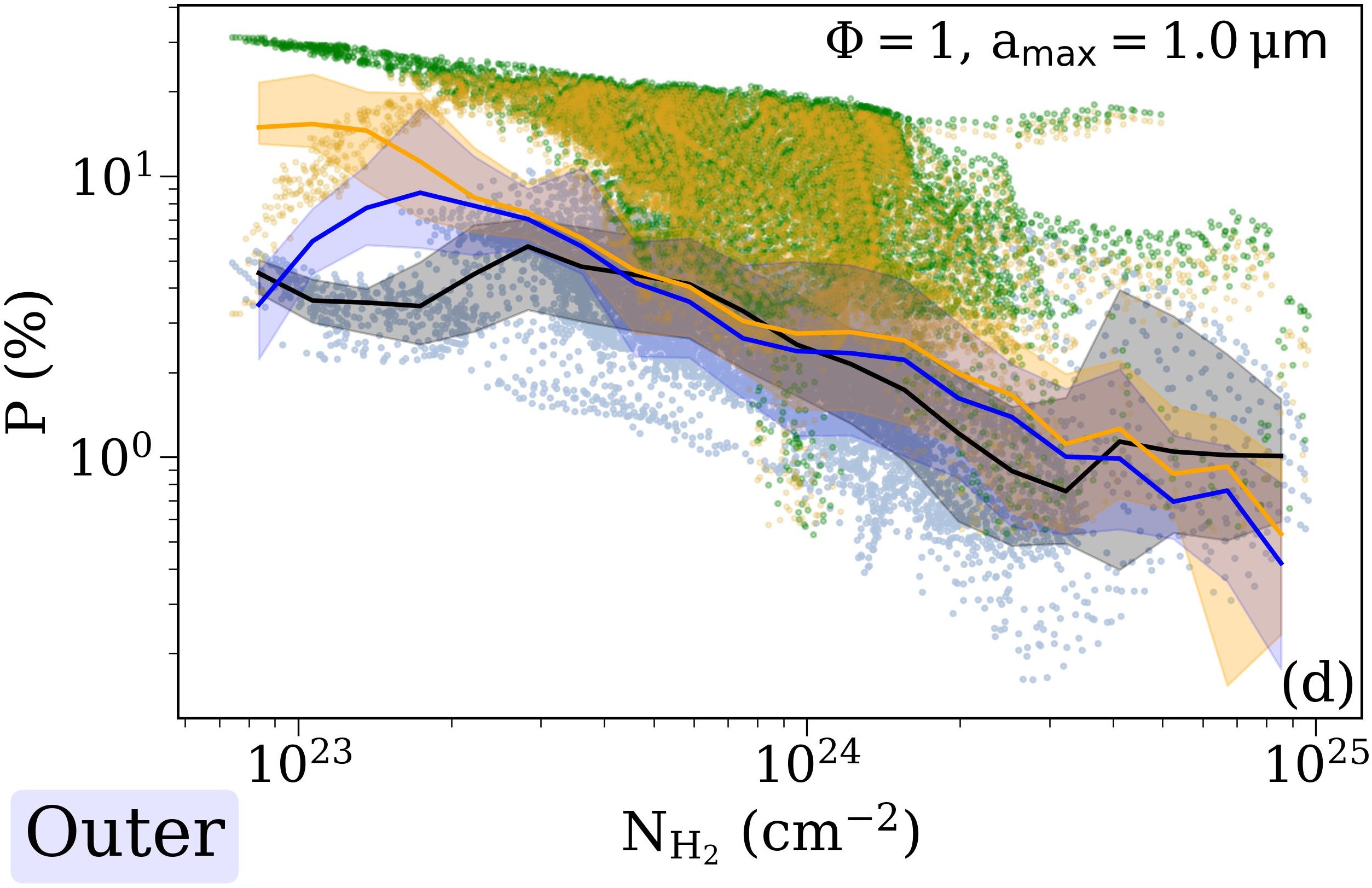}
    \includegraphics[width=5.9cm]{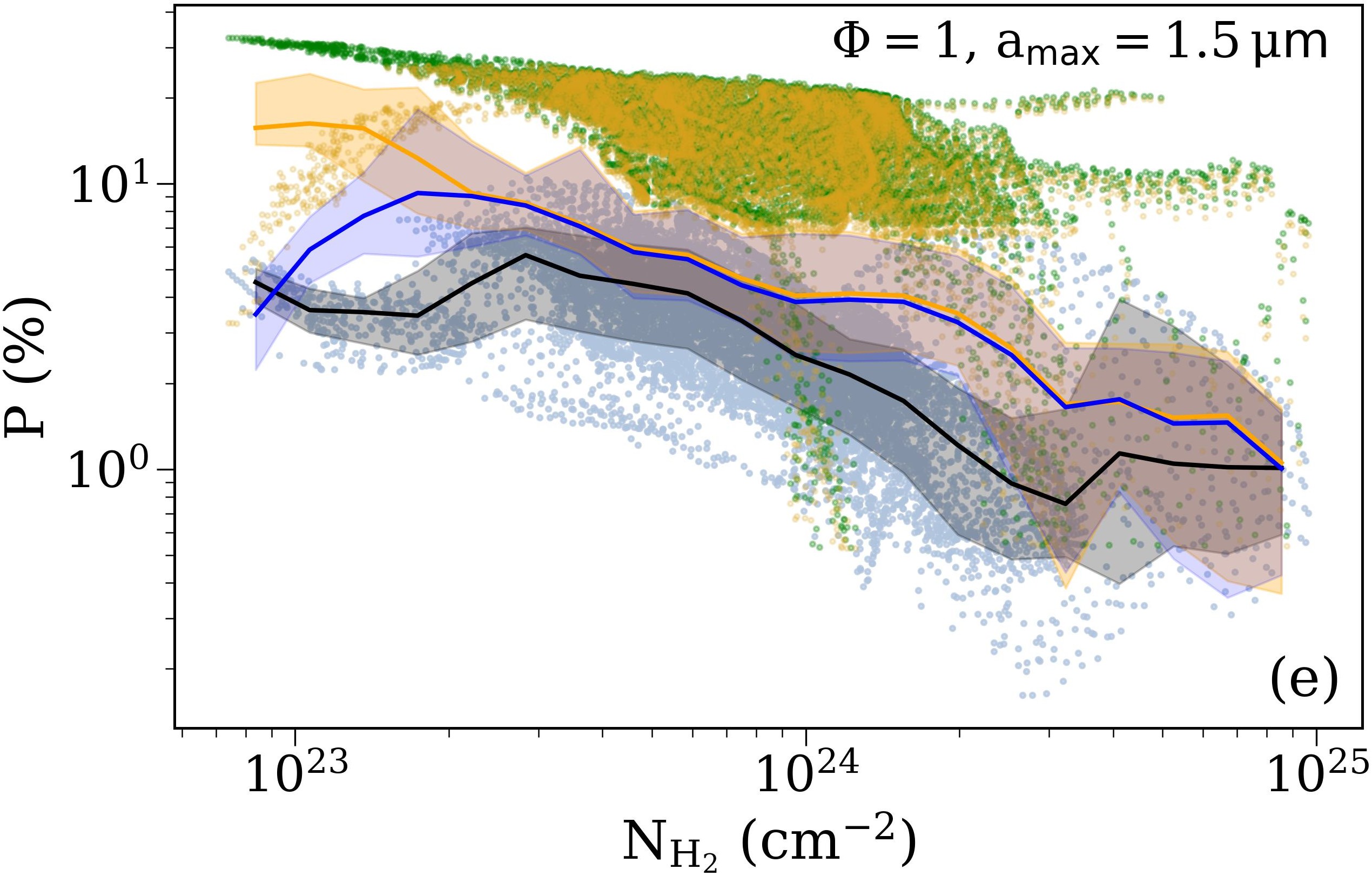}
    \includegraphics[width=5.9cm]{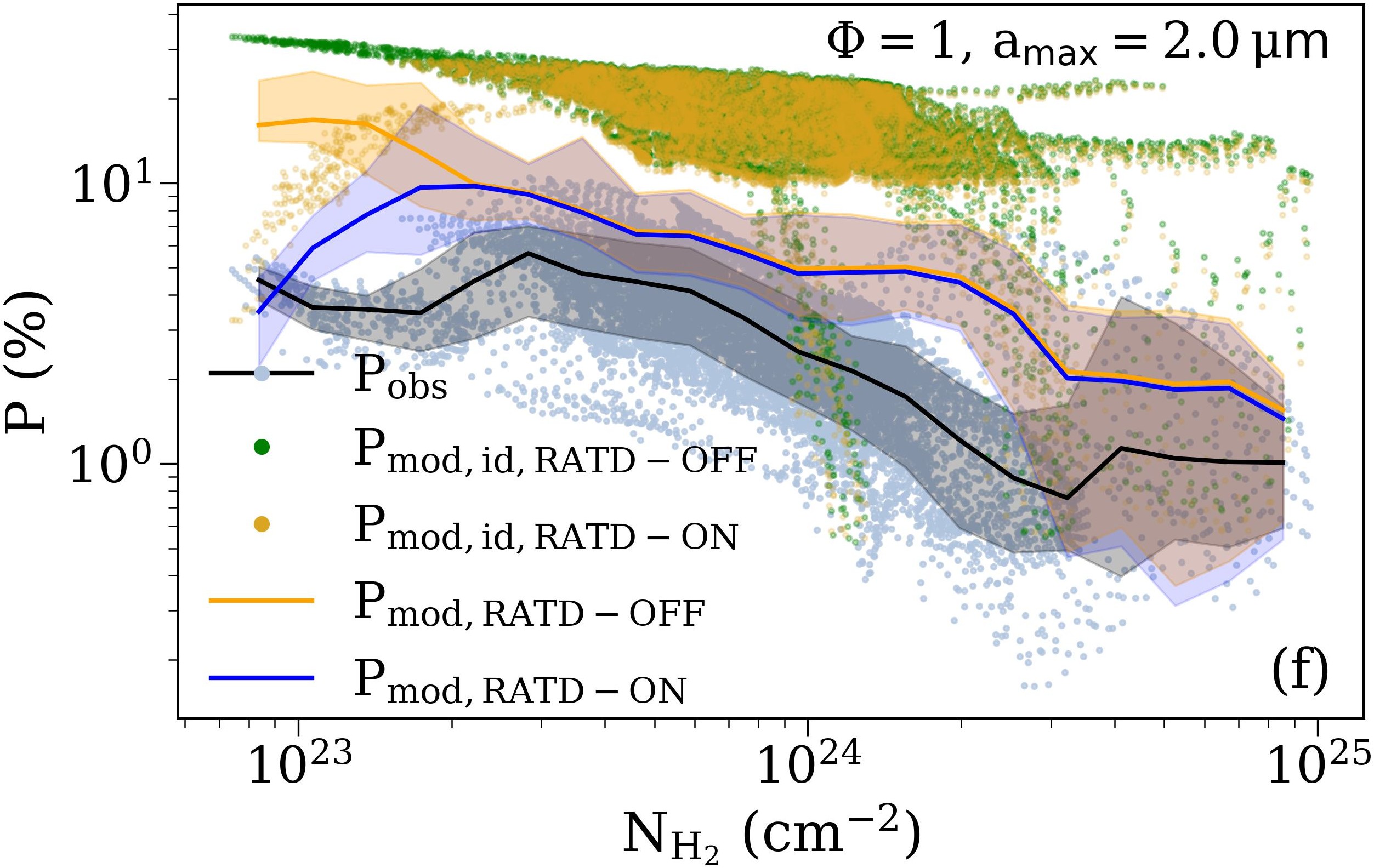}

    \caption{Same as Figure \ref{Model_other_figures}, but for the MM3 and Outer regions.}
    \label{Model_other_figures_2}
\end{figure*}

Figure \ref{Model_other_figures} and \ref{Model_other_figures_2} shows the comparison plot of variation in observed polarization fraction ($P_{\rm obs}$) and modeled polarization fraction ($P_{\rm mod}$), including ideal and realistic models, with the column density ($N_{\rm H_2}$) for different maximum aligned grain sizes ($a_{\mathrm{max}}$). 


\begin{figure*}
    \centering
    \includegraphics[width=8.5cm]{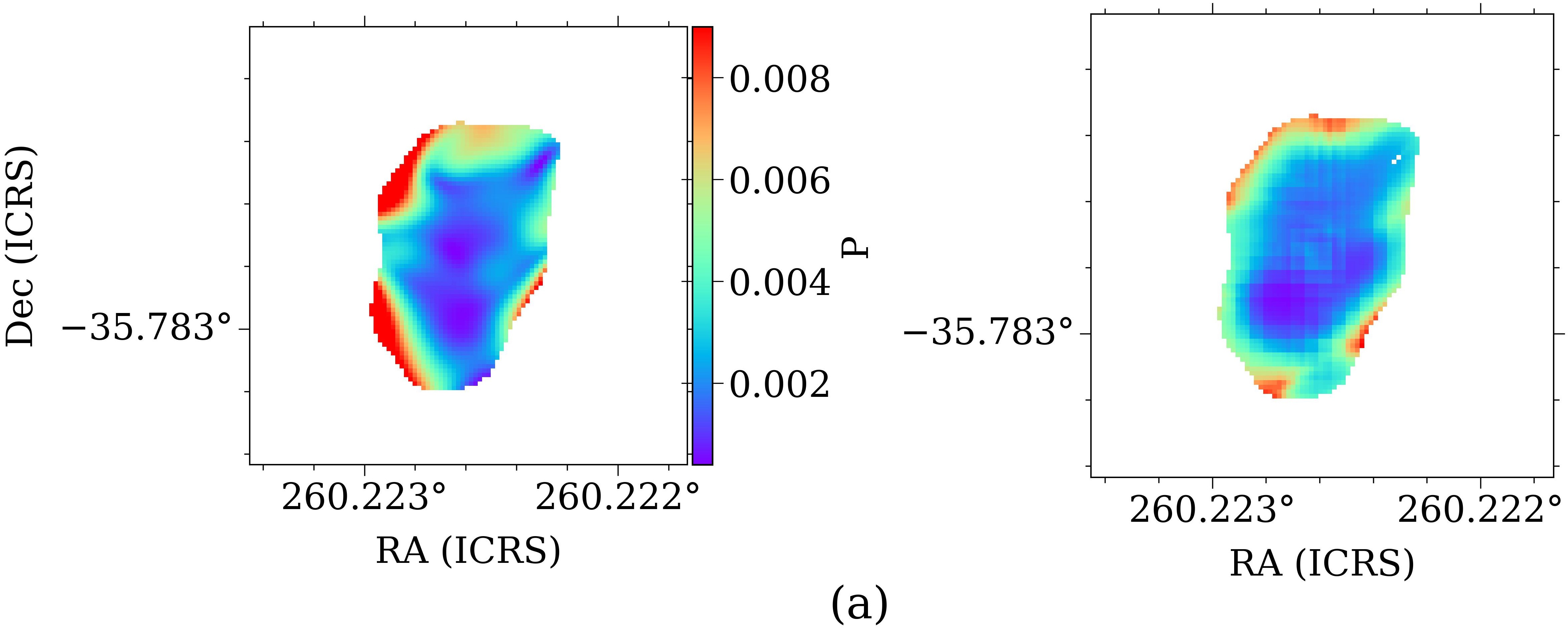}  
     \includegraphics[width=8.5cm]{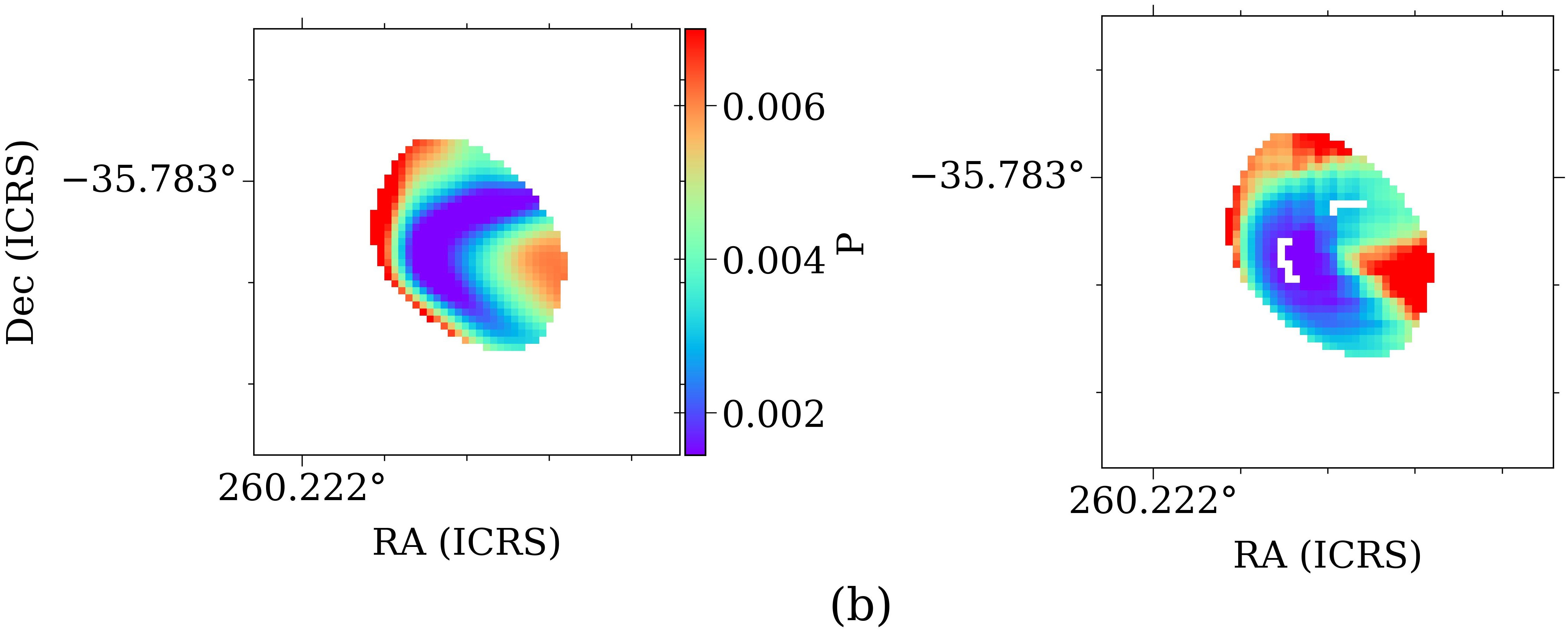}  
 \caption{Comparison of the observed ($\emph{left}$) and modeled ($\emph{right}$) polarization fractions for (a) MM1 and (b) MM2, using $a_{\rm max} = 0.50~\mu$m, which corresponds to the lowest RMSE in the polarization modeling (see Table \ref{tab:results}).}
    \label{P_model_comp_MM1_MM2}
\end{figure*}

Figure \ref{P_model_comp_MM1_MM2} presents images that show a comparison between the polarization fractions measured from the ALMA 1.2 mm observations and those predicted by the \texttt{DustPOL\_py} models for $a_{\mathrm{disr}} = 0.50\,\mu$m in MM1 and MM2 (see Section \ref{model_w_B-tang}). These modeled maps are presented only for the maximum aligned grain size that provides the best match to the observations, as determined by the minimum RMSE (see Table \ref{tab:results}).

\section{Optical depth in \ngc}
\label{optical_depth}

\begin{figure}
    \centering
    \includegraphics[width=8.5cm]{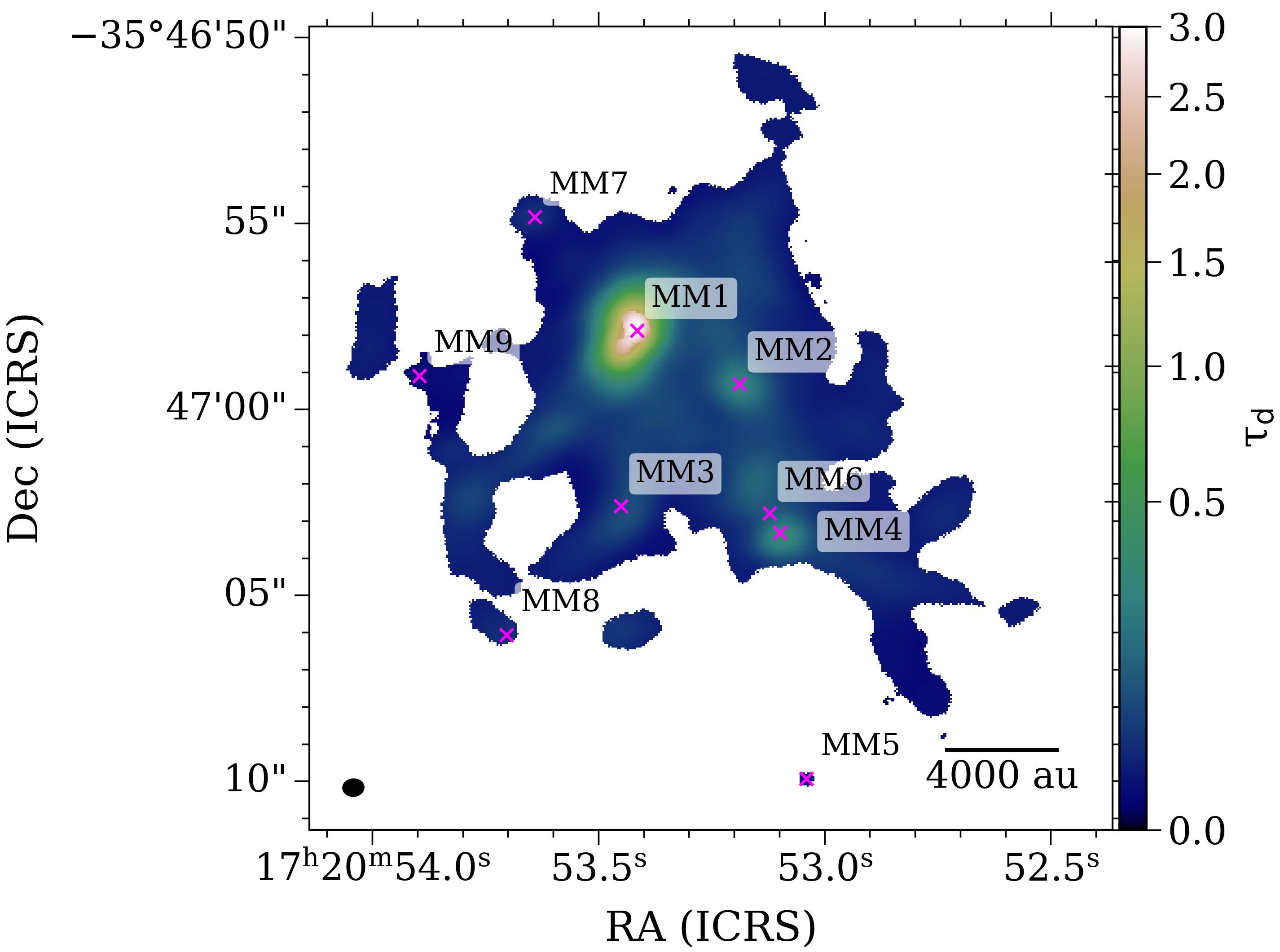}  
 \caption{Optical depth map of \ngc~at 1.2 mm.}
    \label{tau_map}
\end{figure}

The optical depth is calculated using the brightness temperature ($T_{\rm b}$) inferred from the dust emission flux from ALMA and the dust temperature based on $^{13}$CH$_3$OH rotational temperature, following $T_{\rm b} = J_\nu(T_{\rm d}) (1-e^{-\tau_\nu})$. Here, $J_\nu (T_{\rm d}) = \frac{h\nu/k}{(e^{h\nu/kT_{\rm d}} - 1)}$. First, we used the MagMaR ALMA Band 6 data to derive the optical depth map at 1.2 mm. In the central dense regions, however, Band 6 optical depths could not be reliably determined because $T_{\rm b}$ approaches $T_{\rm d}$, causing the optical-depth inversion to diverge. To recover the optical depths in these regions, we utilized ALMA Band 4 (2.2 mm) observations with a beam size of $\sim0.^{\prime\prime}90 \times 0.^{\prime\prime}64$, BPA of $\sim -68^\circ$, and a pixel size of $\sim0.^{\prime\prime}13$. The NaN pixels (only $\sim0.6\%$) in the Band 6 optical depth map were replaced with the corresponding values from the Band 4 optical depth map. The Band 4 optical depths were then scaled to 1.2 mm using a dust opacity spectral index of $\beta = 1.7$ found in MM1 \citep{Brogan_2016} to account for the difference in observing frequencies. We assumed a conservative 30\% uncertainty to account for uncertainties in temperature, opacity assumptions, and differences between the two datasets (e.g., $\emph{uv}$-sampling). Figure \ref{tau_map} shows the final optical depth map at 1.2 mm, in which the MM1 region is found to be optically thick ($\tau > 1$), while other regions are mostly optically thin ($\tau < 1$).

\section{Rotational disruption and grain growth timescale}
\label{timescales}
The timescale for rotational disruption (RAT-D) can be estimated as the time required for dust grains to spin up to the critical angular velocity for disruption, $\omega_{\rm{disr}}$ \citep{Hoang_2019b}

\begin{equation}
\begin{aligned}
t_{\rm disr} \simeq
10^{5}\,
U^{-1}
\left(\frac{\bar{\lambda}}{0.5~\mu{\rm m}}\right)^{1.7}
\left(\frac{S_{\rm max}}{10^{7}~{\rm erg~cm^{-3}}}\right)^{1/2}\\ \times
\left(\frac{a_{\rm disr}}{0.1~\mu{\rm m}}\right)^{-0.7}
~{\rm yr}.\\
\end{aligned}
\end{equation}
We estimated $t_{\rm disr}$ by taking $\bar{\lambda} = 1.0$ $\mu$m and the maximum tensile strength, $S_{\rm{max}}$ of $\sim10^5$ erg \cmq. The ratio of grain growth timescale to the free-fall timescale for peak grain sizes of 1 $\mu$m can be calculated as \citep{Hirashita_2013} 

\begin{equation}
    \frac{t_{\rm{grow}}}{t_{\rm{ff}}} = 5.5 \left(\frac{5}{S_{\rm{cross}}}\right) \left(\frac{n_{\rm H}}{10^5 \cmq}\right)^{-1/4},
\end{equation}
where $S_{\rm cross}=5$ is the enhancement factor for the collisional cross-section in the maximal coagulation model of grain growth \citep[for details, see][]{Hirashita_2013}. For the median density of $n_{\rm H}\sim6.4\times10^{8}~{\rm cm^{-3}}$ in MM1, we calculated $\frac{t_{\rm grow}}{t_{\rm ff}}$ to be around 0.61. By estimating the free-fall time in MM1 to be $\sim1.7\times10^{3}$ years, we derived the grain-growth timescale to be of the order of $10^{3}$ years.

\bibliography{myref}{}
\bibliographystyle{aasjournal}



\end{document}